\documentclass[aps,twocolumn,nofootinbib,groupedaddress,longbibliography]{revtex4-1}

\usepackage{amsmath,amssymb}
\usepackage{graphicx}
\usepackage{multirow,makecell}
\usepackage{bm}
\usepackage{mathtools}
\usepackage{dsfont}
\usepackage{amsfonts}
\usepackage{xcolor}
\usepackage[normalem]{ulem}
\usepackage{url}
\usepackage[makeroom]{cancel}
\usepackage[colorlinks=true,linkcolor=magenta,citecolor=magenta,hypertexnames=false]{hyperref}
\newcommand{\nocontentsline}[3]{}
\newcommand{\tocless}[2]{\bgroup\let\addcontentsline=\nocontentsline#1{#2}\egroup}
\def\ba#1\ea{\begin{align}#1\end{align}}
\def\bastar#1\eastar{\begin{align*}#1\end{align*}}
\def\bg#1\eg{\begin{gather}#1\end{gather}}
\def\bpm{\begin{pmatrix}}
\def\epm{\end{pmatrix}}
\newcommand{\nn}{\nonumber \\ }
\newcommand{\trm}[1]{\textrm{#1}}

\newcommand{\bs}[1]{\boldsymbol{#1}}
\newcommand{\bb}[1]{\mathbf{#1}}
\newcommand{\bx}{\bb x}
\newcommand{\bR}{\bb R}
\newcommand{\bk}{\bb k}
\newcommand{\kinv}{\cm{\bb k}}

\newcommand{\cm}[1]{\overline{#1}}
\newcommand{\Z}{\mathbb{Z}}
\newcommand{\N}{\mathbb{N}}
\newcommand{\mc}[1]{\mathcal{#1}}

\newcommand{\dg}{\dagger}

\newcommand{\sg}{\sigma}

\newcommand{\vev}[1]{\langle #1 \rangle}
\newcommand{\ket}[1]{| #1 \rangle}
\newcommand{\bra}[1]{\langle #1 |}

\newcommand{\ourtitle}{
Global and Local Topological Crystalline Markers for Rotation-Symmetric Insulators}
\allowdisplaybreaks

\begin{document}
\title{\textbf{\ourtitle}}
%%%%%%

%%%%%%Author
\author{Saavanth Velury}
\email{svelury2@illinois.edu}
\affiliation{Department of Physics and Anthony J. Leggett Institute for Condensed Matter Theory, University of Illinois at Urbana-Champaign, Urbana, Illinois 61801-3080, USA}

\author{Yoonseok Hwang}
\affiliation{Department of Physics and Anthony J. Leggett Institute for Condensed Matter Theory, University of Illinois at Urbana-Champaign, Urbana, Illinois 61801-3080, USA}

\author{Taylor L. Hughes}
\affiliation{Department of Physics and Anthony J. Leggett Institute for Condensed Matter Theory, University of Illinois at Urbana-Champaign, Urbana, Illinois 61801-3080, USA}
%%%%%%

%%%%%%Abstract
\begin{abstract}
Crystalline symmetry can be used to predict bulk and surface properties of topological phases.
For non-interacting cases, symmetry-eigenvalue analysis of Bloch states at high symmetry points in the Brillouin zone simplifies the calculation of topological quantities.
However, when open boundaries are present, and only the point group part of the symmetry group remains, it is unclear how to utilize crystalline symmetries to diagnose band topology.
In this work, we introduce topological crystalline markers to characterize bulk topology in $C_n$-symmetric ($n=2,3,4,6$) crystalline insulators and superconductors with and without translation symmetry.
These markers are expressed using a crystalline symmetry operator and the ground state projector, and are defined locally in position space.
First, we provide a general method to calculate topological markers in periodic systems with an arbitrary number of unit cells.
This includes cases where momentum quantization does not span all necessary high-symmetry points for computing the topological quantities, which we address using twisted boundary conditions.
Second, we map these markers to the Chern number, bulk polarization, and sector charge for two-dimensional $C_n$-symmetric insulators in symmetry classes A, AI, AII, and superconductors in class D.
Finally, we show how to numerically calculate the markers in finite-size systems with translation-symmetry (and even rotation-symmetry) breaking defects, and how to diagnose the bulk topology from the marker.
Our results demonstrate how to compute bulk topological crystalline invariants locally in position space, thereby providing broader scope to diagnosing bulk crystalline topology that works even in inhomogeneous systems where there is no global rotation symmetry.
\end{abstract}
%%%%%%

\maketitle

%%%%%%
\section{Introduction}
\label{sec:intro}
%%%%%%
Recently, the classification of non-interacting topological crystalline insulators has been established for the 230 nonmagnetic, and 1651 magnetic, space groups~\cite{Bradlyn2017,Po2017,Kruthoff2017,Bradlyn2018,Cano2018a,Song2018a,Khalaf2018,Ono2018,Po2020,Cano2021,Elcoro2021,Ono2021,Zhang2021,Shiozaki2022,Wieder2022,Fu2007a,Turner2010,Hughes2011,Fu2011,Fang2012,Hsieh2012,Chiu2013,Jadaun2013,Slager2013,Benalcazar2014,Fang2014,Shiozaki2014,Fang2015,Alexandradinata2016,Chiu2016}.
For the efficient diagnosis of such insulators, lattice translation symmetry plays an essential role.
In fact, symmetry indicator methods and topological quantum chemistry use the irreducible, crystalline-symmetry representations of the electronic bands at high symmetry momenta within the Brillouin zone~\cite{Kruthoff2017,Bradlyn2017,Po2017,Po2020,Cano2021}.
For example, for an inversion and time-reversal symmetric insulator in two dimensions, the number of occupied, odd-parity at each high symmetry momentum (modulo 2) indicates the $\Z_2$ topological invariant~\cite{Fu2007a}.
We call such information {\it momentum-space data}.
Momentum-space data is also useful for predicting the boundary signatures of crystalline insulators, e.g., the appearance of edge or corner states, or fractional boundary charge~\cite{Benalcazar2017,Benalcazar2017a,Song2017,Miert2018,Benalcazar2019,Schindler2019,Hwang2019,Hirayama2020,Tanaka2020a,Fang2021,Takahashi2021,Schindler2022,Tanaka2022,Naito2022,Tanaka2023,Vaidya2023,Wada2024}.

Given the success of these classification approaches that leverage momentum-space data to determine position-space, e.g., boundary, properties, one natural question is whether it is possible to characterize the topological properties of the ground state of an insulating phase {\it solely} in position-space.
This is an important question because determining momentum space data requires translation symmetry, and translation symmetry can be broken in many cases.
For example, a given system can be confined within a domain, have open boundary conditions, or be subjected to
crystal defects such as vacancies or dislocations.
Even with periodic boundary conditions and perfect translation symmetry, there are cases where one cannot access enough momentum-space data to compute the symmetry indicators for finite-sized systems.
For example, the momentum quantization grid for finite systems depends on the linear sizes of the system, and hence some high-symmetry points may not be defined for particular system sizes.

Considering these challenges, we plan to develop a position-based method that can classify crystalline topology and connect to boundary signatures.
There have been several approaches in this direction.
For two-dimensional Chern insulators, the real-space Chern marker can determine the Chern number by computing the local Chern number in each unit cell~\cite{Kitaev2006,Bianco2011,Bau2024}.
Another approach is the spectral localizer technique~\cite{Loring2019,LozanoViesca2019,Cerjan2022,Cerjan2023a,Cerjan2024}, which can predict the appearance of topological surface states~\cite{Cerjan2024}.
However, these techniques do not address, or fully exploit, the crystalline symmetries.
Furthermore, it is unclear how other bulk topological quantities protected by crystalline symmetry can be determined using these methods.

In this work, we instead formulate bulk topological crystalline invariants in terms of basis-independent, operator-based quantities known as {\it topological crystalline markers}.
As the name suggests, topological crystalline markers~\cite{MondragonShem2024} are topological markers, which locally measure bulk topological invariants on the position-space lattice.
The focus of this work will be to construct topological markers that incorporate the crystalline symmetry, namely the $C_n$ rotation symmetry for $n=2,3,4$ and $6$, for gapped phases with spinless or spin-1/2 fermions, with or without time-reversal symmetry or particle-hole symmetry (i.e., Altland-Zirnbauer (AZ) symmetry classes A, AI, AII, and D~\cite{Altland1997,Kitaev2009}).

The topological crystalline markers we discuss in this work are constructed from projected symmetry operators, i.e., crystalline symmetry operators projected onto the subspace of occupied energy states (i.e., the full ground state) of a given system.
By expressing the polarization and sector charge of obstructed atomic insulators, and, separately, the Chern number in Chern insulators and topological crystalline superconductors in terms of these projected symmetry operators, we are able to spatially resolve these quantities on the crystalline lattice, and show that their support is localized at the symmetry centers of the lattice.
Importantly, compared to the existing Chern markers~\cite{Kitaev2006,Bianco2011,Bau2024} that can diagnose the exact Chern number, our crystalline markers can efficiently determine the polarization, corner charge, etc., but can determine the Chern number only modulo $n$ for $C_n$ symmetry.
So, while our method has advantages over previous spatially-resolved topological markers (which we will discuss in more detail within), we sacrifice precision in the determination of the Chern number.

Remarkably, we will see that our method also provides a new way of understanding previously established results.
For example, the bulk Chern number and the bulk polarization can be determined precisely in the presence of interactions and/or disorders when employing twisted boundary conditions\cite{Niu1985,Watanabe2018b}.
However, these methods require global consideration of the phase space spanned by the angles parameterizing the twisted boundary condition.
Furthermore, they do not leverage the crystalline symmetries of the Hamiltonian, if present.
Our method involving the topological crystalline markers provides a simpler representation for both the bulk Chern number and the bulk polarization when that can be resolved locally in position-space.
Indeed, our method is applicable to systems that do not have translational symmetry, and even cases that are not globally point-group symmetric.

This work is organized as follows.
First, in Sec.~\ref{sec:notation}, we layout the notation used in this article for the rotation operators and topological crystalline markers.
Then we discuss the properties of $C_n$-symmetric lattices and Brillouin zones.
In Sec.~\ref{sec:tcm_defs}, we define the topological crystalline markers and detail their basis-independent properties.
We first demonstrate how the markers can be decomposed in terms of momentum-space based topological invariants when translation symmetry is present (e.g., irreducible representation multiplicities of the Bloch states, and rotation invariants).
We then describe how to spatially resolve these quantities on the position-space lattice.
Following this, in Sec.~\ref{sec:map_perfect}, we construct a mapping between bulk quantities and the topological crystalline markers for lattices having periodic boundary conditions and a number of unit cells that allows for the maximal number of high-symmetry momenta to exist in the Brillouin zone.
In Sec.~\ref{sec:map_general}, we generalize this mapping to lattices which have no restriction on the number of unit cells in each lattice direction, and thus could have quantized-momentum grids that miss high symmetry momenta.
For such lattices, the traditional diagnosis of bulk crystalline topology via momentum-space based symmetry indicators does not hold.
However, we show that the marker formalism we develop can still be applied to any lattice by considering twisted boundary conditions in addition to untwisted periodic boundary conditions.
Next, in Sec.~\ref{sec:example}, we consider some applications of the topological crystalline markers to tight-binding models realizing phases having Chern number quantized by $C_{2}$ symmetry, as well as obstructed atomic limit phases that have well-defined bulk polarization.
Importantly, we show how the markers can be used to diagnose bulk crystalline topology in systems hosting spatial inhomogeneities such as domain walls separating regions having different bulk topological properties.
This is an exciting application since it allows us to treat systems that are not globally crystalline symmetric, but which have a sense of crystalline symmetry only locally.
Finally, in Sec.~\ref{sec:conclusion}, we conclude this work by reviewing our results, as well as mentioning possible applications of topological crystalline markers to strongly interacting systems and other phases such as amorphous systems and quasicrystals.
%%%%%%

%%%%%%
\section{Overview of notation and properties of $C_n$-symmetric lattices and Brillouin zone}
\label{sec:notation}
\begin{figure}[t!]
\centering
\includegraphics[width=0.45\textwidth]{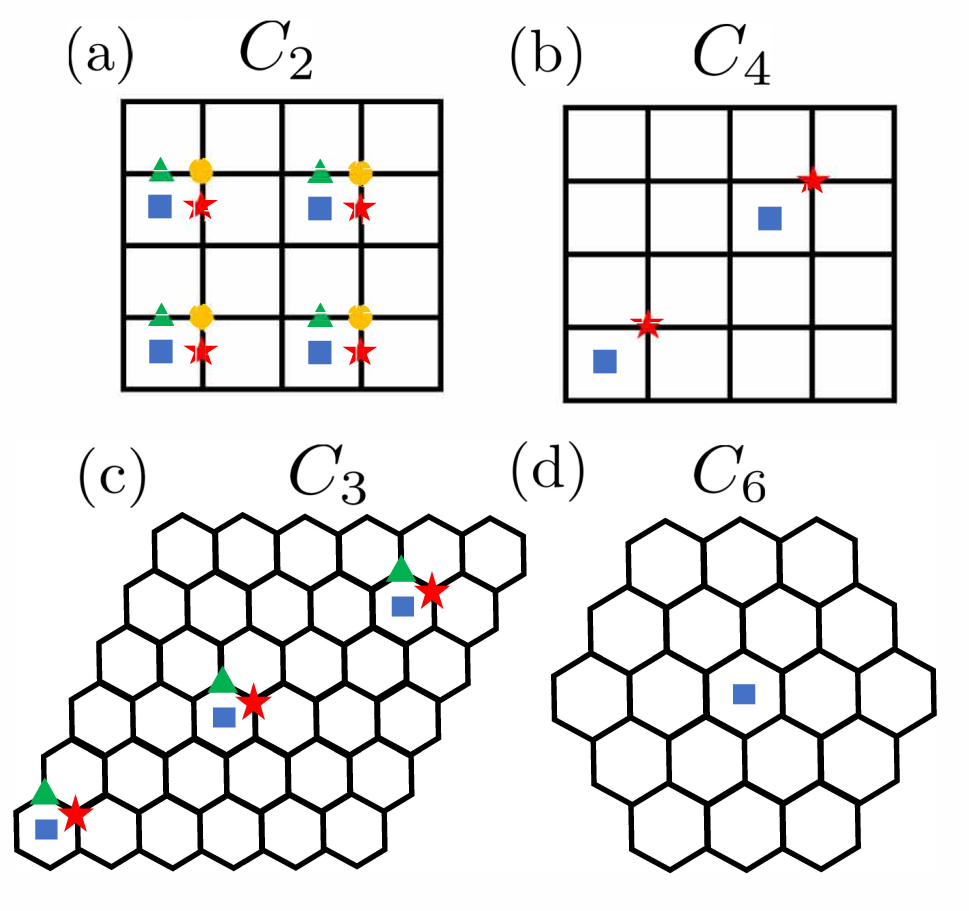}
\caption{
Illustration of (a) $C_2$-symmetric, (b) $C_4$-symmetric, (c) $C_3$-symmetric, and (d) $C_6$-symmetric lattices and the corresponding sets of invariant positions $\mc{X}[c_n(\bb r_0)]$ given by Eq.~\eqref{eq:invariant_positions} for lattices with dimensions $N_1 \times N_2$ specified by Eq.~\eqref{eq:perfect_constraint}.
The set of invariant positions $\mc{X}[c_n(\bb r_0)]$ for each $C_n$-symmetric lattice is determined with respect to the unit cell at $\bb R = \bb 0$ located at the lower left corner for the $C_2$, $C_3$, and $C_4$ lattices shown in (a)-(c) respectively, while the unit cell at $\bb R = \bb 0$ is located at the center of the $C_6$-symmetric lattice in (d).
For each lattice depicted in (a)-(d), the invariant positions correspond to WPs of multiplicity $1$, and when the dimensions $N_1 \times N_2$ satisfy Eq.~\eqref{eq:perfect_constraint}, the set $\mc{X}[c_n(\bb r_0)]$ is formed from Wyckoff positions of the same symbol (i.e., same shape and color) as per Eq.~\eqref{eq:invariant_positions}.
The different choices of $\bb r_0$ correspond to different symbols.
}
\label{fig:invariant_positions}
\end{figure}

\begin{figure}[t!]
\centering
\includegraphics[width=0.48\textwidth]{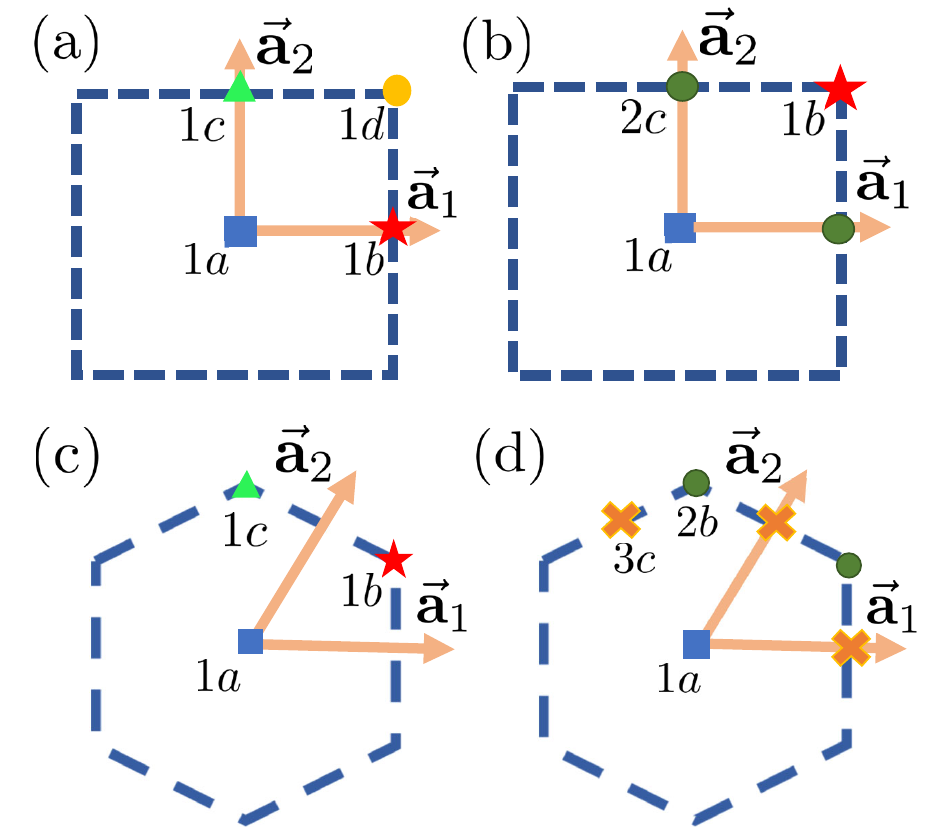}
\caption{
Illustration of a unit cell for each $C_n$-symmetric lattice and the corresponding Wyckoff positions (WPs).
(a) $C_2$ symmetry.
The WPs $1a,1b,1c,1d$ are located at $\bx_{1a} = \bb 0$, $\bx_{1b} = \frac{1}{2} \bb a_1$, $\bx_{1c} = \frac{1}{2} \bb a_2$, and $\bx_{1d} = \frac{1}{2} (\bb a_1 + \bb a_2)$, respectively.
(b) $C_4$ symmetry.
The WPs $1a$ and $1b$ are located at $\bx_{1a} = \bb 0$ and $\bx_{1b} = \frac{1}{2}(\bb a_1 + \bb a_2)$, respectively.
The WP $2c$ is composed of two $C_4$-related positions, $\bx_{2c,1} = \frac{1}{2} \bb a_1$ and $\bx_{2c,2} = \frac{1}{2} \bb a_2$.
(c) $C_3$ symmetry.
The WPs $1a,1b,1c$ are located at $\bx_{1a} = \bb 0$, $\bx_{1b} = \frac{1}{3} (\bb a_1 + \bb a_2)$, and $\bx_{1c} = \frac{1}{3} (-\bb a_1 + 2 \bb a_2)$, respectively.
(d) $C_6$ symmetry.
The WP $1a$ is located $\bx_{1a} = \bb0$.
The WP $2b$ is composed of $\bx_{2b,1} = \frac{1}{3} (\bb a_1 + \bb a_2)$, $\bx_{2b,2} = \frac{1}{3} (-\bb a_1 + 2 \bb a_2)$.
The WP $3c$ is composed of $\bx_{3c,1} = \frac{1}{2} \bb a_1$, $\bx_{3c,2} = \frac{1}{2} \bb a_2$, and $\bx_{3c,3} = \frac{1}{2} (- \bb a_1 + \bb a_2)$.
The primitive lattice vectors $\bb a_{1,2}$ are set as $\bb a_1 = (1,0)$ and $\bb a_2 = (0,1)$ for $C_{2,4}$, and $\bb a_1 = (1,0)$ and $\bb a_2 = (\tfrac{1}{2}, \tfrac{\sqrt{3}}{2})$ for $C_{3,6}$.
}
\label{fig:unitcells}
\end{figure}

%%%%%%
We begin by reviewing some properties of lattices that have rotational symmetry, and the notation that will be used throughout this work.
For now we will focus on 2D lattices of finite-size that are $C_n$-symmetric ($n=2,3,4,6$) and have periodic boundary conditions.
Figure~\ref{fig:invariant_positions} illustrates the different $C_n$-symmetric lattices.
The lattices shown in Fig.~\ref{fig:invariant_positions} are spanned by a choice of lattice vectors $\bb a_1$ and $\bb a_2$.
The location of any unit cell on the position-space lattice is specified by a Bravais lattice vector $\bR = n_1 \bb a_1 + n_2 \bb a_2$ where $n_{1,2}$ are non-negative integers.
The dimensions $N_1, N_2$ of the lattice are specified by the number of unit cells along each direction spanned by the primitive lattice vectors $\bb a_1$ and $\bb a_2$ respectively.
The Wyckoff positions (WPs) within each unit cell will play an important role in our discussion and are shown in Fig.~\ref{fig:unitcells}.
%

%%%%%%
%%%%%%

For a given $C_n$-symmetric lattice having periodic boundary conditions along the directions spanned by $\bb a_{1,2}$, there is a set of positions that are invariant under the $C_n$ rotation as shown in Fig.~\ref{fig:invariant_positions}.
To define the set of invariant positions, let us consider a $C_n$ rotation axis located at $\bb r_o$, and denote the rotation operation as $c_n(\bb r_o)$.
The position $\bb r_o$ which specifies the location of a $C_n$ rotation axis is always given by a multiplicity $1$ WP, hence there are multiple, inequivalent choices marked with different symbols/colors in Fig.~\ref{fig:invariant_positions}.
Because of periodic boundary conditions, there can be more than one fixed point of the lattice under rotation, as shown by the matching symbols near different unit cells in Fig.~\ref{fig:invariant_positions}.
The set of invariant positions corresponding to $n=2,3,4$ and $6$ $C_n$-symmetric lattices respectively, is given by:
\ba
\mc{X}[c_2(\bb r_o)]
=& \Bigg\{\bb r_o, \frac{N_1}{2} \bb a_1 + \bb r_o, \frac{N_2}{2} \bb a_2 + \bb r_o,
\nn
\quad & \frac{N_1}{2} \bb a_1 + \frac{N_2}{2} \bb a_2 + \bb r_o \Bigg\},
\nn
\mc{X}[c_3(\bb r_o)]
=& \left\{ \bb r_o, \frac{N}{3} (\bb a_1 + \bb a_2) + \bb r_o, \frac{2N}{3} (\bb a_1 + \bb a_2) + \bb r_o \right\},
\nn
\mc{X}[c_4(\bb r_o)]
=& \left\{ \bb r_o, \frac{N}{2} (\bb a_1 + \bb a_2) + \bb r_o \right\},
\nn
\mc{X}[c_6(\bb r_o)]
=& \{ \bb r_o \},
\label{eq:invariant_positions}
\ea
where we assumed $N_1=N_2=N$ for $C_{3,4,6}$.

As a simple example to demonstrate how the sets of invariant positions can be determined, we briefly explain how to determine $\mc{X}[c_2(\bb r_o)]$.
Under the action of $c_2(\bb r_o)$, a position $\bb r$ on the lattice is mapped to $-\bb r + 2\bb r_o$.
The fixed points $\bb r_*$ comprising the set of invariant positions are determined from the condition $\bb r_* = -\bb r_* + 2\bb r_o$.
Solving this equation for a $C_2$-symmetric lattice having periodic boundary conditions along both directions yields the set of four positions specified by $\mc{X}[c_2(\bb r_o)]$.
Similar reasoning holds for determining $\mc{X}[c_n(\bb r_o)]$ for $n=3,4,6$.
Noting that $\bb r_o$ always denotes the location of a multiplicity $1$ WP, one can determine the allowed values of $\bb r_o$ for each $C_n$-symmetric lattice.
When restricted to a unit cell at $\bR = \bb 0$, the allowed $\bb r_o$ for each $C_n$-symmetric lattice are
\ba
C_2:& \, \bb r_o \in \{ \bx_{1a}, \bx_{1b}, \bx_{1c}, \bx_{1d} \},
\nn
C_3:& \, \bb r_o \in \{ \bx_{1a}, \bx_{1b}, \bx_{1c} \},
\nn
C_4:& \, \bb r_o \in \{ \bx_{1a}, \bx_{1b} \},
\nn
C_6:& \bb r_o \in \{ \bx_{1a} \},
\label{eq:WPs}
\ea
where $\bx_{W}$ with $W=1a,1b,1c,\dots$ are the position offsets for the multiplicity 1 WPs as illustrated in, and detailed in the caption of, Fig.~\ref{fig:unitcells}.
Note that from Fig.~\ref{fig:unitcells}, we use the standard convention that the $1a$ WP is the origin of the unit cell, i.e., $\bx_{1a} = \bb 0$.

Although the set of invariant positions given in Eq.~\eqref{eq:invariant_positions} is determined with respect to the unit cell located at $\bb R = \bb 0$, it is possible to determine an equivalent set of invariant positions with respect to a unit cell located at any $\bb R$ on the periodic lattice.
In general, for a rotation axis located at $\bb R+\bb r_o$, the corresponding rotation operator is denoted by $c_n(\bb R+\bb r_o)$, and the set of invariant positions is denoted by $\mc{X}[c_n(\bb R+\bb r_o)]$.
This detail will be important for the discussion in Sec.~\ref{subsec:domain}.
However, for the majority of this work, $\mc{X}[c_n(\bb r_0)]$ will be specified with respect to $\bb R = \bb 0$.

It is also important to note that the positions forming $\mc{X}[c_n(\bb r_o)]$ in Eq.~\eqref{eq:invariant_positions} depend on the number of unit cells, $N_{1,2}$ and $N$.
For example, for $C_2$-symmetric lattices with $N_{1,2} \in 2 \Z$, all four positions in $\mc{X}[c_2(\bb r_o)]$ correspond to the same WP $\bb r_o$ up to a Bravais lattice translation.
However, when $N_{1,2} \in 2 \Z+1$, $\mc{X}[c_2(\bb r_o)]$ is composed of four different WPs, $1a$, $1b$, $1c$, and $1d$.
Until Sec.~\ref{sec:map_general}, we will consider only the simplest case where all invariant positions of rotation axis $c_n(\bb r_o)$, i.e., all elements of $\mc{X}[c_n(\bb r_o)]$, represent the same WPs (but within different unit cells).
Hence, we impose the following conditions for each lattice:
\ba
& C_2: \, (N_1, N_2) = (0, 0) \pmod 2
\nn
& C_4: \, N = 0 \pmod 2
\nn
& C_{n=3,6}: \, N = 0 \pmod n.
\label{eq:perfect_constraint}
\ea
In Sec.~\ref{sec:map_general}, we will generalize our results to lattices that have unit cell numbers that do not satisfy the constraint given by Eq.~\eqref{eq:perfect_constraint}.
%

%%%%%%
%%%%%%

\begin{figure}[t!]
\centering
\includegraphics[width=0.48\textwidth]{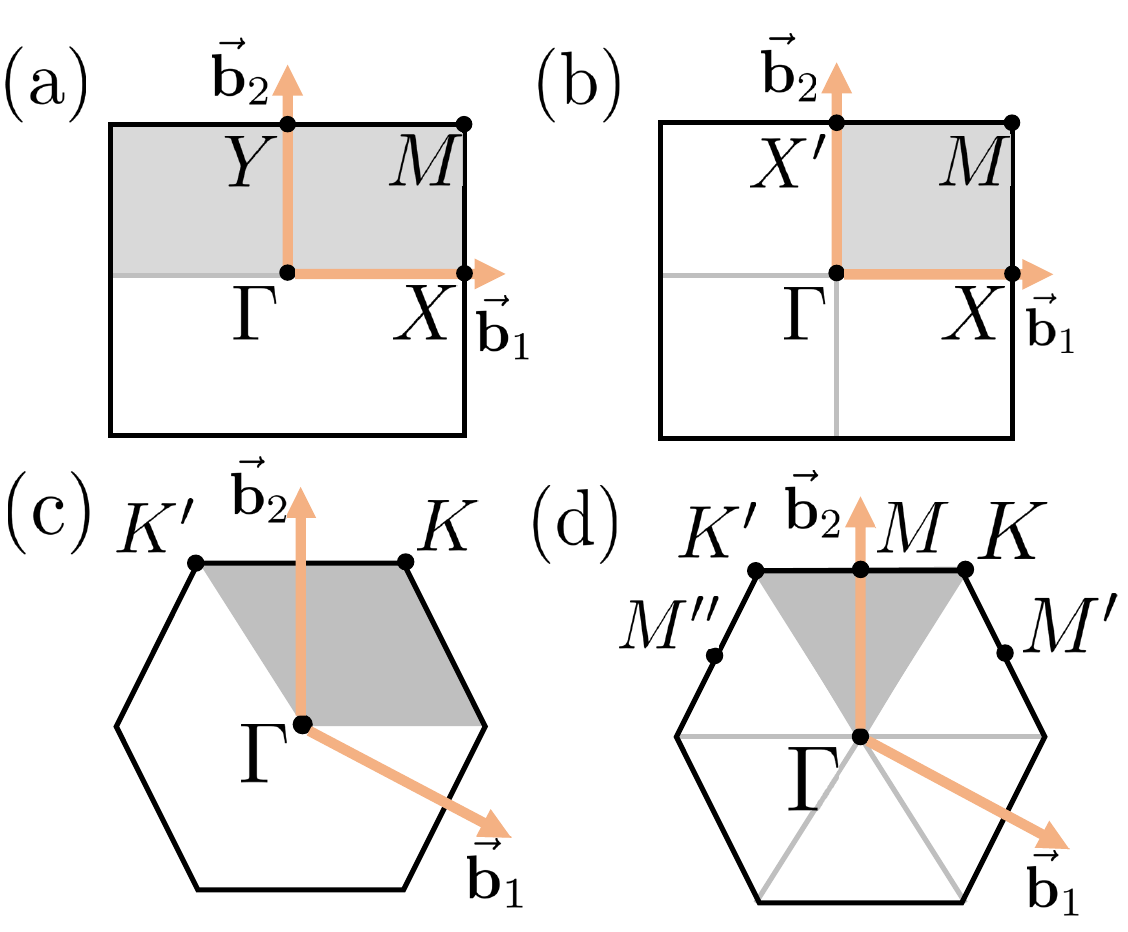}
\caption{
Brillouin zones and their high symmetry momenta for the corresponding $C_n$-symmetric lattices shown in Fig.~\ref{fig:invariant_positions} with dimensions $N_1 \times N_2$ given by Eq.~\eqref{eq:perfect_constraint}.
(a) $C_2$-symmetric BZ (b) $C_4$-symmetric BZ (c) $C_3$-symmetric BZ, and (d) $C_6$-symmetric BZ.
The reciprocal lattice vectors $\bb b_{1,2}$ that span each Brillouin zone are set as $\bb b_1 = (2\pi,0)$ and $\bb b_2= (0,2\pi)$ for $C_{2,4}$, and $\bb b_1 = \frac{2\pi}{\sqrt{3}} (\sqrt{3}, -1)$ and $\bb b_2=\frac{4\pi}{\sqrt{3}} (0,1)$ for $C_{3,6}$.
Shaded regions indicate the domains that generate the entire Brillouin zone upon rotation around the fixed point at the center $\Gamma$ of the Brillouin zones.
}
\label{fig:BZs}
\end{figure}

Before proceeding to the next section it will be useful to connect our position-space discussion to momentum-space.
In conventional studies of topological crystalline insulators based on momentum-space symmetry eigenvalues/indicators, the constraints on the lattice dimensions are crucial, but often implicitly assumed to obey the constraints in Eq.~\eqref{eq:perfect_constraint}.
Indeed, satisfying these constraints allows for the maximal number of high-symmetry points to exist in the Brillouin zone.
This is important, because it is at these points where one can efficiently compute topological crystalline invariants from symmetry eigenvalues of Bloch states.
However, the existence of the high symmetry points depends on the dimensions of the lattice, because the crystalline momentum wavevector $\bk$ is quantized as:
\bg
\bk = \frac{n_1}{N_1} \bb b_1 + \frac{n_2}{N_2} \bb b_2,
\label{eq:momentum}
\eg
for suitable reciprocal lattice basis vectors $\bb b_{1,2}$ such that $\bb b_i \cdot \bb a_j = 2\pi \delta_{ij}$, and $n_i \in \{0, \dots, N_i-1\}$ for $i, j = 1,2$.
The high symmetry momenta $\kinv$ are defined to be the momenta in the Brillouin zone (BZ) that are invariant under the $C_n$ rotation (up to a reciprocal lattice vector).
The maximal set of high symmetry points for each $C_n$-symmetric BZ is shown in Fig.~\ref{fig:BZs}, which require the dimensions given by Eq.~\eqref{eq:perfect_constraint}.
For lattices that do not satisfy this constraint, a subset of high symmetry points will be missing depending on the values of $N_{1,2}$ and $N$.
As mentioned above, we will revisit this complication in Sec.~\ref{sec:map_general}.
%%%%%%

%%%%%%
%%%%%%

%%%%%%
\section{Definition and properties of topological crystalline markers}
\label{sec:tcm_defs}
%%%%%%
Having reviewed the properties of the $C_n$-symmetric lattices in position-space and their corresponding BZs in momentum-space, we now proceed to define the topological crystalline markers (TCMs) (as discussed in Ref.~\onlinecite{MondragonShem2024}) and discuss their properties.
In order to do so, we first set some conventions for the lattice Hamiltonians we consider.
The single-particle Hamiltonian $H$ can be expressed as
\bg
H = \sum_{\bR, \bR'} \, \sum_{\alpha, \beta} \, t_{(\bR,\alpha),(\bR',\beta)} \, \ket{\bR, \alpha} \bra{\bR', \beta},
\label{eq:hamiltonian_def}
\eg
where $\ket{\bR, \alpha}$ is the orthonormal position-space basis, and $t_{(\bR, \alpha),(\bR',\beta)}$ are the hopping amplitudes.
Also, $\alpha$ denotes internal degrees of freedom, such as spin, orbital, sublattice, etc., defined within a unit cell $\bR$.
To construct a TCM we will need the ground-state projector $P_{GS}$, which is a sum of projectors over occupied single-particle states.
The TCM is then given by the product of $P_{GS}$ and $c_n(\bb r_o)$ measured by the position-space basis $\ket{\bR, \alpha}$:
\bg
\vev{c_n(\bb r_o)}_\bR
= \sum_\alpha \, \langle \bR, \alpha | c_n(\bb r_o) \, P_{GS} | \bR, \alpha \rangle.
\label{eq:tcm_def}
\eg

From this definition, the TCM is a function of the unit cell position $\bR$ for a fixed choice of rotation center $\bb r_o$.
It was shown in Ref.~\onlinecite{MondragonShem2024} that for gapped and non-interacting systems, the TCM is sharply localized around the invariant points $\mc{X}[c_n(\bb r_o)]$ of the position-space lattice.
The sharpness of the localization is controlled by the correlation length $\zeta$, which is determined by the inverse of the gap in the single-particle energy spectrum.
As such, the ground-state projector for a gapped, rotation-symmetric ground state exhibits a short-range correlation behavior:
\bg
\sum_\alpha \, \langle \bR', \alpha | P_{GS} | \bR, \alpha \rangle \sim \mc{O} \left( e^{-|\bR' - \bR|/\zeta} \right),
\label{eq:projector_decay}
\eg
for $|\bR' - \bR| \gg \zeta$.
After a short calculation one observes that this leads to the exponential localization of the TCMs around the invariant points $\bb r \in \mc{X}[c_n(\bb r_o)]$ as defined in Eq.~\eqref{eq:invariant_positions}~\cite{MondragonShem2024}.
That is,
\bg
\vev{c_n(\bb r_o)}_\bR \sim \sum_{\bb r \in \mc{X}[c_n(\bb r_o)]} \mc{O} \left( e^{-2 \left| \sin \frac{\pi}{n} \right| |\bb r - \bR|/\zeta} \right).
\label{eq:tcm_decay}
\eg

Since the TCM exhibits exponential localization around rotation-invariant positions $\bb r \in \mc{X}[c_n(\bb r_o)]$ (c.f. Fig.~\ref{fig:tcm}), we can compute the weight of the distribution of the TCM in some region $\mc{S}$ near $\bb r$.
For this, we define {\it a traced TCM} for a given support $\mc{S}$,
\bg
\vev{c_n(\bb r_o)}_\mc{S} = \sum_{\bR \in \mc{S}} \, \vev{c_n(\bb r_o)}_\bR.
\label{eq:tcm_traced_def}
\eg
For the special case when the support $\mc{S}$ covers the entire periodic lattice, then the traced TCMs become basis-independent quantities expressed solely in terms of operators, $P_{GS}$ and $c_n(\bb r_o)$.
In that case, we define the {\it global, fully traced TCM}
\bg
\vev{c_n(\bb r_o)}_F
= \trm{Tr} [ c_n(\bb r_o) \, P_{GS}].
\label{eq:tcm_traced_full}
\eg
By noticing that $P_{GS} = \sum_{a \in occ} \ket{\psi_a} \bra{\psi_a}$ and
\bg
\vev{c_n(\bb r_o)}_F = \sum_{a \in occ} \, \bra{\psi_a} c_n(\bb r_o) \ket{\psi_a},
\eg
for the wave functions $\ket{\psi_a}$ comprising the occupied states, the fully traced TCM $\vev{c_n(\bb r_o)}_F$ yields the group-theory character of a ground state for the rotation symmetry $c_n(\bb r_o)$.

While it is not apparent at this stage, we will show in Secs.~\ref{sec:map_perfect} and \ref{sec:map_general} that topological invariants such as the Chern number and bulk polarization can be determined using the fully traced TCMs.
Furthermore, using the developments of Ref.~\onlinecite{MondragonShem2024}, we explicitly demonstrate how the TCMs that are not fully traced can be used to spatially resolve topological properties of inhomogeneous systems in Sec.~\ref{sec:example}.
This application can be anticipated from Eq.~\eqref{eq:tcm_decay}.
That is, since the TCMs are exponentially localized around $\mc{X}[c_n(\bb r_o)]$, the fully traced TCMs for periodic boundary conditions are approximately the sum of partially traced TCMs around each of the individual invariant positions,
\bg
\vev{c_n(\bb r_o)}_F = \sum_{\bb r \in \mc{X}[c_n(\bb r_o)]} \, \vev{c_n(\bb r_o)}_\mc{S(\bb r)} + O(e^{-\xi_\mc{S}/\zeta}),
\label{eq:tcm_local_and_global}
\eg
if the linear size $\xi_\mc{S}$ of the support $\mc{S(\bb r)}$ around the rotation-invariant position $\bb r$ is sufficiently larger than the correlation length $\zeta$.
The result simplifies for open boundary conditions since the sum receives contributions from only the neighborhood near the unique rotation center $\bb r_0$.
Hence, by scanning the choice of rotation center $\bb r_0$, we can form a mesh of invariants defined across the entire lattice that will allow us to spatially resolve topological properties.
We will demonstrate this approach in much more detail in Sec.~\ref{subsec:domain}.

As a complementary picture to our position-space discussion, we now consider the decomposition of the fully traced TCMs in the momentum-space eigenstate basis.
This is not just a formal exercise.
Indeed, in the next section we will use this decomposition to map previous results for momentum-space symmetry indicators to the fully traced TCMs.
To begin, let us consider a $C_n$-symmetric lattice having periodic boundary conditions and lattice dimensions given by the constraint in Eq.~\eqref{eq:perfect_constraint}.
Starting with a lattice Hamiltonian expanded in momentum-space, we denote the Bloch Hamiltonian as $H(\bk)$, and the momentum-space representation of the rotation operator with origin at a WP $\bb r_o$ as $c_n(\bb r_o,\bk)$.

Now, suppose that a high symmetry momentum (HSM) $\kinv$ is left invariant under $C_n$ rotation (modulo reciprocal lattice vectors), i.e.,
\bg
c_n(\bb r_o,\kinv): \kinv \to C_n \, \kinv = \kinv + \bb b,
\eg
regardless of $\bb r_o$.
Then, $H(\kinv)$ commutes with $c_n(\bb r_o, \kinv)$, and we can define simultaneous eigenstates for $H(\kinv)$ and $c_n(\bb r_o,\kinv)$.
%

%%%%%%
{
\renewcommand{\arraystretch}{1.5}
\begin{table}[t!]
\caption{
Character table for $C_2$.
The rows and columns denote the irreps and symmetry operations respectively.
$E$, $\cm{E}$ are the trivial element and $2\pi$ spin rotation.
Note that $(C_2)^2 = \cm{E}$.
The irreps with positive (negative) $\cm{E}$ character correspond to spinless (spin-1/2) electrons.}
\label{table:C2_character}
\begin{tabular*}{0.3\textwidth}{@{\extracolsep{\fill}} c| c c c c}
\hline \hline
& $E$ & $C_2$ & $\cm{E}$ & $\cm{E} C_2$ \\
\hline
$A$ & 1 & 1 & 1 & 1 \\
$B$ & 1 & $-1$ & 1 & $-1$ \\
$^{1} \bar{E}$ & 1 & $i$ & $-1$ & $-i$ \\
$^{2} \bar{E}$ & 1 & $-i$ & $-1$ & $i$ \\
\hline \hline
\end{tabular*}
\end{table}
}
%%%%%%

Since $(C_n)^n$ is identical to a $2\pi$ spin rotation for generic $C_n$ rotations, the eigenvalues with respect to $c_n(\bb r_o, \kinv)$ (i.e., the rotation eigenvalues) are quantized as $\exp( {\frac{2\pi i}{n} \Z} )$ for spinless electrons and $\exp( {\frac{2\pi i}{n} (\Z + \frac{1}{2})} )$ for spin-1/2 electrons.
Hence, we can label an energy eigenstate at a $C_n$-invariant HSM $\kinv$ with an irreducible representation (irrep) $\kinv_p$ ($p=1,2,\dots,n$) of, say, $c_n(\bx_{1a} = \bb 0, \bk)$.
Where our choice of $\bb r_o = \bx_{1a}$ will serve as a reference for labeling the irreps at HSM.
We define an irrep $\kinv_p$ at the HSM $\kinv$ if it carries the rotation eigenvalue of $c_n(\bb 0, \kinv)$ equal to $e^{\frac{2\pi i}{n} \ell}$.
Here, $\ell$ is the orbital angular momentum of irrep $\kinv_p$, which is defined by $\ell = p-1$ for spinless electrons and $\ell = p-1/2$ for spin-1/2 electrons.
As an example, Table~\ref{table:C2_character} shows the spinless and spin-1/2 representations for $C_2$ rotations.
So, for instance, when a band in a spinless $C_2$-symmetric lattice corresponds to irrep $A$ ($B$) at $\Gamma$, we label it as $\Gamma_1$ ($\Gamma_2$).

In some cases we will need to know the irrep multiplicities for more than one rotation center.
When needed, we can infer the rotation eigenvalue of irrep $\kinv_p$ with respect to $c_n(\bb r_o, \kinv)$ from the relation between $c_n(\bb r_o)$ and $c_n(\bb 0)$:
\ba
c_n(\bb r_o) = T(\bb r_o)c_n(\bb 0) T (\bb r_o)^{-1}
\nn
= c_n(\bb 0) \, T \left( c_n(\bb 0)^{-1} \, \bb r_o - \bb r_o \right),
\ea
where $T(\bx)$ is a translation operator mapping $\bb r$ to $\bb r+ \bx$.
For example, $c_2(\bx_{1b}) \, \bb r = - \bb r + 2 \bx_{1b} = -\bb r + (\bx_{1b} - (-\bx_{1b}))$, i.e., $c_2(\bx_{1b}) = c_2(\bb 0) \, T ( c_2(\bb 0)^{-1} \, \bx_{1b} - \bx_{1b})$ because $c_2(\bb 0)$ acts on real-space coordinate as $-\mathds{1}_{2 \times 2}$.
Thus, we find the relation between momentum-space representation of rotation operators,
\bg
c_n(\bb r_o, \bk) = c_n(\bb 0, \bk) \, \exp \left( - i \bk \cdot (c_n(\bb 0)^{-1} \, \bb r_o - \bb r_o) \right).
\eg
Here, we utilized the momentum-space representation of the translation operator, $t(\bx) = e^{-i\bk \cdot \bx}$ at momentum $\bk$.
Thus, the irrep $\kinv_p$ carries the rotation eigenvalue of $c_n(\bb r_o, \kinv)$,
\ba
e^{i \phi_{n,p}(\bb r_o, \kinv)}
=& \exp \left( \frac{2\pi i}{n}\ell - i \kinv \cdot (c_n(\bb 0)^{-1} \, \bb r_o - \bb r_o) \right)
\nn
=& \exp \left( \frac{2\pi i}{n}\ell - i (C_n \, \kinv - \kinv) \cdot \bb r_o \right),
\label{eq:rot_eigv}
\ea
where in the last step we used the transpose property of orthogonal rotation matrices with respect to the dot product.

For insulators, the irrep multiplicities $m(\kinv_p) \in \N_0$\footnote{Here $\N_0$ is the set of natural numbers including zero, while $\N$ is the set of natural numbers.}, i.e., the number of each representation comprising the ground state, are well-defined and are topological crystalline invariants.
Crucially, we find a relationship between the fully-traced TCMs and the irrep multiplicities $m(\kinv_p)$, which shows that the traced TCMs encode bulk topological data.
By expanding the ground-state projector $P_{GS}$, and the rotation operator $c_n(\bb r_o)$, in terms of momentum-space eigenstates, the fully traced TCM becomes
\bg
\vev{c_n(\bb r_o)}_F
= \sum_{\kinv \in \trm{HSM}_n} \, \sum_{p=1}^n \, e^{i \phi_{n,p}(\bb r_o, \kinv)} \, m(\kinv_p)
\label{eq:tcm_decomp_k}
\eg
where $\trm{HSM}_n$ is the set of momenta in the first BZ that are left invariant under $C_n$ rotation.
Thus, the fully traced TCM can be decomposed into the irrep multiplicities in momentum space.
While this is not difficult to derive (see Supplemental Material~\cite{supple} for details), this result is the essential element to connect the fully traced TCMs to bulk topological invariants.

At this point, we have considered only periodic boundary conditions, and in the next section, we will detail the mapping between the TCMs, bulk topological invariants, and physical observables for periodic lattices that have dimensions $N_1 \times N_2$ which satisfy Eq.~\eqref{eq:perfect_constraint}.
However, in Sec.~\ref{sec:map_general}, we will extend our discussion to lattices whose dimensions do not satisfy Eq.~\eqref{eq:perfect_constraint}, where one must consider lattices with {\it twisted} periodic boundary conditions.
Consequently, it is useful to extend the TCMs discussed above to lattices with twisted boundary conditions.
Denoting $\bs \theta = \theta_1 \bb a_1 + \theta_2 \bb a_2 \equiv (\theta_1, \theta_2)$ as the parameters of the boundary condition, we introduce the following modified notation for the TCM when twisted boundary conditions are present:
\bg
\vev{c_n(\bb r_o)}_\bR^{\bs \theta}
= \sum_\alpha \, \langle \bR, \alpha | c_n(\bb r_o) \, P_{GS}^{(\bs \theta)}| \bR, \alpha \rangle.
\label{eq:tcm_traced_twisted}
\eg
This modified notation extends in a similar manner to the traced TCM (e.g., $\vev{c_n(\bb r_o)}_\mc{S}^{\bs \theta}$) and the fully traced TCM (e.g., $\vev{c_n(\bb r_o)}_F^{\bs \theta}$).
Throughout this work, when the superscript $\bs \theta$ is omitted from the TCM, it can be assumed that the TCM is applied to a lattice having untwisted periodic or open boundaries.
%%%%%%

%%%%%%
\section{Mapping topological crystalline markers to bulk topological invariants and physical observables}
\label{sec:map_perfect}
%%%%%%
In the previous section, we exemplified the basis-independent properties of the fully traced TCMs by explicitly decomposing them in both momentum-space and position-space.
In this section, we will use these results to show how the Chern number in $C_n$-symmetric Chern insulators, and the bulk polarization and sector charge for $C_n$-symmetric atomic insulators, can be expressed in terms of the markers, starting with the momentum-space decomposition of the fully traced TCMs.
For this section, we restrict our focus to $C_n$-symmetric lattices with constraints on the dimensions given by Eq.~\eqref{eq:perfect_constraint}; these lattices contain the maximal set of high symmetry momentum points in the BZ.
The other types of lattices will be discussed in Sec.~\ref{sec:map_general}.
%%%%%%

%%%%%%
\subsection{Chern number and mapping procedure}
\label{subsec:map_chern}
%%%%%%
First, we discuss the mapping between the fully traced TCMs and the Chern number $\mc{C}$ for $C_n$-symmetric Chern insulators in Altland-Zirnbauer (AZ) class A~\cite{Altland1997}.
Then, we will proceed to lay out the general procedure for mapping the fully traced TCMs to the momentum-space symmetry data for each AZ symmetry class.
From Ref.~\onlinecite{Fang2012}, the Chern number can be expressed as follows for each $C_n$ symmetry
\ba
C_2: \, \mc{C} &= [X_1] + [Y_1] + [M_1] \pmod 2,
\nn
C_3: \, \mc{C} &= [K_1] + 2[K_2] + [K'_1] + 2[K'_2] \pmod 3,
\nn
C_4: \, \mc{C} &= [M_1] + 2[M_2] + 3[M_3] + 2[X_1] \pmod 4,
\nn
C_6: \mc{C} &= 2[K_1] - 2[K_2] + 3[M_1] \pmod 6,
\label{eq:chern_rot_invs}
\ea
where we have introduced the notation for momentum-space rotation invariants $[\kinv_p]$ at each HSM given by $\kinv$.
The rotation invariants are defined as follows,
\ba
C_2: \, & [\kinv_{p=1,2}] = m(\kinv_p) - m(\Gamma_p) \, \trm{for } \kinv \in \{X, Y, M\},
\nn
C_3: \, & [\kinv_{p=1,2,3}] = m(\kinv_p) - m(\Gamma_p) \, \trm{for } \kinv \in \{K, K'\},
\nn
C_4: \, & [M_{p=1,2,3,4}] = m(M_p) - m(\Gamma_p),
\nn
& [X_{p=1,2}] = m(X_p) - m(\Gamma_p) - m(\Gamma_{p+2}),
\nn
C_6: \, & [K_{p=1,2,3}] = m(K_p) - m(\Gamma_p) - m(\Gamma_{p+3}),
\nn
& [M_{p=1,2}] = m(M_p) - m(\Gamma_p) - m(\Gamma_{p+2}) - m(\Gamma_{p+4}).
\label{eq:rot_invs}
\ea

Using these momentum-space rotation invariants, one can re-express the decomposition of the fully traced TCM in momentum-space in Eq.~\eqref{eq:tcm_decomp_k}.
For example, we can rewrite Eq.~\eqref{eq:tcm_decomp_k} as 
\ba
\vev{c_2(\bb 0)}_F
=& \sum_{\kinv \in \trm{HSM}_2} \, \sum_{p=1,2} \, e^{i \phi_{n,p} (\bb r_0, \kinv)}\, m(\kinv_p)
\nn
=& \sum_{\kinv \in \trm{HSM}_2} \, \left(m(\kinv_1) - m(\kinv_2)\right)
\nn
=& [X_1] + [Y_1] + [M_1] - [X_2] - [Y_2] - [M_2] \nn
& + 4 m(\Gamma_1) - 4 m(\Gamma_2),
\label{eq:C2_example}
\ea
where $\trm{HSM}_2 = \{\Gamma,X,Y,M\}$ is the set of HSM for $C_2$ symmetry.
Thus, we can express the fully traced $C_2$ TCM with rotation center at the $1a$ WP ($\bb r_o=\bx_{1a}=\bb 0$) using a combination of rotation invariants $[\kinv_p]$ and $\Gamma$-point irrep multiplicities $m(\Gamma_p)$.
We gain additional data from the fully traced $C_2$ TCMs obtained for the $1b$, $1c$, and $1d$ WPs, and this generalizes straightforwardly to the other symmetry groups and corresponding WPs.
%

%%%%%%
{
\renewcommand{\arraystretch}{1.8}
\begin{table*}[t!]
\centering
\caption{
Chern number for class A spinless and spin-1/2 $C_n$-symmetric insulators and class D crystalline superconductors expressed in terms of fully traced TCMs $\vev{c_n(\bx_W)}_F$ evaluated over multiplicity $1$ WPs $W$.
These labels are given as $W \in \{1a,1b,1c,1d\}$ for $C_2$, $W \in \{1a,1b,1c\}$ for $C_3$, $W \in \{1a,1b\}$ for $C_4$, and it is simply the $1a$ WP for $C_6$.
The Chern number for $C_{4,6}$-symmetric insulators has other expressions in terms of TCMs (see Supplemental Material~\cite{supple}).
For class D superconductors we assume that the particle-hole and the rotation operator in its BdG form commute, which includes a wide class of models.
}
\label{table:chern}
\begin{tabular*}{0.9\textwidth}{@{\extracolsep{\fill}} c| c c}
\hline \hline
 & A and D (spinless) & A and D (spin-1/2)
\\ \hline
$C_2$ & $\frac{1}{2} \vev{c_2(\bx_W)}_F$ mod 2 & $-\frac{i}{2} \vev{c_2(\bx _W)}_F$ mod 2
\\
$C_3$ & $\frac{2}{\sqrt{3}} \trm{Im} \vev{c_3(\bx_W)}_F$ mod 3 & $-\frac{2}{\sqrt{3}} \trm{Im} \vev{c_3(\bx_W)}_F$ mod 3
\\
$C_4$ & $\pm \sqrt{2} \trm{Re} [ e^{-\frac{i\pi}{4}} \vev{c_4(\bx_W)}_F ] - \frac{1}{2} \vev{c_2(\bx_W)}_F$ mod 4 & $\pm \sqrt{2} \trm{Re} \vev{c_4(\bx_W)}_F - \frac{i}{2} \vev{c_2(\bx_W)}_F$ mod 4
\\
$C_6$ & $- \frac{4}{\sqrt{3}} \trm{Im} \vev{c_3(\bx_W)}_F + \frac{3}{2} \vev{c_2(\bx_W)}_F$ mod 6 & $\frac{4}{\sqrt{3}} \trm{Im} \vev{c_3(\bx_W)}_F + \frac{3i}{2} \vev{c_2(\bx_W)}_F$ mod 6
\\ \hline \hline
\end{tabular*}
\end{table*}
}
%%%%%%

Our goal of expressing the Chern number formulae in Eq.~\eqref{eq:chern_rot_invs} in terms of the fully traced TCMs involves mapping the fully traced TCMs to a combination of momentum-space rotation invariants and irrep multiplicities at the $\Gamma$-point.
As illustrated in the example above, this will require multiple applications of the decomposition given by Eq.~\eqref{eq:tcm_decomp_k}.
To construct the mapping, we must first determine both the set of independent, fully traced TCMs, and the independent momentum-space symmetry data.
The latter consists of the set of independent momentum-space rotation invariants and irrep multiplicities at $\Gamma$.
The procedure to determine these sets can be summarized in three steps: 

\begin{itemize}
\item Step 1: For a given wallpaper ($d=2$) or space ($d=3$) group, list the WPs $W$ and their onsite symmetry group $G_W$.
The onsite symmetry group $G_W$ is composed of symmetries that leave the location $\bx_W$ of WP $W$ invariant, i.e., $g_W \bx_W = \bx_W$ for all $g_W \in G_W$.
When a WP $W$ is composed of multiple positions, $\bx_{W,1}, \bx_{W,2}, \dots$, we consider the onsite symmetry group of one representative position, say $\bx_{W,1}$.
(The reason for this is explained below.)
A set of all elements in the {\it unitary subset} of $G_{\rm TCM} = \cup_W G_W$ are considered operators for defining the TCMs.
Note that it is not necessary to consider a generic WP which has a trivial onsite symmetry group since the trivial group is always a subgroup of $G_W$ and will be included in $G_{\rm TCM}$.
If $g \in G_{\rm TCM}$ is antiunitary, then the corresponding TCM $\vev{g}_F$ is not invariant under a basis transformation of the orbitals comprising the ground state.
\item Step 2: Construct a set of fully traced TCMs for the group elements in $G_{\rm TCM}$ defined in Step 1.
One can further identify certain relations between the TCMs of some group elements.
For example, for $g$ and its inverse $g^{-1}$, the corresponding fully traced TCMs are complex conjugates of each other, i.e., $\vev{g^{-1}}_F = \vev{g}_F^*$.
This means that when we consider a $C_n$-symmetric system and compute the fully traced TCM of $c_{m \le n}(\bb r_o)$, ${\vev{c_m(\bb r_o)}_F}^* = \vev{c_m(\bb r_o)^{-1}}_F$.
\item Step 3: Utilize symmetries of the Hamiltonian (e.g., charge conservation and/or time-reversal symmetry and/or particle-hole symmetry) to determine an independent set of momentum-space irrep multiplicities.
This set can be translated into the set of independent momentum-space rotation invariants and irrep multiplicities at $\Gamma$.
Then, we construct the mapping between this momentum-space symmetry data and the set of fully traced TCMs constructed in Step 2, based on Eq.~\eqref{eq:tcm_decomp_k}.
\end{itemize}

To demonstrate the first two steps of constructing the fully traced TCMs for a particular $C_n$-symmetry group, we consider the specific example of a $C_6$-symmetric system.
Let $\bb a_1, \bb a_2$ be defined as in Fig.~\ref{fig:unitcells}.
For Step 1, let us first list the WPs having nontrivial onsite symmetry groups.
As shown in Fig.~\ref{fig:unitcells}, a $C_6$-symmetric unit cell has three sets of maximal WPs distinguished by the labels $1a$, $2b$, and $3c$.
The $1a$ WP is the origin of the unit cell given by $\bx_{1a} = \bb 0$, the $2b$ WPs are given by $\bx_{2b,1} = (\bb a_1 + \bb a_2)/3$ and $\bx_{2b,2} = (-\bb a_1 + 2 \bb a_2)/3$, and the $3c$ WPs are given by $\bx_{3c,1}=\bb a_1/2$, $\bx_{3c,2}=\bb a_2/2$, and $\bx_{3c,3}=(-\bb a_1 + \bb a_2)/2$.
Now we consider the various unitary symmetries that are elements of onsite symmetry groups of the $1a$, $2b$, $3c$ WPs.
The onsite symmetry group of the $1a$ WP, $G_{1a}$, is point group $C_6$, which is generated by $c_6(\bx_{1a})$.
Similarly, we define the onsite symmetry groups $G_{2b,i=1,2}$, $G_{3c,j=1,2,3}$ for the WP positions $\bx_{2b,i}$ and $\bx_{3c,j}$.
Then, $G_{2b,i}$ is generated by $C_3$ rotation $c_3(\bx_{2b,i})$, and $G_{3c,j}$ by $C_2$ rotation $c_2(\bx_{3c,j})$.
In this way, we define $G_{\rm TCM} = G_{1a} \cup G_{2b,1} \cup G_{3c,1}$.

As an aside, note that we do not include $G_{2b,2}$, $G_{3c,1}$, $G_{3c,2}$ in $G_{\rm TCM}$.
Indeed, for a WP with multiplicity greater than 1, the fully traced TCMs evaluated at the positions composing the WP give the same result.
This means that, for example, in a $C_6$-symmetric lattice, evaluating the fully traced TCM for the $C_3$ rotation operator with origin at either $\bx_{2b,1}$ or $\bx_{2b,2}$ is equivalent.
This also holds true for the fully traced TCMs for the $C_2$ rotation operators with origin at $\bx_{3c,1}$, $\bx_{3c,2}$, or $\bx_{3c,3}$.
To see this, let us consider the relation between $G_{2b,1}$ and $G_{2b,2}$.
First, note that $\bx_{2b,2} = c_6(\bx_{1a}) \bx_{2b,1}$.
By definition, $g \bx_{2b,1} = \bx_{2b,1}$ for any $g \in G_{2b,1}$, and thus $h g h^{-1} \bx_{2b,2} = \bx_{2b,2}$ where $h=c_6(\bx_{1a})$.
In other words, $G_{2b,2} = \{g'=h g h^{-1}| \forall g \in G_{2b,1}\}$.
In this case, the fully traced TCM of $g'$ must be same as the one for $g$ since $\vev{g'}_F = {\rm Tr} [g' P_{SG}] = {\rm Tr} [h g h^{-1} P_{GS}] = {\rm Tr} [g P_{GS}] = \vev{g}_F$.
Here, we use the fact that the ground state is symmetric under $h$, i.e. $h P_{GS} = P_{GS} h$.

For Step 2, we compute the TCMs for the elements in $G_{\rm TCM}=G_{1a} \cup G_{2b,1} \cup G_{3c,1}$.
There are $8$ group elements of $G_{\rm TCM}$ besides the trivial group element.
They are $c_{2,3,6}(\bx_{1a})$, $c_3(\bx_{2b,1})$, $c_2(\bx_{3c,1})$, and the (independent) inverses $c_{3,6}(\bx_{1a})^{-1}$ and $c_3(\bx_{2b,1})^{-1}$.
(Note that the inverse of $C_2$ is $C_2$ up to a $2\pi$ spin rotation.)
Thus, we find that the complete set of fully traced TCMs for a $C_6$-symmetric system is given as $\vev{c_{2,3,6}(\bx_{1a})}_F$, $\vev{c_3(\bx_{2b})}_F$, $\vev{c_2(\bx_{3c})}_F$ and their complex conjugations, which represent the inverses of the rotation operators.
(Note that $\vev{g^{-1}}_F = {\rm Tr}[g^{-1} P_{GS}] = {\rm Tr} [(P_{GS} g)^\dg] = \vev{g}_F^*$ from the properties of trace and the Hermiticity of $P_{GS}$.)
Therefore, for a $C_6$-symmetric system, there are a total of $8$ independent fully traced TCMs that can be constructed (or 9 independent fully traced TCMs if one includes a fully traced TCM for the trivial group element, which is associated to the particle number and filling factor).

For Step 3, we need to identify both the minimal set of momentum-space rotation invariants and irrep multiplicities at the $\Gamma$-point (denoted as $m(\Gamma_p)$) that must be considered.
To achieve this we must take into account the internal symmetries of the Hamiltonian.
The internal symmetries of the Hamiltonian (e.g., charge conservation, time-reversal, particle-hole, chiral) impose constraints on the momentum-space rotation invariants and the $m(\Gamma_p)$.

The most fundamental symmetry we consider is charge conservation, which allows one to define a filling constraint.
The filling constraint implies that the total number of occupied Bloch states is equal to a filling $\nu$.
Hence, at each HSM, the irrep multiplicities must satisfy a filling constraint sum rule, because the total number of occupied Bloch states is fixed throughout the BZ for a gapped system.
To illustrate the filling constraint, suppose that $\Gamma$ and some HSM $\kinv$ are $C_n$- and $C_{m \le n}$-invariant points respectively.
Then, we can set
\bg
\nu = \sum_{p=1}^m \, m(\kinv_p) = \sum_{p=1}^n \, m(\Gamma_p)\,\to\sum_{p=1}^m \, [\kinv_p] = 0.
\label{eq:filling_constraint}
\eg
In the last line, we used the definition of the rotation invariant $[\kinv_p]$ in Eq.~\eqref{eq:rot_invs}.
For instance, in a $C_2$-symmetric system, $\nu = m(\Gamma_1) + m(\Gamma_2) = m(X_1) + m(X_2)$ implies $[X_1] + [X_2] = m(X_1) -m(\Gamma_1) + m(X_2) - m(\Gamma_2) = 0$.

Other internal symmetries, such as time-reversal symmetry for class AI and AII insulators, can introduce additional constraints between the rotation invariants.
A rotation eigenvalue $e^{\frac{2\pi i}{n}\ell}$ is determined by the angular momentum $\ell$, which can be expressed either as $\ell=p-1$ for spinless electrons, or $\ell=p-1/2$ for spin-1/2 electrons, where $p\in\{1,\dots, n\}$.
Time-reversal symmetry maps the rotation eigenvalue $e^{\frac{2\pi i}{n}\ell}$ of an occupied Bloch state at an HSM $\kinv$ to its complex conjugate $e^{-\frac{2\pi i}{n}\ell}$ of an occupied Bloch state at $-\kinv$.
This imposes the constraint that the total number of occupied Bloch states at an HSM $\kinv$ with rotation eigenvalue $e^{\frac{2\pi i}{n}\ell}$ must equal the total number of occupied Bloch states at $-\kinv$ with the corresponding complex conjugated rotation eigenvalue.
For example, for class AII systems having time-reversal symmetry and $\ell=p-1/2$, i.e., the spin-1/2 case, the rotation invariants are constrained to obey:
\ba
C_2: \, & [\kinv_p] = [\kinv_{3-p}] \trm{ for } \kinv \in \{X, Y, M\} \trm{ and } p \in \{1,2\},
\nn
C_3: \, & [K_{p}] = [K'_{4-p}] \trm{ for } p \in \{1,2,3\},
\nn
C_4: \, & [M_p] = [M_{5-p}] \trm{ for } p \in \{1,2,3,4\},
\nn
& [X_p] = [X_{3-p}] \trm{ for } p \in \{1,2\},
\nn
C_6: \, & [K_p] = [K_{4-p}] \trm{ for } p \in \{1,2,3\},
\nn
& [M_p] = [M_{3-p}] \trm{ for } p \in \{1,2\}.
\label{eq:trs_constraint}
\ea

In addition to time-reversal symmetry, one can consider constraints induced by the particle-hole symmetry for class D superconductors, described using the Bogoliubov-de Gennes (BdG) formalism.
For such BdG Hamiltonians the viability of symmetry indicators, such as rotation invariants, to diagnose the bulk topological properties of superconductors depends on the pairing symmetry~\cite{ono2019symmetry,Geier2020,Ono2020b}.
For the results shown in Table~\ref{table:chern}, we study spinless and spin-1/2 superconductors where the particle-hole and $C_n$ rotation operators commute (e.g, as in Ref.~\onlinecite{Benalcazar2014}).
(For a discussion of pairing symmetry possibilities, see the Supplemental Material~\cite{supple}.)

Finally, rotational symmetry can also impose a constraint between rotation invariants; specifically for $C_4$ and $C_6$-symmetric systems.
As shown in Fig.~\ref{fig:BZs}, the $C_4$-symmetric BZ has two $C_2$-symmetric HSMs labeled by $X$ and $X'$ that are related to each other via a $C_4$ rotation.
The rotation invariants for these points are equal to each other (e.g., $[X_{p}]=[X_{p}']$ for $p=1,2$).
Similarly, for the $C_6$-symmetric BZ, it is clear that $[K_{p}]=[K_{p}']$ for $p=1,2,3$ and $[M_{p}]=[M_{p}']=[M_{p}'']$ for $p=1,2$.

To illustrate how the symmetry constraints can be applied, let us return to the example of the $C_6$-symmetric system and consider the symmetries of class A for spinless fermions.
For a class A insulator the relevant symmetries that impose constraints between the momentum-space rotation invariants and $\Gamma$ point irrep multiplicities are only charge conservation and rotation symmetry.
The filling constraint leads to the following relations,
\bg
\nu = \sum_{p=1}^6 m(\Gamma_p),
\quad
\sum_{p=1}^{3}[K_{p}] = \sum_{p=1}^{3}[K_{p}'] = 0,
\nn
\sum_{p=1}^2 [M_p] = \sum_{p=1}^2 [M_p'] = \sum_{p=1}^2 [M_p''] = 0.
\label{eq:C6_example_filling_constraint}
\eg
Additionally, rotational symmetry imposes the equivalence of the rotation invariants at the $K$ and $K'$ points, and likewise for the $M$, $M'$, and $M''$ points.

Taking these constraints into account, we find that the minimal set of rotation invariants and $\Gamma$ irrep multiplicities one needs to consider is given by $m(\Gamma_1)$, $m(\Gamma_2)$, $m(\Gamma_3)$, $m(\Gamma_4)$, $m(\Gamma_5)$, $[K_1]$, $[K_2]$, and $[M_1]$, for a total of $8$ momentum-space quantities, or equivalently, $9$ momentum-space quantities if one chooses to include either the filling $\nu$ (or alternatively $m(\Gamma_6)$).
Thus, for a class A $C_6$-symmetric insulator, this indicates there can be a bijective map between the fully traced TCMs and the momentum-space symmetry data.
Indeed, we expect this to be these case since Eq.~\eqref{eq:tcm_decomp_k} is a linear map and the number of independent, fully-traced TCMs matches the number of independent momentum-space invariants.
To confirm this intuition we found that the map between the fully traced TCMs and momentum-space data is invertible for every $C_n$-symmetry group.
The computational details of all such mappings are provided in the Supplemental Material~\cite{supple}.

As a result of the mapping, we find that the Chern number modulo $n$ can be expressed {\it solely} in terms of the fully traced TCMs.
The results for class A $C_n$-symmetric insulators, for both spinless and spin-1/2 fermions, as well as class D crystalline superconductors with trivial pairing symmetry, are shown in Table~\ref{table:chern}.
These results reveal that for any $C_n$-symmetric Chern insulator ($n=2,3,4,6$), the bulk Chern number can be determined modulo $n$ by evaluating the TCMs of the rotation operators with an origin at {\it any} multiplicity $1$ WP.
Particularly for $C_2$, $C_3$, and $C_4$-symmetric Chern insulators, the Chern number can be {\it equally} determined at any multiplicity $1$ WP, which demonstrates that the Chern number does not depend on the origin of the rotation operator.
Equivalently, this result demonstrates that in the bulk, redefining the unit cell origin does not change the Chern number, which corroborates with the Chern number being a strong topological invariant.
%%%%%%

%%%%%%
{
\renewcommand{\arraystretch}{1.2}
\begin{table}[t!]
\centering
\caption{
Representation of bands induced from Wannier orbitals for a $C_2$-symmetric lattice with spinless electrons.
Each Wannier orbital type is labeled by a maximal WP, $W=1a,1b,1c,1d$, and an angular momentum $l=0$ or $1$.
The 2-5th columns denote the $C_2$ eigenvalues of bands at high-symmetry points in momentum space.
The 6-8th columns represent the rotation invariants.
}
\label{table:C2_bandrep}
\begin{tabular*}{0.42\textwidth}{ @{\extracolsep{\fill}} c | c c c c c c c }
\hline \hline
$(l,W)$ & $\Gamma$ & $X$ & $Y$ & $M$ & $[X_1]$ & $[Y_1]$ & $[M_1]$
\\ \hline
$(0,1a)$ & $+1$ & $+1$ & $+1$ & $+1$ & 0 & 0 & 0
\\
$(1,1a)$ & $-1$ & $-1$ & $-1$ & $-1$ & 0 & 0 & 0
\\
$(0,1b)$ & $+1$ & $-1$ & $+1$ & $-1$ & $-1$ & 0 & $-1$
\\
$(1,1b)$ & $-1$ & $+1$ & $-1$ & $+1$ & $+1$ & 0 & $+1$
\\
$(0,1c)$ & $+1$ & $+1$ & $-1$ & $-1$ & 0 & $-1$ & $-1$
\\
$(1,1c)$ & $-1$ & $-1$ & $+1$ & $+1$ & 0 & $+1$ & $+1$
\\
$(0,1d)$ & $+1$ & $-1$ & $-1$ & $+1$ & $-1$ & $-1$ & 0
\\
$(1,1d)$ & $-1$ & $+1$ & $+1$ & $-1$ & $+1$ & $+1$ & 0
\\ \hline \hline
\end{tabular*}
\end{table}
}
%%%%%%

%%%%%%
\subsection{Bulk polarization and sector charge}
\label{subsec:map_atomic}
%%%%%%%
In atomic insulators, the Chern number is always zero and therefore, it is possible to construct exponentially localized Wannier functions.
The Wannier functions allow for an efficient position-space description in terms of the Wannier-orbital configuration on the position-space lattice.
For point-group symmetric insulators, the symmetry representation and multiplicity of each Wannier orbital is well defined.
Furthermore, with this knowledge, one can determine topological quantities that characterize atomic insulators in position-space.
Here, let us focus on the polarization and sector charge~\cite{Benalcazar2019} as representative examples.
%%%%%%

%%%%%%
\subsubsection{Electronic polarization}
%%%%%%
For each $C_n$ symmetry, it is possible to determine the bulk polarization by considering the number of Wannier orbitals localized at maximal WPs away from the $1a$ position.
For class A systems, these are given as follows,
\ba
C_2: \, \bb P &= \frac{e}{2} \left[ (n_{1b}+n_{1d}) \, \bb a_1 + (n_{1c}+n_{1d}) \, \bb a_2 \right],
\nn
C_3: \, \bb P &= \frac{e}{3} (n_{1b}-n_{1c}) (\bb a_1 + \bb a_2),
\nn
C_4: \, \bb P &= \frac{e}{2} (n_{1b}+n_{2c}) (\bb a_1 + \bb a_2),
\nn
C_6: \, \bb P &= \bb 0,
\label{eq:pol}
\ea
where we have used the convention $e=-|e|$ for the electron charge.
Each component of the polarization $\bb P = p_1 \bb a_1 + p_2 \bb a_2$ is defined modulo $e$, and $n_W$ refers to the total number of Wannier orbitals at Wyckoff position $W$.

Equation~\eqref{eq:pol} can be understood as follows.
Without loss of generality, we can assume that all Wannier orbitals are localized at maximal WPs.
Indeed, any Wannier orbital on non-maximal WPs can be adiabatically moved to maximal WPs while preserving all the symmetries of the system.
At each maximal WP each Wannier orbital carries rotation eigenvalue $e^{\frac{2\pi i}{n} \ell}$, where $\ell$ is their angular momenta\footnote{Recall that $\ell=p-1$ ($p-1/2$) for spinless (spin-1/2) cases for $p = 1, \dots, n$.}.
Thus, we can define $n_W^{(\ell)}$, the multiplicity of each Wannier orbital type at maximal position $W$.
Note that the total number of Wannier orbitals at $W$ is given by
\ba
n_W = \sum_\ell \, n_W^{(\ell)}.
\label{eq:wannier_multi}
\ea
Note that $\sum_\ell$ means $\sum_{\ell=0}^{n-1}$ for spinless electrons and $\sum_{\ell=1/2}^{n-1/2}$ for spin-1/2 electrons.
Since the polarization is the sum of Wannier centers of occupied electrons~\cite{KingSmith1993}, we set
\bg
\bb P = e \sum_W \, n_W \, \bx_W,
\label{eq:VanderbiltWannier}
\eg
which yields Eq.~\eqref{eq:pol}.
For example, in a $C_2$-symmetric system, there are four maximal WPs $W=1a,1b,1c,1d$ having positions $\bx_{1a}=(0,0)$, $\bx_{1b} = 1/2 \bb a_1$, $\bx_{1c} = 1/2 \bb a_2$, and $\bx_{1d} = 1/2(\bb a_1 + \bb a_2)$.
Substituting these quantities into Eq.~\eqref{eq:VanderbiltWannier} directly leads to the first line in Eq.~\eqref{eq:pol}.

In this context we will now show that Eq.~\eqref{eq:pol} can be obtained by computing the fully traced TCMs in position space, instead of using the usual momentum-space data.
First, a linear combination of Wannier orbital multiplicities can be mapped to momentum-space data, using the induced band representation~\cite{Kruthoff2017,Bradlyn2017,Po2017}.
Second, Eq.~\eqref{eq:tcm_decomp_k} implies the fully traced TCM can be represented in terms of momentum-space data including the rotation-invariants and $\Gamma$ irrep multiplicities.
Combining these two, we construct a mapping from the set of fully traced TCMs to the Wannier orbital multiplicities for each $C_n$ symmetry and symmetry class A, AI, and AII.

To demonstrate how this mapping is carried out with a simple example, consider a $C_2$-symmetric, spinless, class A atomic insulator.
Table~\ref{table:C2_bandrep} lists the HSM $C_2$ eigenvalue data of the band representations induced from Wannier orbitals at maximal WPs~\cite{Kruthoff2017,Bradlyn2017,Po2017}.
From Table~\ref{table:C2_bandrep}, and the definition of the rotation invariants $[\kinv_p]$ ($p=1,2$) in Eq.~\eqref{eq:rot_invs}, we find
\bg
[X_1] = -[X_2] = n_{1b}^{(1)} - n_{1b}^{(0)} + n_{1d}^{(1)} - n_{1d}^{(0)},
\nn
[Y_1] = -[Y_2] = n_{1c}^{(1)} - n_{1c}^{(0)} + n_{1d}^{(1)} - n_{1d}^{(0)},
\nn
[M_1] = -[M_2] = n_{1b}^{(1)} - n_{1b}^{(0)} + n_{1c}^{(1)} - n_{1c}^{(0)},
\nn
m(\Gamma_p) = \sum_{W \in \trm{WP}_2} \, n_W^{(p-1)},
\label{eq:C2_band_rep}
\eg
where $\trm{WP}_2 = \{ 1a,1b,1c,1d \}$ is the list of maximal WPs for a $C_2$-symmetric lattice.
These relations are the outcome of the first step.

For the second step we use Eq.~\eqref{eq:tcm_decomp_k}, which is the mapping between the fully traced TCMs for each WP and a combination of momentum-space rotation invariants and the $\Gamma$ irrep multiplicities.
Noting that the fully traced TCM for the $1a$ WP is already given by Eq.~\eqref{eq:C2_example}, we can use Eqs.~\eqref{eq:rot_eigv} and \eqref{eq:tcm_decomp_k} to find
\ba
\vev{c_2(\bx_{1a})}_F =& 2([X_1] + [Y_1] + [M_1]) + 4(m(\Gamma_1) - m(\Gamma_2)),
\nn
\vev{c_2(\bx_{1b})}_F =& 2(-[X_1] + [Y_1] - [M_1]),
\nn
\vev{c_2(\bx_{1c})}_F =& 2([X_1] - [Y_1] - [M_1]),
\nn
\vev{c_2(\bx_{1d})}_F =& 2(-[X_1] - [Y_1] + [M_1]).
\label{C2_TCM_rot_inv_mapping}
\ea
With the second step complete we can combine the mappings in Eqs.~\eqref{eq:C2_band_rep} and \eqref{C2_TCM_rot_inv_mapping} to yield the following mapping between the fully traced TCMs and the Wannier orbital irrep multiplicities:
\bg
\vev{c_2(\bx_W)}_F = 4( n_W^{(0)}-n_W^{(1)} ),
\label{eq:C2_TCM_Wannier_orb_mapping}
\eg
for $W \in \trm{WP}_2 = \{1a,1b,1c,1d\}$.

This is an important result, and we can derive it from an alternative perspective.
For a generic $C_n$-symmetric atomic insulator, the Wannier-orbital multiplicity at each WP is well defined.
Then, the fully traced TCM in Eq.~\eqref{eq:tcm_traced_full} can be computed with the position-space Wannier orbital basis.
In this case, we have
\bg
\vev{c_n(\bb r_o)}_F
= \sum_{W^* \in \mc{X}[c_n(\bb r_o)]} \, e^{2 \pi i \ell/n} \, n_{W^*}^{(\ell)}.
\label{eq:fully_tcm_Wannier}
\eg
In the previous example of $C_2$-symmetric insulators with unit-cell constraint $(N_1,N_2)=(0,0)$ mod 2 [Eq.~\eqref{eq:perfect_constraint}], all four $C_2$-invariant positions in $\mc{X}[c_2(\bb r_o)]$ correspond to same WP type, say $W$.
This explains the factor 4 in Eq.~\eqref{eq:C2_TCM_Wannier_orb_mapping}.
Also, at the WP $W$, Wannier orbital types having angular momentum $\ell=0,1$ are defined.
Thus, their multiplicities $n^{(\ell)}_W$ and angular momentum factor $e^{2\pi i \ell/n}$ contribute to Eq.~\eqref{eq:C2_TCM_Wannier_orb_mapping} as $e^{i0}n^{(0)}_W + e^{i\pi} n^{(1)}_W$.
So, in this way, the fully traced TCM is related to the momentum-space data and the position space data contained in the Wannier orbital multiplicities of atomic insulators.

With the mapping between the fully traced TCMs and Wannier-orbital multiplicities in hand, we can express the bulk polarization in Eq.~\eqref{eq:pol} solely in terms of the fully traced TCMs.
For example, we can combine Eqs.~\eqref{eq:pol} and \eqref{eq:C2_TCM_Wannier_orb_mapping} to express the polarization for $C_2$-symmetric, spinless, class A atomic insulators in terms of the fully traced TCMs.
Since the polarization is defined modulo $e$, and $n^{(\ell)}_W \in \Z$ for atomic insulators, the polarization along the $\bb a_1$ direction [Eq.~\eqref{eq:pol}] is equivalent to $(\bb P)_x = \frac{e}{2} [(-n_{1b}^{(0)} + n_{1b}^{(1)}) + (- n_{1d}^{(0)} + n_{1d}^{(1)})]$.
By combining this and Eq.~\eqref{eq:C2_TCM_Wannier_orb_mapping}, we find $(\bb P)_x = -\frac{e}{8} ( \vev{c_2(\bx_{1b})}_F + \vev{c_2(\bx_{1d})}_F )$, and a similar result for $(\bb P)_y$.
We can generalize these results to other symmetry classes, and we summarize the results for $C_n$-symmetric insulators in classes A, AI, and AII in Tables~\ref{table:pol_1} and \ref{table:pol_2}.
%%%%%%

%%%%%%
\begin{figure}[t!]
\centering
\includegraphics[width=0.48\textwidth]{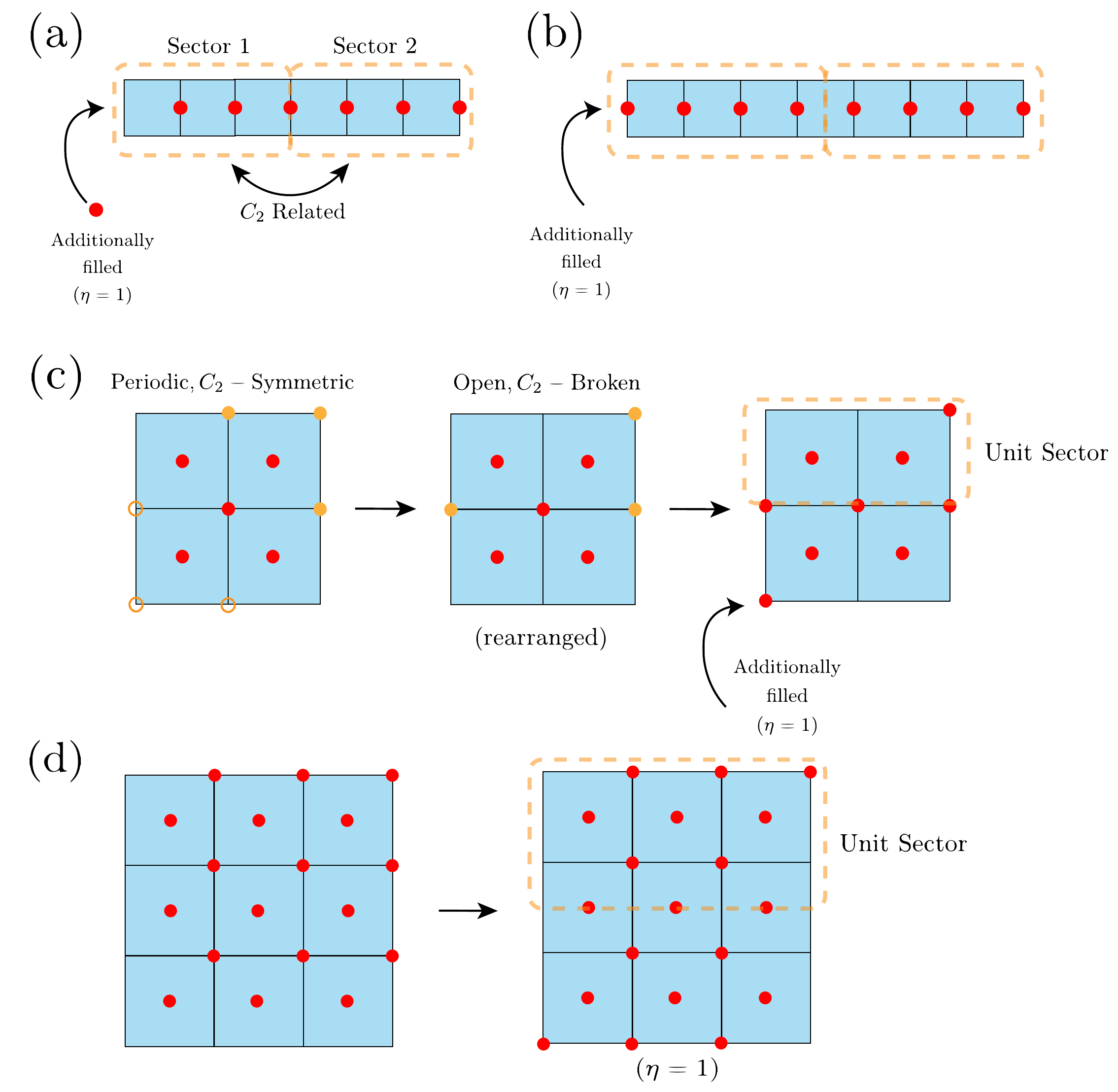}
\caption{
Illustration of the filling anomaly in a 1D obstructed atomic insulator (e.g., SSH model) with (a) an even number of unit cells ($N=6$), (b) an odd number of unit cells ($N=7$), and a 2D $C_2$ obstructed atomic insulator with (c) an even number of unit cells ($N_1 = N_2 = 2$), and (d) an odd number of unit cells ($N_1 = N_2 = 3$).
The yellow dashed lines in all the figures indicate a fundamental unit $C_2$ sector.
In (a) and (b), the 1D obstructed atomic insulator has a filling anomaly of $\eta = 1$ since an additional electron must be filled to restore the $C_2$ symmetry, at the expense of losing charge neutrality.
Similar reasoning holds for (c) and (d), in which the 2D $C_2$ obstructed atomic insulator has a filling anomaly of $\eta = 1 \pmod 2$.
In open boundary conditions, the neutral configuration of localized electrons cannot satisfy the $C_2$ symmetry even if the boundary electrons are rearranged.
This can also be seen from the fact that the number of electrons on the boundary is an odd integer, however, $C_2$ symmetry requires an even number of boundary electrons.
To respect $C_2$ symmetry, we must deviate from neutrality, i.e., a filling anomaly must occur.
}
\label{fig:sector_charge}
\end{figure}
%%%%%%

%%%%%%
\subsubsection{Fractional sector charge and filling anomaly}
%%%%%%
In addition to the bulk polarization, another important property of atomic insulators is the sector charge.
The sector charge is essentially the fractional charge in a $2\pi/n$ sector of a $C_n$-symmetric, finite lattice having open boundaries~\cite{Benalcazar2019,Peterson2020}.
It is a property that generalizes the notion of fractional corner charge when rotation symmetry is present, but perhaps not translation symmetry.

To compute the sector charge, a unit sector must be defined first.
To illustrate how the unit sector is defined, let us consider the 1D Su-Schrieffer-Heeger (SSH) model~\cite{Su1979} having $C_2$ symmetry.
Figure~\ref{fig:sector_charge}(a) illustrates a finite-length SSH chain having an even number of unit cells ($N_\trm{cells} \in 2 \N$).
This system has $C_2$ rotation which flips the sign of 1D coordinate $x$, i.e., $C_2: x \to -x$.
Using the $C_2$ rotation center, $x=0$, as a reference point, the system can be divided into two sectors: one sector is the region where $x \ge 0$, and the other one where $x \le 0$ [see Fig.~\ref{fig:sector_charge}(a)].
Note that these sectors are related to each other by $C_2$, they do not have any overlapping area, and together they cover the whole system.

Analogously, we can define $C_n$-related sectors in $C_n$-symmetric systems in any spatial dimension.
For example, a sector of a $C_n$-symmetric lattice is the region subtended by an angle of $2\pi/n$ radians around a fixed (unique for open boundaries) rotation center.
The other $(n-1)$ sectors can be identified with $C_n$-rotated copies of the first sector.
Figures~\ref{fig:sector_charge}(c)-(d) illustrate how $C_2$-related sectors are defined in a 2D $C_2$-symmetric lattice.

Once a $C_n$-symmetric lattice has a clear division into its $n$ sectors, the sector charge $Q_{\rm sector}$ is defined as follows.
The total system has charge $Q_{\rm tot}$.
Because $n$ sectors are related by $C_n$ without overlapping area, the sector charge is
\bg
Q_{\rm sector} = \frac{Q_{\rm tot}}{n}.
\label{eq:sector_q_def1}
\eg
For a system having periodic boundary conditions, the total charge is simply defined by the filling per unit cell $\nu$, and the total number of unit cells $N_{\rm cells}$: $Q_{\rm total} = e \nu N_{\rm cells}$.
Thus $Q_{\rm sector} = e \nu N_{\rm cells}/n$ for periodic boundary conditions\footnote{Note that we are not considering ionic contributions to the total charge or the sector charge.}.

However, the situation is more interesting when we apply open boundary conditions.
In open boundary conditions, the filling $\nu N_{\rm cells}$ defined in periodic boundary conditions may not be compatible with keeping the crystalline symmetry of the system.
In fact, this is the case in the obstructed atomic insulator phase of the SSH model, which has a quantized polarization $e/2$.
In the obstructed phase, one electron is filled per unit cell $(\nu=1)$, and the Wannier functions are centered at the unit cell boundary ($x=1/2$ mod 1) as shown in Fig.~\ref{fig:sector_charge}(a).
When periodic boundary conditions are applied, the electron configuration with total number of electrons $\nu N_{\rm cells}=N_{\rm cells}$ shown in Fig.~\ref{fig:sector_charge} respects the $C_2$ rotation.
(Note that the left and right edges should be identified in this case because of periodic boundary conditions.)

In contrast, a configuration of $N_{\rm cells}$ electrons in the obstructed phase must break $C_2$ rotation when open boundary conditions are implemented (see Ref.~\onlinecite{Benalcazar2019} for more detail).
In our example shown in Fig.~\ref{fig:sector_charge}(a), one electron occupies the right edge, but the left edge is empty.
While which edge is occupied and which is empty is a choice, the two ends will always be unbalanced at this filling.
Crucially, the $C_2$ symmetry can be restored if we additionally fill one electron at the left edge or unfill one electron at the right edge.
Now, the system respects the $C_2$ symmetry but the total number of electrons is changed by $\eta=\pm 1$ from the total number of electrons needed to make a symmetric, insulating configuration in periodic boundary conditions.
This phenomena is called the filling anomaly~\cite{Benalcazar2019}.
Importantly, the filling anomaly is not present in the trivial atomic insulator phase of the SSH chain.
This implies that the filling anomaly depends on the locations of the Wannier centers of the occupied bands.
For some configurations of Wannier centers, a system having open boundary conditions cannot respect the crystal symmetry at the same filling required for a symmetric insulator in periodic boundary conditions.
The filling anomaly occurs in other symmetry classes and higher-dimensional systems as well.
For example, the filling anomaly in a 2D $C_2$-symmetric insulator is illustrated in Figs.~\ref{fig:sector_charge}(c)-(d).

The point of reviewing the filling anomaly was to motivate contributions to the sector charge for rotation invariant insulators having open boundaries.
As such, we find that, for open boundary conditions, the contribution from the filling anomaly must be considered when computing the total and sector charges.
Let us denote the filling-anomaly as $\eta$.
If $\eta>0$ ($\eta<0$) then, to keep the symmetry in an open boundary system, $|\eta|$ additional electrons must be filled (unfilled) with respect to the symmetric, insulating filling defined for a translation symmetric system having periodic boundary conditions.
We note that $\eta=0$ means there is no filling anomaly.
The total charge for a system having open boundary conditions is now given by $Q_{\rm tot} = e \nu N_{\rm cells} + e \eta$, and thus the sector charge is
\bg
Q_{\rm sector} = e \nu N_{\rm cells}/n + e \eta/n.
\label{eq:sector_q_def2}
\eg
Note that the sector charge depends on $\nu, N_{\rm cells}$ and $\eta$.

As an example, in Fig.~\ref{fig:sector_charge}(a) the SSH model has an even number of unit cells, a periodic filling factor $\nu=1$, and a filling anomaly factor $\eta = 1$.
In this case, $Q_{\rm sector} = e \nu N_{\rm cells}/2 + e/2 = e\N + e/2$, and thus $Q_{\rm sector} = e/2$ mod $e$.
When the number of unit cells is odd, as shown in Fig.~\ref{fig:sector_charge}(b), the electronic contribution to the sector charge is computed as $Q_{\rm sector} = e N_{\rm cells}/2 + e/2 = e(\N +1/2) + e/2$.
Thus $Q_{\rm sector} = 0$ mod $e$.

To find the mapping between the fully traced TCMs and the sector charge, we need to find the relationship between the sector charge and Wannier-orbital multiplicities, which can then be subsequently related to the fully traced TCMs as we just carried out for the polarization.
To achieve this, let us determine the total charge $Q_{\rm total}$ from a slightly different perspective~\cite{Takahashi2021}.
Suppose that a ground state with $Q_{\rm tot}$ in Eq.~\eqref{eq:sector_q_def2} respects $C_n$ symmetry.
Note that the rotation center must coincide with a maximal WP $X$ with $C_n$ on-site symmetry.
However, there is a subtlety we now point out.
A $C_2$-symmetric system having open boundary conditions has a unique $C_2$ center that depends on the linear sizes $N_{1}, N_2$.
The rotation center coincides with (i) the WP $1a$ if $(N_1,N_2)=(1,1) \pmod 2$, (ii) the WP $1b$ if $(N_1,N_2)=(0,1) \pmod 2$, (iii) the WP $1c$ if $(N_1,N_2)=(1,0) \pmod 2$, and (iv) the WP $1d$ if $(N_1,N_2)=(0,0) \pmod 2$.
The cases (i) and (iv) are shown in Figs.~\ref{fig:sector_charge}(d) and (c) respectively.
Note that although the center WP is determined by $(N_1,N_2)$ mod 2, the on-site symmetry of the rotation center is always $C_2$ in all four cases.

For atomic insulators where the electronic Wannier orbital functions are exponentially localized, the total charge is determined by two contributions, (i) the electrons localized at the rotation center $X$ and (ii) those away from $X$.
Fortunately, there is a simplification because the electrons away from $X$ form $C_n$ symmetric multiplet configurations, and thus their number is $n \Z$.
For example, let us consider the SSH chain with an even number of unit cells shown in Fig.~\ref{fig:sector_charge}(a).
There is a single electron at the $C_2$ center, which corresponds to the contribution (i).
In addition to the electron at the center, there are 4 [6] other electrons in our diagram when the filling-anomaly factor $\eta$ is $-1$ [1].
These electrons correspond to the contribution (ii).
They respect the $C_2$ symmetry, and thus their number is $2\Z$ since they clearly come in $C_2$-related pairs.
Similar arguments hold for an SSH chain having an odd number of unit cells [Fig.~\ref{fig:sector_charge}(b)].
Furthermore, the arguments immediately apply to the $2D$ $C_2$-symmetric insulators illustrated in Figs.~\ref{fig:sector_charge}(c)-(d), and any other $C_n$-symmetric atomic insulators.

With this in mind, we can immediately conclude that for generic $C_n$-symmetric atomic insulators, $Q_{\rm tot} \pmod{ne}$ is solely determined by the number of electrons localized at the rotation center.
This is simply expressed as
\bg
Q_{\rm tot} = e \sum_\ell \, n_X^{(\ell)} \pmod{ne}.
\eg
Combining this and Eq.~\eqref{eq:sector_q_def2}, we find the expression for the sector charge (modulo $e$) and the filling-anomaly factor (modulo $n$ for $C_n$ symmetry):
\bg
Q_{\rm sector} = \frac{e}{n} \sum_\ell \, n_X^{(\ell)} \pmod e,
\nn
\eta = \sum_\ell \, n_X^{(\ell)} - \nu N_{\rm cells} \pmod n.
\label{eq:sector_q_and_fa}
\eg
In general, both the sector charge and the filling-anomaly factor $\eta$ depend on the values of $N_{1,2}$, the filling $\nu$ per unit cell, and the orbital multiplicities $n^{(\ell)}_X$.
Indeed, for the SSH chain example above we saw that the sector charge flipped from half-integer to integer when the number of unit cells flipped from even to odd.
Note that the sector charge depends on $N_{1,2}$ implicitly, as the types of WPs $X$ corresponding to the rotation centers are determined by $N_{1,2}$.
%

%%%%%%
{
\renewcommand{\arraystretch}{1.8}
\begin{table*}[t!]
\centering
\caption{
Bulk polarization for AZ class A (for both spinless and spin-1/2) $C_n$-symmetric atomic insulators expressed in terms of fully traced TCMs evaluated over maximal WPs.
The bulk polarization for $C_4$-symmetric atomic insulators has alternative expressions in terms of fully traced TCMs (see Supplemental Material~\cite{supple}).
Each component of the polarization is defined modulo $e$, i.e., $\bb P = p_1 \bb a_1 + p_2 \bb a_2$ with $p_{1,2}$ (mod $e$) where $e=-|e|$.
For $C_4$, $s_{1,2} \in \{-1,+1\}$.
}
\label{table:pol_1}
\begin{tabular*}{0.9\textwidth}{@{\extracolsep{\fill}} c| c c}
\hline \hline
 & A (Spinless) & A (Spin-1/2)
\\ \hline
\multirowcell{2}{$C_2$} & $p_1 = -\frac{e}{8} (\vev{c_2(\bx_{1b})}_F + \vev{c_2(\bx_{1d})}_F)$ & $p_1 = i \frac{e}{8} (\vev{c_2(\bx_{1b})}_F + \vev{c_2(\bx_{1d})}_F)$
\\
& $ p_2 = -\frac{e}{8} (\vev{c_2(\bx_{1c})}_F + \vev{c_2(\bx_{1d})}_F)$ & $p_2 = i \frac{e}{8} (\vev{c_2(\bx_{1c})}_F + \vev{c_2(\bx_{1d})}_F)$
\\
$C_3$ & $p_1=p_2 = \frac{4e}{9} \trm{Re} [ \vev{c_3(\bx_{1b})}_F - \vev{c_3(\bx_{1c})}_F ]$ & $p_1=p_2 = \frac{2e}{9} \trm{Re} [ \vev{c_3(\bx_{1b})}_F - \vev{c_3(\bx_{1c})}_F ]$
\\
$C_4$ & $p_1=p_2 = \frac{e}{8} ( s_1 \vev{c_2(\bx_{1b})}_F + s_2 \vev{c_2(\bx_{2c})}_F )$ & $p_1=p_2 = i \frac{e}{8} ( s_1 \vev{c_2(\bx_{1b})}_F + s_2 \vev{c_2(\bx_{2c})}_F )$
\\
$C_6$ & $p_1=p_2=0$ & $p_1=p_2=0$
\\ \hline \hline
\end{tabular*}
\end{table*}
}
%%%%%%

%%%%%%
{
\renewcommand{\arraystretch}{1.8}
\begin{table*}[t!]
\centering
\caption{
Bulk polarization for AZ classes AI and AII $C_n$-symmetric atomic insulators expressed in terms of fully traced TCMs evaluated over maximal WPs.
The bulk polarization for $C_4$-symmetric atomic insulators has other expressions in terms of fully traced TCMs (see Supplemental Material~\cite{supple}).
Each component of the polarization is defined modulo $e$, i.e., $\bb P = p_1 \bb a_1 + p_2 \bb a_2$ with $p_{1,2}$ (mod $e$) where $e=-|e|$.
For $C_4$, $s_{1,2} \in \{-1,+1\}$.
}
\label{table:pol_2}
\begin{tabular*}{0.9\textwidth}{@{\extracolsep{\fill}} c| c c}
\hline \hline
 & AI & AII
\\ \hline
\multirowcell{2}{$C_2$} & $p_1 = -\frac{e}{8} (\vev{c_2(\bx_{1b})}_F + \vev{c_2(\bx_{1d})}_F)$, & \multirowcell{2}{$p_1=p_2=0$}
\\
& $ p_2 = -\frac{e}{8} (\vev{c_2(\bx_{1c})}_F + \vev{c_2(\bx_{1d})}_F)$ &
\\
$C_3$ & $p_1=p_2 = \frac{e}{9} ( \vev{c_3(\bx_{1b})}_F - \vev{c_3(\bx_{1c})}_F )$ & $p_1=p_2 = \frac{2e}{9} ( \vev{c_3(\bx_{1b})}_F - \vev{c_3(\bx_{1c})}_F )$
\\
$C_4$ & $p_1=p_2 = \frac{e}{8} ( s_1 \vev{c_2(\bx_{1b})}_F + s_2 \vev{c_2(\bx_{2c})}_F )$ & $p_1=p_2 = 0$
\\
$C_6$ & $p_1=p_2=0$ & $p_1=p_2=0$
\\ \hline \hline
\end{tabular*}
\end{table*}
}
%%%%%%

The final step is to connect the sector charge and filling-anomaly factor formulas in Eq.~\eqref{eq:sector_q_and_fa} to the fully traced TCMs.
Even though the sector charge and filling-anomaly factor are defined under open boundary conditions, they are still determined by linear combinations of orbital multiplicities, as shown in Eq.~\eqref{eq:sector_q_and_fa}.
These linear combinations can be computed by evaluating the fully traced TCMs under periodic boundary conditions as follows.
First, the factor $\nu N_{\rm cells}$ can be determined by $N_{1,2}$ and the fully traced TCM for the trivial symmetry operation $\mathds{1}$; $N_{\rm cells}= N_1 N_2$ and $\nu = \vev{\mathds{1}}_F = {\rm Tr}[P_{GS}]$.
Second, Eq.~\eqref{eq:fully_tcm_Wannier} maps the fully traced TCM $\vev{c_n(\bb r_o)}_F$ to a linear combination of orbital multiplicities, $\sum_{W \in \mc{X}[c_n(\bb r_o)]} \, e^{2 \pi i \ell/n} \, n_W^{(\ell)}$, at several WPs in $\mc{X}[c_n(\bb r_o)]$.
Hence, under the constraint Eq.~\eqref{eq:perfect_constraint}, the sector charge can be expressed solely in terms of the fully traced TCMs, as we can simply apply the mapping between the fully traced TCMs and Wannier orbital multiplicities.
In this case, for a system having open boundaries, the rotation center $X$ corresponds to a WP $1d$, (see Fig.~\ref{fig:sector_charge}(c)).
Then the factor $\sum_{\ell=0}^{n-1} n_X^{(\ell)}$ in Eq.~\eqref{eq:sector_q_and_fa} becomes $n_{1d}^{(0)}+n_{1d}^{(1)}$, which is identical to $-n_{1d}^{(0)}+n_{1d}^{(1)}$ modulo 2, since $n_{1d}^{(\ell)} \in \Z$ for (obstructed) atomic insulators.
Then, from Eq.~\eqref{eq:C2_TCM_Wannier_orb_mapping}, we obtain $\sum_{\ell=0}^{n-1} n_X^{(\ell)} = -\frac{1}{4} \vev{c_2(\bx_{1d})}_F$ mod 2.
This implies that $Q_{\rm sector} = -\frac{e}{8} \vev{c_2(\bx_{1d})}_F$ mod $e$ for a $C_2$-symmetric system.
Accordingly, the filling-anomaly factor $\eta$ is given by $-\frac{1}{4} \vev{c_2(\bx_{1d})}_F$ mod $2$.
(Note that the factor $\nu N_{\rm cell}=\nu N_1 N_2$ in Eq.~\eqref{eq:sector_q_and_fa} becomes 0 mod 2 for any integer, periodic filling $\nu$, when Eq.~\eqref{eq:perfect_constraint} is assumed.)
We computed the TCM expressions of the sector charge for class A insulators (see Table~\ref{table:sector_charge_1}), and for class AI and AII insulators (see Table~\ref{table:sector_charge_2}) for each $C_n$.

Unfortunately, when Eq.~\eqref{eq:perfect_constraint} is not satisfied it is not immediately obvious how to carry out this mapping for a generic lattice type with arbitrary $N_{1,2}$.
This is because when Eq.~\eqref{eq:perfect_constraint} is not satisfied, $\mc{X}[c_n(\bb r_o)]$ is composed of different WPs, while, e.g., the right-hand side of the sector charge formula in Eq.~\eqref{eq:sector_q_and_fa} requires summing the orbital multiplicities $n^{(\ell)}_X$ at the same WP $X$.
We will treat this case separately in Sec.~\ref{sec:map_general}.
(See also the comments in Sec.~\ref{subsec:map_comment} for the case without satisfying Eq.~\eqref{eq:perfect_constraint}.)

Now, let us comment further on the sector charge $Q_{\rm sector}$ and the filling-anomaly factor $\eta$.
First, when Eq.~\eqref{eq:perfect_constraint} is satisfied, $Q_{\rm sector}$ and $\eta$ are simply proportional to each other:
\bg
Q_{\rm sector} = \frac{e}{n}\eta \pmod e.
\label{eq:sector_q_fa_simple}
\eg
We can see this because the factor $\nu N_{\rm cell}$ in Eq.~\eqref{eq:sector_q_and_fa} becomes $n\Z$ when the constraint Eq.~\eqref{eq:perfect_constraint} is satisfied for a $C_n$-symmetric insulator.
Second, for class AII insulators, when only the electronic charges are considered, $Q_{\rm sector}$ ($\eta$) can be defined modulo $2e$ ($2n$) because of Kramers theorem.
However, this is not true if the contribution from the ions is carefully studied~\cite{Watanabe2020}.
Specifically, time-reversal symmetry allows an ion with an integer spin and odd charge, i.e., $e$ mod $2e$.
When such ions are added in a $C_n$-symmetric manner, the total charge of the system is only well defined modulo $ne$ instead of $2ne$, meaning that the sector charge is well defined modulo $e$.
Thus, even when we consider only the electronic contribution for a simpler discussion, we still define $Q_{\rm sector}$ modulo $e$ and $\eta$ modulo $n$ for the classes A, AI, and AII, as in Refs.~\onlinecite{Watanabe2020,Takahashi2021}.
Moreover, defining $Q_{\rm sector}$ modulo $e$ offers an advantage.
In this case, $Q_{\rm sector}$ can be determined by the fully traced TCMs as we have shown.
Note that $Q_{\rm sector}$ (similarly, the corner charges) mod $2e$ is not symmetry-indicated such that the momentum-space data does not determine $Q_{\rm sector}$ uniquely~\cite{Schindler2019,Kooi2021}.
%%%%%%

%%%%%%
{
\renewcommand{\arraystretch}{1.8}
\begin{table*}[t!]
\centering
\caption{
Sector charge for AZ class A (both spinless and spin-1/2) $C_n$-symmetric atomic insulators expressed in terms of fully traced TCMs evaluated over maximal WPs, for lattices with dimensions $N_1 \times N_2$ given by Eq.~\eqref{eq:perfect_constraint}.
The sector charge is defined modulo $e$ where $e=-|e|$.
While the sector charge measures the net charge in a $2\pi/n$ rad sector of an open boundary $C_n$-symmetric lattice, the expressions below are in terms of fully traced TCMs evaluated on $C_n$-symmetric lattices having periodic boundaries.
}
\label{table:sector_charge_1}
\begin{tabular*}{\textwidth}{@{\extracolsep{\fill}} c| c c}
\hline \hline
 & A (Spinless) & A (Spin-1/2)
\\ \hline
$C_2$ & $-\frac{e}{8} \vev{c_2(\bx_{1d})}_F$ & $i \frac{e}{8} \vev{c_2(\bx_{1d})}_F$
\\
$C_3$ & $\frac{4e}{9} \trm{Re} [\vev{c_3(\bx_{1a})}_F]$ & $\frac{2e}{9} \trm{Re} [\vev{c_3(\bx_{1a})}_F]$
\\
$C_4$ & $-\frac{e}{16} ( \vev{c_2(\bx_{1b})}_F + 4 \trm{Re} \vev{c_4(\bx_{1b})}_F )$ & $-\frac{e}{16} ( i \vev{c_2(\bx_{1b})}_F + 4 \trm{Re} [ e^{-\frac{i\pi}{4}} \vev{c_4(\bx_{1b})}_F ] )$
\\
$C_6$ & $-\frac{e}{72} (3 \vev{c_2(\bx_{1a})}_F + 24 \trm{Re} \vev{c_6(\bx_{1a})}_F + 8 \trm{Re} \vev{c_3(\bx_{1a})}_F )$ & $-\frac{e}{72} (3i \vev{c_2(\bx_{1a})}_F + 24 \trm{Im} [\vev{c_6(\bx_{1a})}_F ] - 8 \trm{Re} [\vev{c_3(\bx_{1a})}_F ] )$
\\ \hline \hline
\end{tabular*}
\end{table*}
}
%%%%%%

%%%%%%
{
\renewcommand{\arraystretch}{1.8}
\begin{table*}[t!]
\centering
\caption{
Sector charge for AZ classes AI and AII $C_n$-symmetric atomic insulators expressed in terms of fully traced TCMs evaluated over maximal WPs, for lattices with dimensions $N_1 \times N_2$ given by Eq.~\eqref{eq:perfect_constraint}.
The sector charge is defined modulo $e$ where $e=-|e|$.
While the sector charge measures the net charge in a $2\pi/n$ rad sector of an open boundary $C_n$-symmetric lattice, the expressions below are in terms of fully traced TCMs evaluated on $C_n$-symmetric lattices having periodic boundaries.
}
\label{table:sector_charge_2}
\begin{tabular*}{0.9\textwidth}{@{\extracolsep{\fill}} c| c c}
\hline \hline
 & AI & AII
\\ \hline
$C_2$ & $-\frac{e}{8} \vev{c_2(\bx_{1d})}_F$ & $\bb 0$
\\
$C_3$ & $\frac{e}{9} \vev{c_3(\bx_{1a})}_F$ & $\frac{2e}{9} \vev{c_3(\bx_{1a})}_F$
\\
$C_4$ & $-\frac{e}{16} ( \vev{c_2(\bx_{1b})}_F + 4\vev{c_4(\bx_{1b})}_F )$ & $-\frac{e}{4\sqrt{2}} \vev{c_4(\bx_{1b})}_F$
\\
$C_6$ & $-\frac{e}{72} (3 \vev{c_2(\bx_{1a})}_F + 24 \vev{c_6(\bx_{1a})}_F + 8 \vev{c_3(\bx_{1a})}_F )$ & $\frac{e}{9}\vev{c_3(\bx_{1a})}_F$
\\ \hline \hline
\end{tabular*}
\end{table*}
}
%%%%%%

%%%%%%
\subsection{Some comments on the Chern number, polarization, and sector charge}
\label{subsec:map_comment}
%%%%%%
Interestingly, the expressions for the bulk polarization, sector charge, and Chern number in terms of the fully traced TCMs elucidate fundamental differences in these topological responses in position-space, such as the dependence on unit cell origin, that are otherwise not apparent when each of these quantities are expressed solely in terms of momentum-space irrep multiplicities.
Although it was previously known, our results make manifest that the polarization depends on the unit-cell origin, and the sector charge depends on the rotation center.
Indeed, one must specify the unit-cell origin and rotation center (or linear system sizes $N_{1,2}$ since for open boundaries the rotation symmetry is uniquely determined by the linear dimensions) to define the polarization and the sector charge.
Our results in Tables~\ref{table:pol_1}-\ref{table:pol_2} are specifically for a crystal where the unit cell origin is located at the $1a$ WP.
Redefining the unit cell origin to another WP would change which Wannier orbital multiplicities contribute to the bulk polarization.
Hence, the polarization and sector charge differ from the Chern number, whose value (and manifest expression in terms of TCMs) is unit cell origin-independent, and linear dimension-independent, owing to the strong topological nature of the Chern insulator.
This precisely reflects the differing character of obstructed atomic insulators (which may have polarization and/or sector charge) and topological insulators (which may have Chern number).

Moreover, all of the topological quantities studied so far (which are the Chern number modulo $n$, polarization, sector charge, and filling-anomaly factor) are symmetry indicated, i.e., their values can be determined by momentum-space data.
What we have shown here is that the fully traced TCMs can completely replace the momentum-space data to diagnose the topology.
In other words, the TCMs play the role of position-space symmetry indicators.
Furthermore, this also implies that using the TCMs, topological quantities such as the Chern number and the sector charge still can be defined robustly without momentum-space consideration.

Finally, anticipating what is to come in the next section, we briefly comment on cases where the system is either (i) periodic but not obeying the constraint Eq.~\eqref{eq:perfect_constraint}, or (ii) has open boundaries.
For (i), we need to introduce the TCMs with twisted boundary conditions as we will demonstrate in Sec.~\ref{sec:map_general}.
There, we provide a definitive prescription for how the TCM formulas predicting the topological quantities, e.g., those shown in Tables~\ref{table:pol_1}-\ref{table:sector_charge_2}, change according to the choice of $(N_1,N_2)$.
For (ii), we show instead that the traced TCMs around the local peak at the unique rotation center can diagnose the topological quantities, and we illustrate this with concrete tight-binding calculations in Sec.~\ref{sec:example}.
%%%%%%

%%%%%%
\section{Finite-size lattices and twisting boundary conditions}
\label{sec:map_general}
%%%%%%
In the previous section, we demonstrated how the Chern number, bulk polarization, and sector charge for $C_n$-symmetric Chern insulators and atomic insulators, respectively, could be reformulated in terms of basis-independent, fully traced TCMs which rely on only the rotation operators and the ground state.
This removes the need to utilize symmetry data from the momentum-space band structure to characterize the bulk crystalline topology of $C_n$-symmetric insulators.
Furthermore, the momentum-space based approach of classifying topological crystalline insulators implicitly relies on the assumption that the dimensions of the finite-size lattice allow for the existence of the maximal set of high symmetry momenta in the BZ.
While it is true that in the thermodynamic limit (i.e., $N_{1,2} \to \infty$ such that the spacing between discrete momenta in the BZ vanishes) the maximal set of high symmetry momenta exists in the limit, this does not hold for every finite-size lattice.
Explicitly, a lattice having dimensions $N_1 \times N_2$ and periodic boundary conditions will have a crystal momentum vector quantized as shown in Eq.~\eqref{eq:momentum}.
Thus, unless $N_1$ and $N_2$ satisfy the constraints laid out in Eq.~\eqref{eq:perfect_constraint}, only a subset of all possible high symmetry momenta will exist in the BZ.
Therefore, it remains an open question how to apply the symmetry-based diagnosis of crystalline topology to generic $C_n$-symmetric lattices having arbitrary linear dimensions $N_1, N_2$.
Here we systematically develop a method that accomplishes this challenge by building on earlier results in Refs.~\onlinecite{Teo2008,Tanaka2020,Takahashi2020}.

To overcome this issue, we will show that the mappings between the fully traced TCMs on one side, and the Chern number, bulk polarization, and sector charge on the other, can be modified by adjusting the boundary condition(s) of the lattice along the dimension(s) that do not satisfy the constraint in Eq.~\eqref{eq:perfect_constraint}.
We will demonstrate that this procedure extends the classification of $C_n$-symmetric Chern insulators and (obstructed) atomic insulators using TCMs to any finite-size lattice.
%

%%%%%%
\begin{figure}[t!]
\centering
\includegraphics[width=0.475\textwidth]{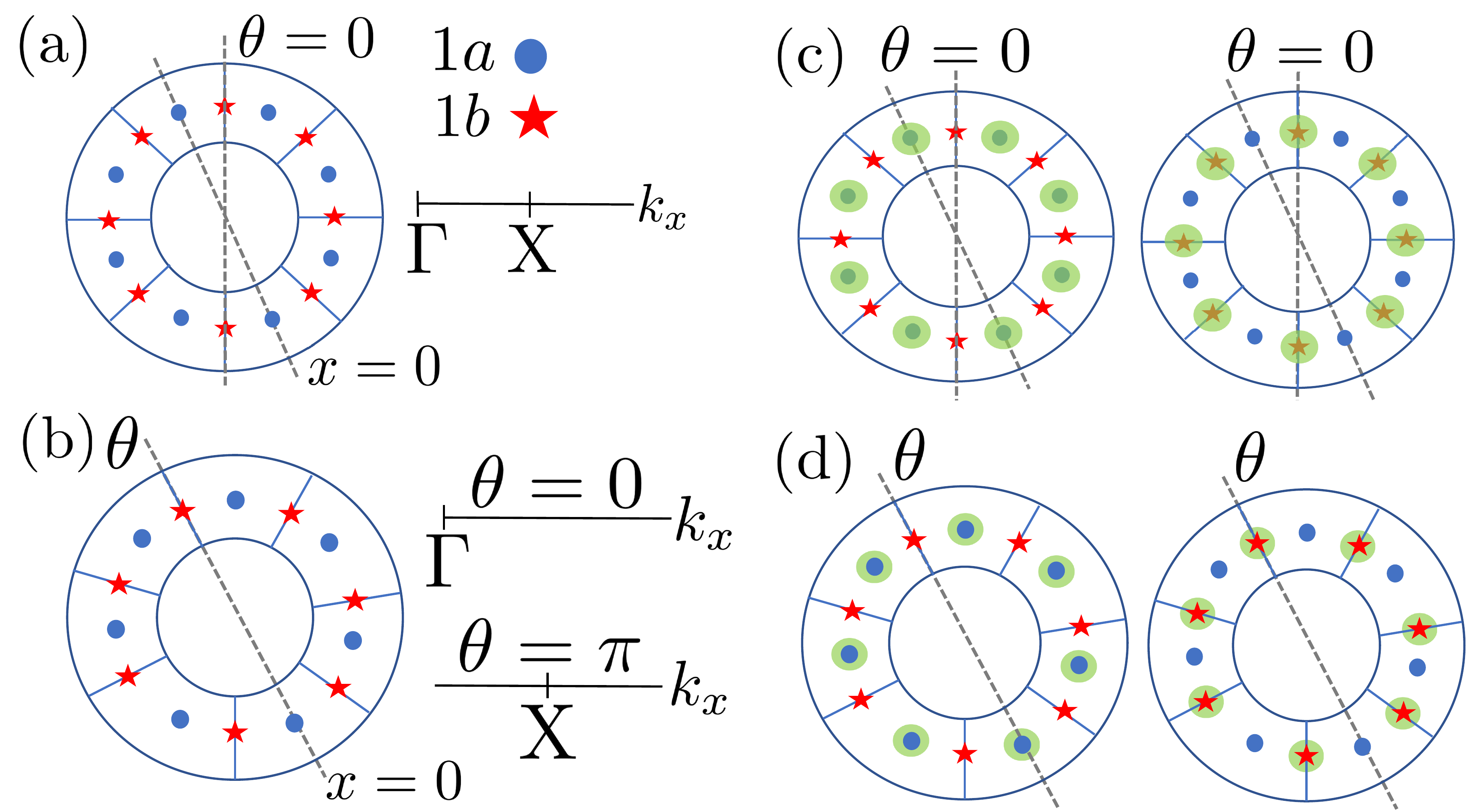}
\caption{
Simplified illustration of a finite-size, 1D atomic insulator such as the SSH chain with $C_2$ symmetry on a periodic lattice shown using a circular geometry to emphasize the periodicity.
Details of the inner structure of the unit cell, such as sublattice sites, have been omitted.
The blue circles denote the $1a$ WPs, and the red stars denote the $1b$ WPs.
The green circles indicate electronic Wannier orbital centers.
The parameter $\theta$ indicates the boundary condition twist/flux at the $C_2$ rotation center ($\theta=0$ periodic, $\theta=\pi$ anti-periodic).
(a) An even number of unit cells and corresponding $C_2$-invariant momenta in BZ.
(b) An odd number of unit cells and the corresponding $C_2$-invariant momenta for each boundary condition.
In both (a) and (b), $x=0$ denotes the coordinates of the unit cell serving as the origin of the lattice.
Positions of electronic Wannier orbital centers in (c) even and (d) odd unit cell cases for both the trivial (left) and obstructed (right) atomic insulator phases.
In order to distinguish the phases of the SSH chain in (d), $\theta=0$ and $\theta=\pi$ must both be considered (i.e., periodic and anti-periodic boundary conditions).
}
\label{fig:ssh}
\end{figure}
%%%%%%

To begin, let us consider a $C_n$-symmetric lattice having arbitrary dimensions $N_1 \times N_2$, where $N_1$ and $N_2$ are such that the $C_n$ symmetry is preserved (e.g., $N_1 \neq N_2$ in general for $C_2$ symmetry, but $N_1 = N_2 =N$ for $C_3$, $C_4$, and $C_6$ symmetry).
We will now consider {\it twisting} the boundary conditions on the lattice, with twists parameterized by $(\theta_1,\theta_2)$, where $\theta_1$ and $\theta_2$ correspond to the flux through the cycles parallel to $\bb a_1$ and $\bb a_2$ respectively.
The twisted boundary conditions can be formulated in terms of the position-space basis as 
\bg
\ket{\bR + N_1 \bb a_1, \alpha} = e^{-i \theta_1}\ket{\bR, \alpha}
\nn
\ket{\bR + N_2 \bb a_2, \alpha} = e^{-i \theta_2}\ket{\bR, \alpha}.
\label{eq:tbc}
\eg
With twisted boundary conditions $\theta_{1,2}$, the quantization of crystalline momentum is modified as:
\bg
\bk = \left( \frac{n_1}{N_1} - \frac{\theta_1}{2\pi N_1} \right) \bb b_1 + \left( \frac{n_2}{N_2} - \frac{\theta_2}{2\pi N_2} \right) \bb b_2.
\label{eq:momentum_tbc}
\eg

Now, for arbitrary, finite $N_1$ and $N_2$, when untwisted periodic boundary conditions are present ($\theta_1 = \theta_2 = 0$ mod $2\pi$), the set of discrete of momenta given by Eq.~\eqref{eq:momentum_tbc} may contain only a subset of high symmetry momenta.
In such cases it is possible to adjust $\theta_1$ and $\theta_2$ such that the set of discrete momenta given by Eq.~\eqref{eq:momentum_tbc} contains a different subset of high symmetry momenta.
Therefore, for arbitrary finite $N_1$ and $N_2$ that do not satisfy Eq.~\eqref{eq:perfect_constraint}, one can consider lattices having different boundary conditions parameterized by $(\theta_1, \theta_2)$ such that the complete set of high symmetry momenta is obtained by considering a set of systems having different twists.
%%%%%%

%%%%%%
\subsection{1D example: SSH chain}
%%%%%%
To demonstrate the necessity of twisting boundary conditions, we first consider a simple model for a one-dimensional atomic insulator with $C_2$ symmetry, such as the Su-Schrieffer-Heeger (SSH) chain~\cite{Su1979}, as shown in Fig.~\ref{fig:ssh}.
Each gapped phase of the SSH model admits a Wannier representation, and there is a clear distinction between the obstructed atomic insulator phase (i.e., Wannier orbitals are localized at WP $1b$), and the trivial atomic insulator phase (i.e., Wannier orbitals localized at WP $1a$).
We know this distinction is captured by the momentum-space data of the occupied energy band at the $\Gamma$ and $X$ points~\cite{Hughes2011,Turner2012}, and thus, we should be able to apply our methods to give an alternative characterization.
Although we describe this procedure for a one-dimensional system for simplicity, this generalizes straightforwardly to two dimensions for any $C_n$ symmetry.

In order to make the discussion presented in this subsection concrete, we introduce the following Hamiltonian for the SSH chain in the tight-binding position-space basis $\ket{x, \sg=A,B}$,
\ba
& H_{\trm{SSH, PBC}} = \sum_{x=1}^N m (\ket{x,B} \bra{x,A} + \ket{x,A} \bra{x,B})
\nn
& +\sum_{x=1}^N t(\ket{x+1,A} \bra{x,B} + \ket{x,B} \bra{x+1,A}).
\label{eq:SSHtb}
\ea
where $A, B$ are the sublattice degrees of freedom, and periodic boundary conditions are specified as $\ket{x+N,\sg}=\ket{x,\sg}$.
For a periodic, translation invariant chain we can Fourier transform to arrive at the Bloch Hamiltonian
\bg
H_{\rm SSH} (k_x) = (m + t \cos k_x) \sg_1 + t \sin k_x \sg_2,
\label{eq:SSHBlochHamiltonian}
\eg 
where $\sg_i$ for $i \in \{1,2,3\}$ are the conventional Pauli matrices with respect to the two-atom sublattice basis in each unit cell.
This model is $C_2$-symmetric, with the $C_2$ operator given by $\sg_1$, e.g., $\sigma_1 H_{\rm SSH} (k_x) \sigma_1 = H_{\rm SSH} (-k_x)$, as well as chiral-symmetric with the chiral operator given by $\sg_3$, e.g., $\sg_3 H_{\rm SSH}(k_x) \sg_3 = - H_{\rm SSH} (k_x)$.
The two bulk bands of this model are therefore symmetric in energy around $E=0$.
Here, we take the ground state to be such that the Fermi level is tuned to $E_F=0$ within the bulk gap so that the lower band is always occupied.

There are two phases of this model.
The first is the so-called ``obstructed atomic limit" phase which occurs for $|m|<|t|$.
Here, the ground state is given by a periodic arrangement of electrons localized at the $1b$ WP of every unit cell, yielding a non-trivial bulk polarization of $\bb P=(e/2)\bb a_1 \pmod e$.
The second phase is the ``trivial atomic limit" phase, which occurs for $|m|>|t|$ where the ground state is given by a periodic arrangement of electrons localized at the $1a$ WP of every unit cell, resulting in a trivial bulk polarization $\bb P=\bb 0 \pmod e$.

Now let us study the properties of finite chains having an even or odd number of unit cells (the former satisfying the constraints Eq.~\eqref{eq:perfect_constraint}, while the latter does not).
First, we consider the conventional case of a periodic SSH chain having an {\it even} number of unit cells $N_1=2q$ where $q \in \N$.
In this case, the momentum is simply given by $k_x = (2\pi/N_1) n_1$ for $n_1 \in \mathbb{Z}$.
Since $N_1$ is even, this implies that both $C_2$-invariant HSM $k_x=0$ (the $\Gamma$-point) and $k_x=\pi$ (the $X$-point) are allowed as discrete momentum points in the BZ, i.e., $n_1=0$ and $n_1 = N_1/2$ respectively.
Hence, in momentum space we have a set of two $\Gamma$-point irrep multiplicities and two rotation invariants given by $\{ m(\Gamma_1), m(\Gamma_2), [X_1],[X_2] \}$\footnote{The lower indices 1 and 2 correspond to $C_2$ eigenvalues of $+1$ and $-1$, respectively, representing angular momentum 0 and 1 mod 2.}.

Another consequence of a having an even number of unit cells, is that the periodic SSH chain has two distinct $C_2$ rotation center choices given by either a pair of invariant $1a$ WPs or a pair of invariant $1b$ WPs.
Thus, using the $C_2$ rotation operators at the $1a$ and the $1b$ WPs, one can construct a corresponding fully traced TCM for each.
Hence, in position space we have two fully traced TCMs given by $\{ \vev{c_2(\bx_{1a})}_F^{\theta=0}, \vev{c_2(\bx_{1b})}_F^{\theta=0} \}$.
Given that a bulk gap exists, and the number of occupied bands is specified by the filling $\nu$, the set of momentum space quantities can be further constrained to just two quantities, one at each $C_2$-invariant momenta, e.g., $\{ m(\Gamma_1), [X_1] \}$.
Hence, for a one-dimensional atomic insulator having an even number of unit cells and $C_2$ symmetry, there exists a one-to-one mapping between the pair of fully traced TCMs and the reduced set of momentum-space irrep multiplicities and rotation invariants.
Both sets of quantities can distinguish the trivial and obstructed phases without having to twist the boundary conditions.

Now we consider the SSH chain having an odd number of unit cells $N_1=2q+1$.
With periodic boundary conditions the momentum is still quantized as $k_x = (2\pi/N_1) n_1$ for $n_1 \in \mathbb{Z}$.
However, since $N_1$ is odd, only one $C_2$-invariant momentum is allowed as a discrete momentum point in the BZ, i.e., $n_1=0$ the $\Gamma$-point.
The $X$ point would require $n_1 = N_1/2 = q+1/2 \notin \mathbb{Z}$.
Hence, the momentum-space symmetry data consists of simply $\{ m(\Gamma_1)_{\theta=0}, m(\Gamma_2)_{\theta=0} \}$, where we have specified the twisted boundary conditions parameterized by $\theta$ in the irrep multiplicity $m(\bar{\bk}_{p})$.
Furthermore, in position space, unlike the even unit cell case, where the $1a$ ($1b$) $C_2$ rotation center leaves two $1a$ ($1b$) WPs at antipodal points of the periodic lattice invariant under $C_2$, this is not true for the odd unit cell case.
In fact, as shown in Figs.~\ref{fig:ssh}(b) and (d), there is only one type of $C_2$ rotation center and it leaves invariant one $1a$ {\it and} one $1b$ WP at antipodal points of the periodic lattice.
For such lattices, we will label the $C_2$ rotation center by the WP with respect to the unit cell located at the origin of the lattice.
In this case, for the one-dimensional lattice shown in Figs.~\ref{fig:ssh}(b) and (d), the $C_2$ rotation center will be denoted by the $1a$ WP.

Unfortunately, neither the irrep multiplicities $m(\Gamma_{1,2})_{\theta=0}$ nor the fully traced TCM $\vev{c_2(\bx_{1a})}_F^{\theta=0}$ contain enough information to distinguish the trivial and obstructed phases of the SSH chain with odd number of unit cells.
For example, if we consider the irrep multiplicities for the occupied band of Eq.~\eqref{eq:SSHBlochHamiltonian}, there is always one occupied Bloch state at $\Gamma$ with a $-1$ $C_2$ eigenvalue, which means $m(\Gamma_1)_{\theta=0}=0$ and $m(\Gamma_2)_{\theta=0}=1$ in both phases.
For comparison, in position space there is either an electronic Wannier orbital localized at the $1a$ WP on the $C_2$ rotation center (trivial phase) with $C_2$ eigenvalue $-1$ or one localized at the $1b$ WP (obstructed atomic limit) phase with $C_2$ eigenvalue of $-1$.
Hence, the fully traced TCM for an odd number of unit cells will receive equal contributions from both configurations and hence cannot distinguish the phases.

We can clearly illustrate the orbital configurations in the obstructed atomic limit (with $m=0$ and $t \neq 0$) and the trivial atomic limit (with $m \neq 0$ and $t=0$) of Eq.~\eqref{eq:SSHBlochHamiltonian}.
In these limits, the single-particle eigenstates of the tight-binding Hamiltonian and the Wannier orbital states are identical.
To be explicit, let the $C_2$ rotation center refer to the $c_2(\bx_{1a})$ operator.
Note that in position-space for $N_1=2q+1$, $c_2(\bx_{1a})$ maps $\ket{x,A} \to \ket{N_1-x,B}$ and $\ket{x,B} \to \ket{N_1-x,A}$.
In the $m=0, t \neq 0$ limit, the electronic Wannier orbital state localized at the $C_2$ rotation center is $\ket{W_{-} (N_1/2)} = \frac{1}{\sqrt{2}}(\ket{(N_1+1)/2, A} - \ket{(N_1-1)/2, B})$, and it follows that $c_2(\bx_{1a}) \ket{W_{-} (N_1/2)} = -\ket{W_{-} (N_1/2)}$, implying that $\ket{W_{-}(N_1/2)}$ has a $C_2$ eigenvalue of $-1$.
Similarly, in the $m\neq 0,t=0$ limit, the electronic Wannier orbital state localized at the $C_2$ rotation center located at the $1a$ WP is now given as $\ket{W_{-} (0)} = \frac{1}{\sqrt{2}}(\ket{0,B} - \ket{0,A})$, which means $c_2(\bx_{1a}) \ket{W_{-}(0)} = -\ket{W_{-}(0)}$.
For reference, in Fig.~\ref{fig:ssh} (d), an illustration of the Wannier orbital centers is provided for the trivial and obstructed atomic limits for $N_1=7$ unit cells.
This analysis extends to all parameter choices within the obstructed atomic limit phase ($|m|<|t|$) and trivial atomic limit phase ($|m|>|t|$) since the limits $m=0, t \neq 0$ and $m\neq 0, t=0$ are adiabatically connected (while preserving $C_2$) to different parameter choices within each of their respective phases due to the existence of a bulk gap.
This means $\vev{c_2(\bx_{1a})}_F^{\theta=0} = -1$ in both phases, and hence does not distinguish the phases.
%

%%%%%%
{
\renewcommand{\arraystretch}{1.8}
\begin{table}
\centering
\caption{
TCMs for the 1D lattice having even ($N=0$ mod 2) unit cells lattice corresponding to each $C_2$ center, i.e., for the $1a$ and $1b$ WPs, as shown in the first column.
The fully traced TCMs for the even unit cell lattice can be mapped to the fully traced TCMs for the odd ($N = 1$ mod 2) unit cell lattice having different boundary conditions (periodic, $\theta=0$ and anti-periodic, $\theta=\pi$), as shown in second column.
}
\label{table:even_odd}
\begin{tabular*}{0.38\textwidth}{@{\extracolsep{\fill}} c c}
\hline \hline
$N=0$ mod 2 & $N=1$ mod 2
\\
\hline
$\vev{c_2(\bx_{1a})}_F^{\theta=0}$ & $\vev{c_2(\bx_{1a})}_F^{\theta=0} + \vev{c_2(\bx_{1a})}_F^{\theta=\pi}$
\\
$\vev{c_2(\bx_{1b})}_F^{\theta=0}$ & $\vev{c_2(\bx_{1a})}_F^{\theta=0} - \vev{c_2(\bx_{1a})}_F^{\theta=\pi}$
\\ \hline \hline
\end{tabular*}
\end{table}
}
%%%%%%

The resolution to this issue is to introduce anti-periodic boundary conditions ($\theta=\pi$) at the $C_2$ rotation center, and to {\it jointly} consider the symmetry data obtained from {\it both} periodic and anti-periodic boundary conditions.
The twisted boundary conditions are introduced by choosing a gauge where the full flux $\theta$ passes through a $C_2$ invariant bond, i.e., where the $1b$ WP is located as shown in Fig.~\ref{fig:ssh}.

For this discussion, we will focus on the tight-binding Hamiltonian with periodic boundary conditions in Eq.~\eqref{eq:SSHBlochHamiltonian}.
To implement twisted boundary conditions we modify this Hamiltonian as follows,
\ba
& H_{\trm{SSH,TBC}} = \sum_x m (\ket{x,B} \bra{x,A} + \ket{x,A} \bra{x,B})
\nn
& + \sum_{x \neq \frac{N_1-1}{2}} t (\ket{x+1,A} \bra{x,B} + \ket{x,B} \bra{x+1,A})
\nn
& + t(e^{-i\theta} \ket{(N_1+1)/2,A} \bra{(N_1-1)/2,B}
\nn
& + e^{i\theta} \ket{(N_1-1)/2,B} \bra{(N_1+1)/2,A}),
\label{eq:twistedSSHmodel}
\ea
where a flux $\theta$ has been inserted on the hopping $t$ between the unit cells located at $x=(N_1-1)/2$ and $x=(N_1+1)/2$ (referring to Fig.~\ref{fig:ssh} (b) where $N_1=7$ unit cells, this would be between the unit cells located at $x=3$ and $x=4$).

The Hamiltonian $H_{\trm{SSH,TBC}}$ given by Eq.~\eqref{eq:twistedSSHmodel} is invariant under $c_2(\bx_{1a})$ which, as an operator, can be expanded in the position-space basis as:
\ba
& c_2(\bx_{1a}) = \sum_{x=0}^{(N_1-1)/2} [\ket{N_1-x,B} \bra{x,A} + \ket{N_1-x,A} \bra{x,B}
\nn
& +\ket{x,B} \bra{N_1-x,A} + \ket{x,A} \bra{N_1-x,B}].
\label{eq:c2operatorpositionspace}
\ea
It can be shown that $[H,c_2(\bx_{1a})]=0$.
Note that the $H_{\trm{SSH,TBC}}$ specified in Eq.~\eqref{eq:twistedSSHmodel} is invariant {\it only} under $c_2(\bx_{1a})$ given by Eq.~\eqref{eq:c2operatorpositionspace}, and not under any other $c_2(\bR+\bx_{1a})$ operator with $\bR\neq\bb 0$ unless the $c_2(\bR+\bx_{1a})$ is modified by an appropriate gauge transformation phase factor that accounts for moving the position of the flux line.

Introducing anti-periodic boundary conditions shifts the momentum quantization to $k_x = (2\pi/N_1) (n_1 - \frac{1}{2})$ where $n_1\in\{1,\ldots,N_1\}$.
Note that only one $C_2$-invariant momentum still exists, but it is no longer the $\Gamma$ point, i.e., the $X$ point is now allowed instead when $n_1 = (N_1+1)/2$ (see Fig.~\ref{fig:ssh}(b)).
Hence, the momentum-space symmetry data is now $\{ m(X_1)_{\theta=\pi}, m(X_2)_{\theta=\pi} \}$.
We can now convert these data to momentum space invariants $\{ [X_1], [X_2] \}$ by redefining the rotation invariants as $[X_p]=m(X_p)_{\theta=\pi}-m(\Gamma_p)_{\theta=0}$ for $p=1,2$.
This redefinition emphasizes how we are combining the momentum-space symmetry data for lattices having both periodic and anti-periodic boundary conditions in order to generate both HSM.

In parallel, we have the fully traced TCMs for the $C_2$ rotation center, but now for anti-periodic boundary conditions, $\vev{c_2(\bx_{1a})}_F^{\theta=\pi}$.
Taking into account the symmetry data from both the periodic and anti-periodic boundary conditions yields $\{ \vev{c_2(\bx_{1a})}_F^{\theta=0}, \vev{c_2(\bx_{1a})}_F^{\theta=\pi} \}$
in position space and $\{ m(\Gamma_1)_{\theta=0}, m(\Gamma_2)_{\theta=0}, [X_1], [X_2] \}$ for momentum space.
Thus, we expect that we now have enough data to diagnose the trivial and obstructed phases of this model, as in the even unit cell case.
Intuitively we can see this as follows.
As shown in Fig.~\ref{fig:ssh}(d), both phases of the odd unit cell SSH model always have an electronic Wannier orbital localized at either the $1a$ or $1b$ WP of the $C_2$ rotation center.
Introducing anti-periodic boundary conditions can flip the $C_2$ eigenvalue of the electronic Wannier orbital at the invariant $1b$ WP in the obstructed atomic limit phase localized from $-1$ to $+1$.
This is enough to show that the trivial and obstructed limits will generate different fully traced TCMs if we include both twisted and untwisted boundary conditions.

More precisely, while we have seen that we can generate a pair of position space quantities by twisting boundary conditions, we have not shown that the twisted, fully traced TCMs capture the relevant symmetry data.
To do this we need to construct the mapping between the twisted, fully traced TCMs and the momentum-space symmetry data.
This can be made explicit by expanding the fully traced TCMs in momentum-space.
First, we have $\vev{c_2(\bx_{1a})}_F^{\theta=0} = m(\Gamma_1)_{\theta=0} - m(\Gamma_2)_{\theta=0}$ and $\vev{c_2(\bx_{1a})}_F^{\theta=\pi} = m(X_1)_{\theta=\pi} - m(X_2)_{\theta=\pi}$.
This implies that
\bg
\vev{c_2(\bx_{1a})}_F^{\theta=0} + \vev{c_2(\bx_{1a})}_F^{\theta=\pi} = 4 m(\Gamma_1)_{\theta=0} -2 \nu + 2[X_1],
\nn
\vev{c_2(\bx_{1a})}_F^{\theta=0} - \vev{c_2(\bx_{1a})}_F^{\theta=\pi} = -2[X_1],
\label{eq:tcm_decompose_ssh}
\eg
where $\nu$ denotes the filling, i.e., $m(\Gamma_1)_{\theta=0} + m(\Gamma_2)_{\theta=0} = m(X_1)_{\theta=\pi} + m (X_2)_{\theta=\pi} = \nu$.
We recall in this context that that $[X_1] \equiv m(X_1)_{\theta=\pi} - m(\Gamma_1)_{\theta=0}$.

The correspondence between the TCMs for the even and odd unit cell cases, is summarized in Table~\ref{table:even_odd}.
Using Eq.~\eqref{eq:tcm_decompose_ssh}, when the system has an odd number of unit cells we can now easily determine quantities such as the bulk polarization, or, since our model has chiral symmetry, the parity of the chiral winding number $w$.
The bulk polarization $\bb P$ for the SSH chain is determined by the number of Wannier orbitals localized at the $1b$ WP.
That is,
\bg
\bb P = \frac{e}{2} n_{1b} \, \bb a_1.
\eg
For the even unit cell SSH chain, it is established that the relationship between the chiral winding number, bulk polarization, and the fully traced TCM (even in the absence of translation symmetry) is given as follows~\cite{Velury2021},
\bg
2ew = 4 |\bb P|
= e \, \vev{c_2(\bx_{1b})}_F^{\theta=0} \pmod 4.
\label{eq:ssh_relation}
\eg
Now for the odd unit cell case, having shown that both periodic and anti-periodic boundary conditions are necessary to distinguish the trivial and obstructed atomic limits, we can now modify the relationship given by Eq.~\eqref{eq:ssh_relation} in the following manner,
\bg
2ew = 4 |\bb P|
= e \, \left( \vev{c_2(\bx_{1a})}_F^{\theta=0} - \vev{c_2(\bx_{1a})}_F^{\theta=\pi} \right) \pmod 4.
\eg
Hence, in order to determine the bulk polarization and the parity of the chiral winding number from the symmetry data for an odd number of unit cells, we take a linear combination of the fully traced TCMs for periodic ($\theta=0$) and anti-periodic ($\theta=\pi$) boundary conditions.
%%%%%%

%%%%%%
\begin{figure*}[t!]
\centering
\includegraphics[width=\textwidth]{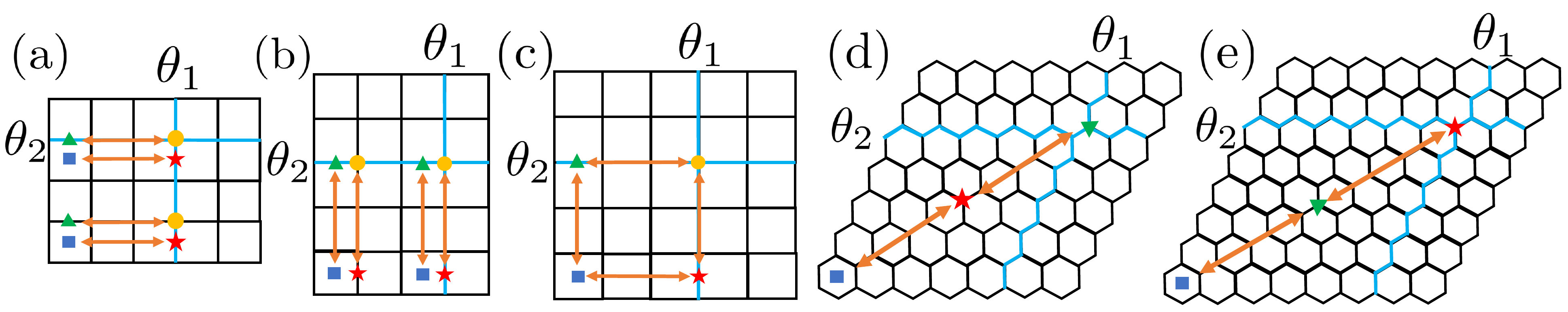}
\caption{
Set of invariant positions $\mc{X}[c_n(\bb r_o)]$ for finite lattices that do not obey the constraints on the linear dimensions $(N_1, N_2)$ given by Eq.~\eqref{eq:perfect_constraint}.
The origin $\bR = \bb 0$ is located at the unit cell at the lower left corner of each lattice shown in (a)-(e).
Each rotation center is specified by a multiplicity $1$ WP for two-dimensional lattices of various dimensions having $C_2$ symmetry [(a)-(c)] and $C_3$ symmetry [(d)-(e)] The dimensions of the lattices are given by (a) $(N_1, N_2)=(1,0)$ mod 2, (b) $(N_1, N_2)=(0,1)$ mod 2, (c) $(N_1, N_2)=(1,1)$ mod 2, (d) $N=1$ mod 3, (e) $N=2$ mod 3.
Note that $N_1=N_2=N$ for $C_3$-symmetric lattices.
Each of the lattices in (a)-(e) have periodic boundary conditions, which means that the top and bottom edges are identified with each other, and separately, the left and right edges are identified with each other.
The light blue lines indicate the location of the bonds modified by Peierls factors to generate fluxes $\theta_1$ and $\theta_2$ that parameterize the twisted boundary conditions.
The location of these fluxes is specifically chosen such that the resulting Hamiltonian with twisted boundary conditions commutes with the rotation operator, i.e., $[H_{\trm{TBC}},c_n(\bb r_o)]=0$.
The orange double-headed arrows are a visual indicator that two positions in $\mc{X}[c_n(\bb r_o)]$ lie on the same rotation axis when periodic boundary conditions are maintained.
}
\label{fig:invariant_positions_tbc}
\end{figure*}
%%%%%%

%%%%%%
\subsection{Generalizing the procedure to 2D}
%%%%%%
As demonstrated on a one-dimensional lattice, twisting boundary conditions is a useful way generalizing the mapping between symmetry data in momentum-space and position-space in order to properly distinguish the phases of rotation-symmetric insulators.
The procedure discussed for the one-dimensional atomic insulator can be generalized to both $C_n$-symmetric Chern insulators and atomic insulators in two dimensions by twisting boundary conditions via Eq.~\eqref{eq:tbc} and utilizing Eq.~\eqref{eq:momentum_tbc}.

Before discussing the procedure for two-dimensional lattices, we need to understand the geometric features of periodic lattices when they do not satisfy the constraints given by Eq.~\eqref{eq:perfect_constraint}.
Figure~\ref{fig:invariant_positions_tbc} illustrates the invariant, multiplicity $1$ WPs of $C_2$ and $C_3$-symmetric lattices that have dimensions $N_1 \times N_2$ which do not respect the constraints, and therefore do not support the maximal set of high symmetry points.
Comparing these lattices to those that were shown in Fig.~\ref{fig:invariant_positions} which do follow the constraints, there is a crucial difference between the set of invariant positions for $C_2$, $C_3$, and $C_4$-symmetric lattices.
In Fig.~\ref{fig:invariant_positions}, each rotation center leaves invariant the multiplicity $1$ WPs of the same type, which is depicted as those corresponding to the same color and shape.
That is, if our rotation center is a $1a$ WP then the other invariant points of the periodic lattice are also $1a$ WPs.

This is no longer true for the lattices shown in Fig.~\ref{fig:invariant_positions_tbc}; in these cases the rotation center(s) can leave invariant different types of multiplicity $1$ WPs.
We saw this already for the SSH chain where in the case of an odd number of unit cells a $1a$ and $1b$ WP are left invariant by the same symmetry center instead of two WPs of the same type.
This is illustrated in Fig.~\ref{fig:invariant_positions_tbc}, where the orange arrows serve to indicate which WPs are fixed by the same rotation center.
Thus, the dimensions of the lattice, specified by the number of unit cells along each primitive lattice vector, determine the invariant, multiplicity $1$ WPs associated with each rotation center.
Since there is not a unique choice of label for the WPs associated to each rotation center, then for the lattices shown in Fig.~\ref{fig:invariant_positions_tbc} we will label the rotation center by the WP located in the unit cell at the origin of the lattice.
The origin is chosen to be the lower left corner of each lattice shown in Figs.~\ref{fig:invariant_positions_tbc}(a)-(e).
%

%%%%%%
{
\renewcommand{\arraystretch}{1.8}
\begin{table*}
\centering
\caption{
The required set of boundary conditions $\bs \theta = (\theta_1, \theta_2)$ for each $C_n$-symmetric lattice with dimensions $N_1 \times N_2$ that do not obey the dimension constraints in Eq.~\eqref{eq:perfect_constraint}.
We have specified the dimensions $(N_1, N_2)$ modulo $n$, where $N_{1,2}>0$.
For $C_4$-symmetric systems, also, note that the boundary conditions $\bs \theta = (\pi,0)$ and $(0,\pi)$ apply only to $C_2$ TCMs.
For $C_6$-symmetric systems, the boundary conditions $\bs \theta = \left( -\frac{2\pi}{3}, \frac{2\pi}{3} \right)$ and $\left( \frac{2\pi}{3}, -\frac{2\pi}{3} \right)$ apply only to $C_3$ TCMs, and the boundary conditions $\bs \theta = (\pi,0)$, $(0,\pi)$, and $(\pi,\pi)$ apply only to $C_2$ TCMs.
}
\label{table:tbc_complete}
\begin{tabular*}{0.6\textwidth}{@{\extracolsep{\fill}} c|cc}
\hline \hline
$C_n$ & $(N_1, N_2)$ & $\left\{ \bs \theta \right\}$
\\ \hline \hline
\multirowcell{3}{$C_2$} & $(1,0)$ mod 2 & $\left\{ (0,0), (\pi,0) \right\}$
\\
& $(0,1)$ mod 2 & $\left\{ (0,0), (0,\pi) \right\}$
\\
& $(1,1)$ mod 2 & $\left\{ (0,0), (\pi,0), (0,\pi), (\pi,\pi) \right\}$
\\ \hline
$C_3$ & $\pm(1,1)$ mod 3 & $\left\{ (0,0), (-\frac{2\pi}{3}, \frac{2\pi}{3}), (\frac{2\pi}{3}, -\frac{2\pi}{3}) \right\}$
\\ \hline
$C_4$ & $(1,1)$ mod 2 & $\left\{ (0,0), (\pi,0), (0,\pi), (\pi,\pi) \right\}$
\\ \hline
\multirowcell{3}{$C_6$} & $\pm (1,1)$ mod 6 & $\left\{ (0,0), (-\frac{2\pi}{3}, \frac{2\pi}{3}), (\frac{2\pi}{3}, -\frac{2\pi}{3}), (\pi,0), (0,\pi), (\pi,\pi) \right\}$
\\
& $\pm (2,2)$ mod 6 & $\left\{ (0,0), (-\frac{2\pi}{3}, \frac{2\pi}{3}), (\frac{2\pi}{3}, -\frac{2\pi}{3}) \right\}$
\\
& $\pm (3,3)$ mod 6 & $\left\{ (0,0), (\pi,0), (0,\pi), (\pi,\pi) \right\}$
\\ \hline \hline
\end{tabular*}
\end{table*}
}
%%%%%%

We can now develop a straightforward generalization of the procedure for the odd unit cell SSH chain.
That is, for each of the 2D lattices that do not satisfy Eq.~\eqref{eq:perfect_constraint}, we can adjust the boundary conditions by threading fluxes $\theta_1$ and $\theta_2$ along each of the cycles along the primitive lattice vectors $\bb a_1$ and $\bb a_2$.
This yields a general momentum quantization condition given by Eq.~\eqref{eq:momentum_tbc}.
The boundary conditions need to be adjusted only for the dimension(s) of the lattice do not satisfy the constraints in Eq.~\eqref{eq:perfect_constraint}.
The necessary boundary conditions that must be considered for each finite-size lattice with $C_n$ symmetry are given in Table~\ref{table:tbc_complete}.
Taking into account the complete set of boundary conditions for each $C_n$-symmetric lattice allows one to distinguish insulating phases.
Furthermore, we can reconstruct the full set of momentum-space symmetry data specified by the irrep multiplicities at all the high symmetry points in the BZ, and map the momentum data to the fully traced TCMs, which themselves are defined for various boundary condition choices.
This allows us to classify insulating phases using momentum or position space data on any finite lattice.

To demonstrate this, we can establish a mapping between (i) the fully traced TCMs over each multiplicity $1$ WP rotation center for $C_n$-symmetric lattices having periodic boundary conditions and satisfying the constraints Eq.~\eqref{eq:perfect_constraint}, and (ii) the fully traced TCMs over each multiplicity $1$ WP corresponding to a rotation center for different twisted periodic boundary conditions for $C_n$-symmetric lattices that does not satisfy the constraints in Eq.~\eqref{eq:perfect_constraint}.
We recall that we have already accomplished this mapping for a 1D $C_2$ symmetric lattice as shown in Table~\ref{table:even_odd}.

As an example, let us consider a $C_2$-symmetric lattice with dimensions $N_1 \times N_2$.
When $(N_1, N_2)=(0,0)$ mod 2, only untwisted periodic boundary conditions need to be considered.
From Fig.~\ref{fig:invariant_positions} we see there should be four distinct, fully traced TCMs; one for each of the four distinct rotation centers at multiplicity $1$ WPs:
\bg
\{ \vev{c_2(\bx_{1a})}_F^{(0,0)}, \vev{c_2(\bx_{1b})}_F^{(0,0)}, \nn \vev{c_2(\bx_{1c})}_F^{(0,0)}, \vev{c_2(\bx_{1d})}_F^{(0,0)} \}.
\label{eq:TCMset1}
\eg
Now, consider the lattice with dimensions such that $(N_1, N_2)=(1,1)$ mod 2.
From Fig.~\ref{fig:invariant_positions_tbc}(c) we see that, for this lattice, there is only one type of fully traced TCM given by $\vev{c_2(\bx_{1a})}_F^{\bs \theta}$.
According to Table~\ref{table:tbc_complete}, we must take into account three additional twisted boundary conditions in addition to periodic boundary conditions: $\bs \theta = (\theta_1, \theta_2) \in \left\{ (0,0), (\pi,0), (0,\pi), (\pi,\pi) \right\}$.
This means the full set of fully traced TCMs for this lattice is 
\bg
\{ \vev{c_2(\bx_{1a})}_F^{(0,0)}, \vev{c_2(\bx_{1a})}_F^{(\pi,0)},
\nn \vev{c_2(\bx_{1a})}_F^{(0,\pi)}, \vev{c_2(\bx_{1a})}_F^{(\pi,\pi)} \}.
\label{eq:TCMset2}
\eg

Now, the key question is how these two sets of TCMs in Eqs.~\eqref{eq:TCMset1} and \eqref{eq:TCMset2} are related to each other.
The mapping of fully traced TCMs corresponding to the lattice with $(N_1,N_2)=(0,0)$ mod 2 to the fully traced TCMs corresponding to the lattice with $(N_1,N_2)=(1,1)$ mod 2 is given as follows,
\ba
\vev{c_2(\bx_{1a})}_F^{(0,0)} &\leftrightarrow
\sum_{\bs \theta} \, \vev{c_2(\bx_{1a})}_F^{\bs \theta},
\nn
\vev{c_2(\bx_{1b})}_F^{(0,0)} &\leftrightarrow
\sum_{\bs \theta} \, e^{i \theta_1} \, \vev{c_2(\bx_{1a})}_F^{\bs \theta},
\nn
\vev{c_2(\bx_{1c})}_F^{(0,0)} &\leftrightarrow
\sum_{\bs \theta} \, e^{i \theta_2} \, \vev{c_2(\bx_{1a})}_F^{\bs \theta},
\nn
\vev{c_2(\bx_{1d})}_F^{(0,0)} &\leftrightarrow
\sum_{\bs \theta} \, e^{i (\theta_1 + \theta_2)} \, \vev{c_2(\bx_{1a})}_F^{\bs \theta},
\label{eq:c2_tbc}
\ea
where $\sum_{\bs \theta}$ means that we sum over the four combinations of $\bs \theta \in \{ (0,0), (\pi,0), (0,\pi), (\pi,\pi)\}$ as specified in Table~\ref{table:tbc_complete}.
Eq.~\eqref{eq:c2_tbc} is a key result that encodes the connection between the four distinct TCMs for $(N_1, N_2)=(0,0)$ mod 2 and the same TCM for four distinct boundary conditions when $(N_1, N_2)=(1,1)$ mod 2.

To illustrate further, for a $C_3$-symmetric system, one can construct a similar mapping between the fully traced TCMs corresponding to the lattice with $N_1=N_2=0$ mod 3 and the TCMs corresponding to the lattice with $N_1=N_2=1$ mod 3,
\ba
\vev{c_3(\bx_{1a})}_F^{(0,0)} & \leftrightarrow
\sum_{\theta \in \{ 0, \pm 2\pi/3 \}} \, \vev{c_3(\bx_{1a})}_F^{(\theta,-\theta)},
\nn
\vev{c_3(\bx_{1b})}_F^{(0,0)} & \leftrightarrow
\sum_{\theta \in \{ 0, \pm 2\pi/3 \}} \, e^{-i \theta} \, \vev{c_3(\bx_{1a})}_F^{(\theta,-\theta)},
\nn
\vev{c_3(\bx_{1c})}_F^{(0,0)} & \leftrightarrow
\sum_{\theta \in \{ 0, \pm 2\pi/3 \}} \, e^{i\theta} \, \vev{c_3(\bx_{1a})}_F^{(\theta,-\theta)}.
\label{eq:c3_tbc}
\ea
The complete details for these maps for every finite-size $C_n$-symmetric lattice with dimensions $(N_1, N_2)$ are provided in the Supplemental Material~\cite{supple}.

With these results, we are able to reformulate the Chern number, bulk polarization, and sector charge shown in Tables~\ref{table:chern}-\ref{table:sector_charge_2} in terms of fully traced TCMs for any finite-size $C_n$-symmetric lattice.
Since we know how the TCMs for lattices satisfying Eq.~\eqref{eq:perfect_constraint} can be used to determine the Chern number, etc., we can use the mappings in, e.g., Eqs.~\eqref{eq:c2_tbc} and \eqref{eq:c3_tbc}, to convert to the data for any finite lattice.
For instance, consider a spinless $C_2$-symmetric class A insulator in a lattice satisfying Eq.~\eqref{eq:perfect_constraint}, i.e., $(N_1,N_2)=(0,0)$ mod 2.
In this case, the Chern number mod 2 is indicated by $\frac{1}{2} \vev{c_2(\bx_W)}_F$ for any maximal WP $W \in \{1a,1b,1c,1d\}$ (see Table~\ref{table:chern}).
For other lattice types, say $(N_1,N_2)=(1,0)$ mod 2, we replace $\vev{c_2(\bx_W)}_F$ in the formula with a linear combination of fully traced TCMs with twisted boundary conditions.
This linear combination is determined by Eq.~\eqref{eq:c2_tbc} and Table~\ref{table:tbc_complete}; if we choose $W=1b$, two different twisted boundary conditions $\bs \theta = (0,0)$ and $(\pi,0)$ are considered to compute $\sum_{\bs \theta} e^{i \theta_1} \, \vev{c_2(\bx_{1a})}_F^{\bs \theta} = \vev{c_2(\bx_{1a})}_F^{(0,0)} -\vev{c_2(\bx_{1a})}_F^{(\pi,0)}$.

This demonstrates that bulk invariants and physical observables can still be evaluated using symmetry data for any finite lattice, so long as the boundary conditions are taken into consideration.
%%%%%%

%%%%%%
\section{Applications of topological crystalline markers}
\label{sec:example}
%%%%%%

%%%%%%
\begin{figure*}[t!]
\centering
\includegraphics[width=0.8\textwidth]{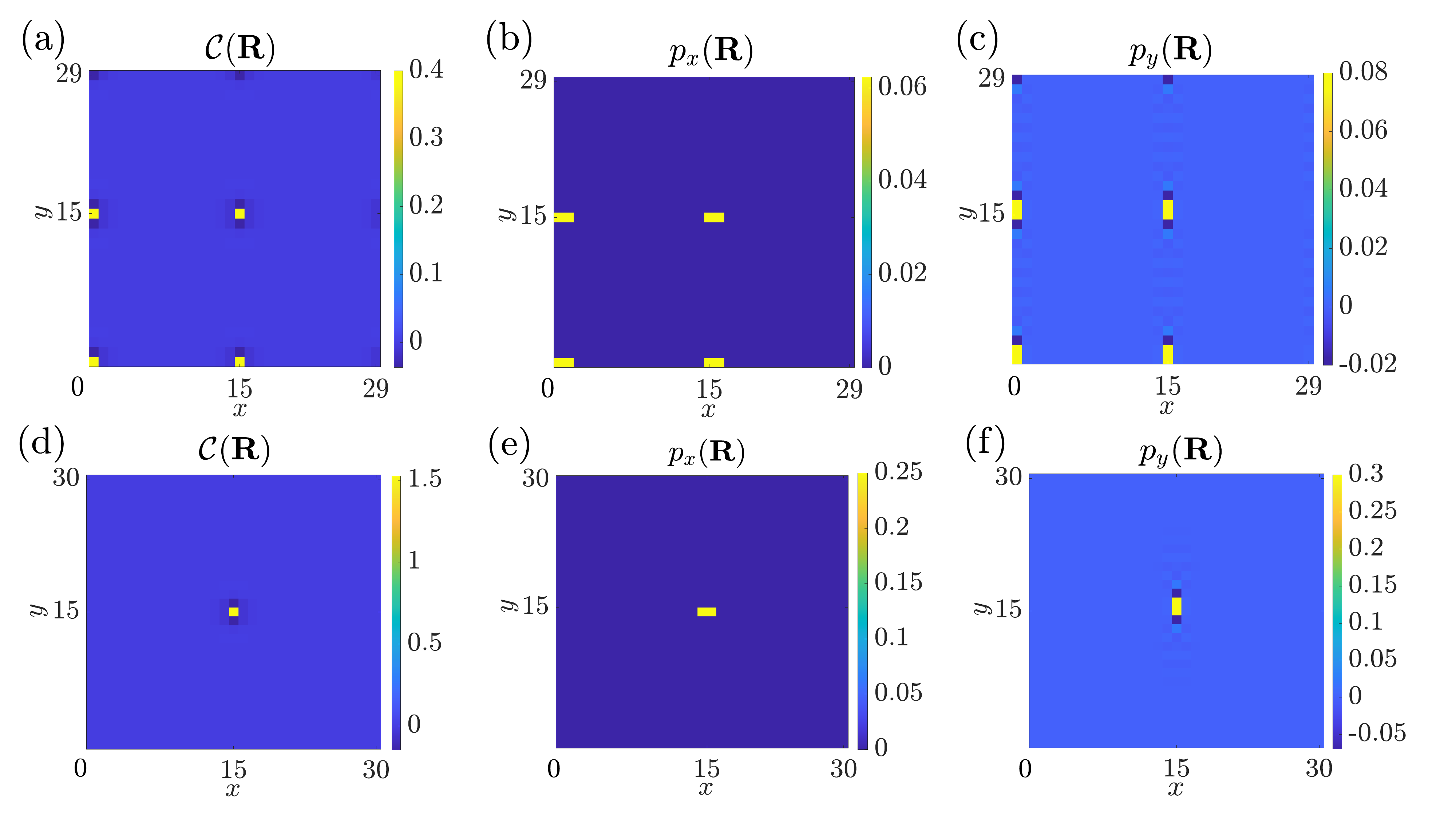}
\caption{
Density plots of the Chern number [(a) and (d)], the $x$-component of the bulk polarization [(b) and (e)], and the $y$-component of the bulk polarization [(c) and (f)] for a $N_x\times N_y=30 \times 30$ lattice [(a)-(c)] and a $N_{x}\times N_{y}=31 \times 31$ lattice [(d)-(f)] (unit cell coordinates $(n_x,n_y)$ are such that $n_i\in\{0,\ldots,N_i-1\}$ for $i=x,y$).
Model parameters $(m,m_H,t_x, t_y)$ of $H_{\rm 2D}(\bk)$ in Eq.~\eqref{eq:stacked_ssh} used were (a) $(3,0.5,1,1.5)$ (Chern insulator phase), (b) $(1,0,1,0.9)$ (obstructed atomic insulator with bulk polarization $\bb P=e (1/2,0)$), and (c) $(1,0.5,1,4)$ (obstructed atomic insulator with bulk polarization $\bb P=e(0,1/2)$).
}
\label{fig:tcm}
\end{figure*}
%%%%%%

Throughout this article we have focused on the global topological information contained in the fully traced TCMs.
We will now show that if the system has rotation symmetry (or more generally, point-group symmetry), even in the absence of translation symmetry, we can use TCMs to characterize crystalline topology.
Along these lines, in this section, we discuss some of the unique applications of TCM {\it densities} for diagnosing bulk crystalline topology in position space.
In addition, we build on the TCM mesh method introduced in Ref.~\onlinecite{MondragonShem2024} to diagnose topological properties in inhomogeneous insulators where translation symmetry, and even point-group symmetry, is broken globally.

To accomplish this we recall from Sec.~\ref{sec:tcm_defs} that it is possible to expand TCMs in the position-space basis as shown in Eq.~\eqref{eq:tcm_def}.
This yields a spatially-resolved density function.
In Sec.~\ref{subsec:example_c2}, we will discuss how the {\it spatially-resolved densities} of bulk invariants such as the bulk Chern number, or the components of the bulk polarization, can be obtained from the TCM densities (c.f., the formulae shown in Tables~\ref{table:chern}-\ref{table:sector_charge_2}).
We will demonstrate that this allows one to plot the distribution of topological crystalline invariants over the position-space lattice and identify which points in space directly contribute to the invariants.

Then, in Sec.~\ref{subsec:domain}, we use the TCM localization property encoded in Eq.~\eqref{eq:tcm_decay} to develop an application of the TCMs for resolving the bulk crystalline topology when spatial inhomogeneities such as domain walls are present.
In such configurations, both translation and rotation symmetry can be broken, but there is a remnant of the crystalline symmetry that survives when we look more than a correlation length away from strong inhomogeneities, e.g., within the bulk of each region separated by the domain wall.
Explicitly, as we detail below, we evaluate a mesh of TCMs at every multiplicity $1$ WP on the entire position-space lattice, not just for one global rotation center.
This mesh encodes the local robustness of topological crystalline phenomena.

In the following subsections where we discuss these applications, the representative tight-binding model we use to evaluate the TCMs is a simple, two-band model that can describe Chern insulators and 2D weak topological insulators (i.e., polarized obstructed atomic limits).
This model can be described as a set of 1D SSH chains extended along the $x$-direction and stacked along the $y$-direction.
Each unit cell consists of two sublattice sites denoted by $A$ and $B$.
When expressed in terms of Pauli matrices $\sg_i$ where $i \in \{ x,y,z \}$ acting on the sublattice sites within each unit cell, the tight-binding Hamiltonian for this model is 
\ba
H_\trm{2D} (\bk)
=& m_H \sin k_y \, \sg_z + t_x \sin k_x \, \sg_y
\nn
+ & (1 - m - t_x \cos k_x - t_y \cos k_y) \sg_x,
\label{eq:stacked_ssh}
\ea
with intra-cell hopping strength $m$, inter-cell hopping strengths $t_x$ and $t_y$ along the $x$ and $y$ directions respectively, and a time-reversal breaking, inter-cell hopping strength $m_H$.
This model is $C_2$ invariant, where $\sg_x$ is the $C_2$ transformation in the sublattice basis.
In the following subsections, we use this model to study Chern insulators and obstructed atomic insulators having non-vanishing polarization.

Our approach of using the densities of TCMs as topological indicators, can be compared to the existing markers for local Chern number from, e.g., Refs.~\onlinecite{Kitaev2006,Bianco2011}.
The local Chern number from these references is defined as a trace per unit cell of the commutator of the position operators projected onto the ground-state.
This quantity indicates the Chern number locally at each unit cell position, has non-vanishing support over essentially the entire lattice in a Chern insulator, and its spatial average evaluates to the Chern number (even when there is spatial inhomogeneity).
On the other hand, the TCMs we present here exhibit sharp peaks (with widths determined by the correlation length) in the neighborhoods of only a finite number of rotation-invariant positions, and can determine $\mc{C}$ mod $n$ instead of the full Chern number.
Interestingly, while there are no previously proposed markers for polarization or sector charge, our TCM densities also capture them.
Hence, there are advantages and disadvantages to the various marker types, and the choice of which marker to use depends on the application.
%%%%%%

%%%%%%
\subsection{Symmetry-based spatial resolution of bulk topological invariants and physical observables}
\label{subsec:example_c2}
%%%%%%
We will begin by studying the spatial localization of bulk crystalline invariants described by TCM densities.
Combining the results shown in Tables~\ref{table:chern}-\ref{table:sector_charge_2} with the TCMs defined in Eq.~\eqref{eq:tcm_def}, one can obtain densities of topological invariants such as the Chern number, and the bulk polarization, in terms of the TCMs.
Consider a periodic, $C_2$-symmetric lattice having an even number of unit cells along each dimension.
If we assume the $C_2$ rotation axis is at the unit cell located at the origin, then the complete set of TCMs\footnote{Note that these are not fully traced as we have been previously considering.} for this lattice is given by
\bg
\{ \vev{c_2(\bx_{1a})}_\bR, \vev{c_2(\bx_{1b})}_\bR, \vev{c_2(\bx_{1c})}_\bR, \vev{c_2(\bx_{1d})}_\bR \}.
\eg
Since the model we plan to study is in class A, we can refer to Tables~\ref{table:chern} and \ref{table:pol_1} to determine the Chern number {\it density}, and the densities of the $x$ and $y$-components of the bulk polarization, as
\bg
\mc{C}(\bR) = \frac{1}{2} \vev{c_2(\bx_W)}_\bR \pmod 2,
\nn
p_x(\bR) = -\frac{e}{2} \left( \vev{c_2(\bx_{1b})}_\bR + \vev{c_2(\bx_{1d})}_\bR \right),
\nn
p_y(\bR) = -\frac{e}{2} \left( \vev{c_2(\bx_{1c})}_\bR + \vev{c_2(\bx_{1d})}_\bR \right).
\label{eq:tcm_c2}
\eg

We can now use these formulae to compute the densities of the Chern number and polarization.
In Fig.~\ref{fig:tcm}(a) we show the density of the bulk Chern number for a Chern insulating phase of our model Eq.~\eqref{eq:stacked_ssh}.
In Figs.~\ref{fig:tcm}(b) and (c), we show and the $x$ and $y$-components of the bulk polarization for two different obstructed phases having quantized, non-vanishing polarization in the $x$ and $y$ direction respectively.
The specific model parameters are listed in the caption of Fig.~\ref{fig:tcm}.

Let us consider these figures in more detail.
The density plot of the Chern number in Fig.~\ref{fig:tcm}(a) was obtained from the TCM for rotation axis at the $1a$ Wyckoff position at the unit cell located at $\bR = \bb 0$.
Since this calculation was done for periodic boundary conditions, this rotation operator fixes the $1a$ WP at the unit cells having coordinates $(0,0)$, $(N_x/2,0)$, $(0,N_y/2)$, and $(N_x/2,N_y/2)$ (i.e., the positions contained in $\mc{X}[c_2(\bb x_{1a})]$).
For our $N_x\times N_y = 30\times 30$ lattice, this would be the coordinates $(0,0)$, $(15,0)$, $(0,15)$, and $(15,15)$.
Indeed, we see that in the neighborhood of these unit cells, the density of the Chern number is sharply localized.
This is a direct consequence of Eq.~\eqref{eq:tcm_decay}.
Evaluating the fully traced TCM by summing the TCM over all the unit cells of the lattice yields $\mc{C}=1$ mod 2, demonstrating that the system has a non-vanishing Chern number.

In the obstructed atomic insulator phases of the model Eq.~\eqref{eq:stacked_ssh}, the system has a quantized bulk polarization and vanishing bulk Chern number $\mc{C}=0$.
Figures~\ref{fig:tcm}(b)-(c) illustrate density plots of the $x$ and $y$-components of the bulk polarization respectively for different obstructed atomic limit phases.
In both plots, the density of the bulk polarization is peaked at the unit cells contained in $\mc{X}[c_2(\bb 0)]$.
Similar to the Chern number, summing the bulk polarization density over all the unit cells of the lattice yields $p_x=e/2$ mod $e$ and $p_y=e/2$ mod $e$ for Figs.~\ref{fig:tcm}(b) and (c) respectively.
%

%%%%%%
\begin{figure*}[t!]
\centering
\includegraphics[width=\textwidth]{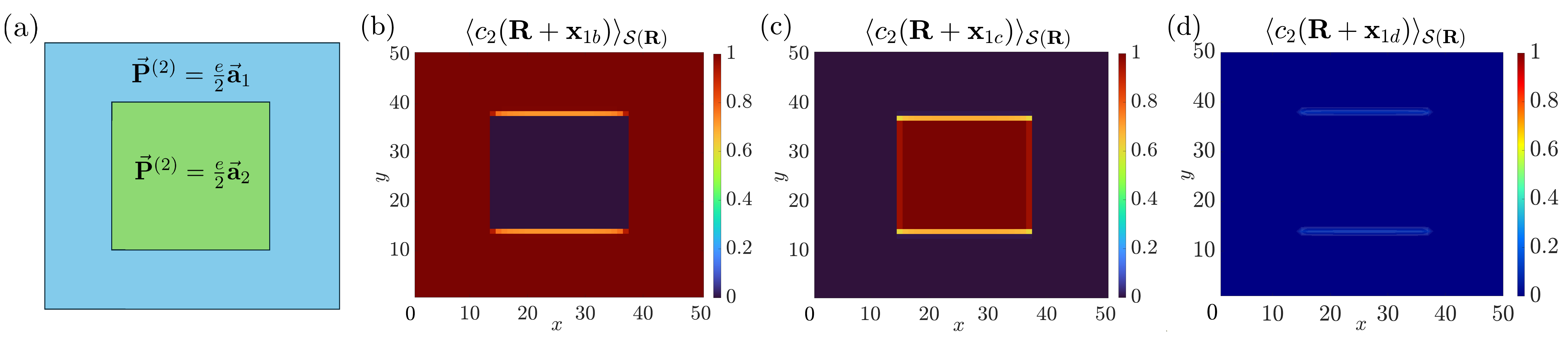}
\caption{
(a) Illustration of a domain wall configuration separating two obstructed atomic insulator regions having different bulk polarizations.
The Hamiltonian is given by Eq.~\eqref{eq:stacked_ssh} with periodic boundary conditions.
Plot of the mesh of traced TCMs for (b) $\{\vev{c_n(\bR + \bx_{1b})}_{\mc{S}(\bR)}\}$,
(c) $\{\vev{c_n(\bR + \bx_{1c})}_{\mc{S}(\bR)}\}$,
and (d) $\{\vev{c_n(\bR + \bx_{1d})}_{\mc{S}(\bR)}\}$.
The outer region is in an obstructed atomic insulator phase with bulk polarization $\bb P=\frac{e}{2} \bb a_1$ for model parameters $(m_1, m_{H,1}, t_{x,1}, t_{y,1})=(0.9,0,1,0)$, whereas the inner region is in an obstructed atomic insulator phase with bulk polarization $\bb P=\frac{e}{2} \bb a_2$ for model parameters $(m_2, m_{H,2}, t_{x,2}, t_{y,2})=(0.8,1,0,1)$.
$\mc{S}(\mathbf{R})$ was chosen to be a $6\times 6$ window centered around a peak for each traced TCM in the mesh.
}
\label{fig:domain_pol}
\end{figure*}
%%%%%%

The calculations so far have been for lattices obeying the dimension constraints in Eq.~\eqref{eq:perfect_constraint}.
Let us now take advantage of our results in Sec.~\ref{sec:map_general} to construct densities of the Chern number and bulk polarization for finite-size lattices having an arbitrary number of unit cells along each direction, including those that do not support all the high symmetry momenta in the BZ.
Specifically, consider a $C_2$-symmetric lattice having an odd number of unit cells along both dimensions.
The mapping between the fully traced $C_2$ TCMs for the lattice with an even number of unit cells in each direction, and the fully traced $C_2$ TCMs for the lattice with an odd number of unit cells in each direction, is given by Eq.~\eqref{eq:c2_tbc}.
This same map directly applies to the TCM densities $\mc{C}(\bR)$, $p_x(\bR)$, and $p_y(\bR)$ resulting in
\bg
\mc{C}(\bR) = \frac{1}{2} \sum_{\theta_1=0,\pi} \sum_{\theta_2=0,\pi} \, \vev{c_2(\bx_{1a})}_\bR^{\bs \theta} \pmod 2,
\nn
p_x(\bR) = -\frac{e}{2} \left( \vev{c_2(\bx_{1a})}_\bR^{(0,0)} - \vev{c_2(\bx_{1a})}_\bR^{(\pi,0)} \right),
\nn
p_y(\bR) = -\frac{e}{2} \left( \vev{c_2(\bx_{1a})}_\bR^{(0,0)} - \vev{c_2(\bx_{1a})}_\bR^{(0,\pi)} \right).
\label{eq:tcm_c2_tbc}
\eg

To confirm this result, in Figs.~\ref{fig:tcm}(d)-(f) we show density plots of the Chern number and the $x$- and $y$-components of the bulk polarization for the same model parameters as Figs.~\ref{fig:tcm}(a)-(c), but for a $C_2$-symmetric lattice having an odd number of unit cells along both dimensions.
To generate the plots shown in Figs.~\ref{fig:tcm}(d)-(f), we implement a (gauge) choice of anti-periodic boundary conditions that commutes with the rotation operator with its origin at the unit cell with coordinates $((N_x-1)/2,(N_y-1)/2)$.
For the $N_x\times N_y = 31\times 31$ lattice we used, this would correspond to the point $(15,15)$ on the lattice.
For this lattice geometry, the rotation center fixes a $1a$, $1b$, $1c$, and $1d$ WP at four different unit cells.
In fact, Figs.~\ref{fig:tcm}(d)-(f) illustrate that the densities of each respective bulk quantity are peaked at unit cells having coordinates $\left( (N_x-1)/2,(N_y-1)/2) \right)$, $\left( (N_x-1)/2,N_y \right)$, $\left( N_x,(N_y-1)/2 \right)$, and $\left( N_x-1,N_y-1 \right)$ (i.e., $(15,15)$, $(15,29)$, $(29,15)$, and $(29,29)$).

The distributions of the Chern number density, and the densities of the components of the bulk polarization, reveal that the largest weight of the distribution is sharply peaked at the unit cell with coordinates $\left( (N_x-1)/2,(N_y-1)/2 \right)$, while the other three unit cells that comprise the invariant positions have small weight.
In particular, as shown in Fig.~\ref{fig:tcm}(e), the density $p_x(\bR)$ is sharply localized at the $1b$ WP, which corresponds to the location of the electronic Wannier orbitals in the obstructed atomic limit phase with bulk polarization $\bb P=(e/2)\bb a_1$.
Similar reasoning holds for Fig.~\ref{fig:tcm}(f), in which the density $p_y(\bR)$ is sharply localized at the $1c$ WP.
We find that integrating each of the densities shown in Figs.~\ref{fig:tcm}(d)-(f) yields $\mc{C}=1$, $p_x=e/2 \pmod e$, and $p_y=e/2 \pmod e$ respectively.

For the plots shown in Fig.~\ref{fig:tcm}(a) and (d), it is worth noting that although the ground state of the Chern insulator presents an obstruction to a Wannier representation, the Chern density itself is exponentially localized at the invariant positions on the lattice.
For example, in Fig.~\ref{fig:tcm}(a), the peaks occur at the invariant positions contained in $\mc{X}[c_2(\bb 0)]$.
Therefore, when expressed in terms of the traced TCMs, the densities of $\mc{C}$, $p_x$, and $p_y$ all contain {\it exponentially} localized peaks in their distributions despite the contrasting Wannierizable nature of their ground state wavefunctions.
This is because the traced TCMs rely solely on the ground state projector, which always exhibits short-ranged, exponentially decaying behavior when a single-particle gap is present, i.e., when there is a finite correlation length.
Hence, using the TCM densities, the bulk properties of strong topological insulators (e.g., Chern insulators) and atomic insulators can be resolved in position-space despite the differences in the Wannierizability of the ground states of these systems.
In the next subsection, we will leverage the sharp localization of the densities to spatially resolve topological properties in inhomogeneous systems.
%%%%%%

%%%%%%
\begin{figure*}[t!]
\centering
\includegraphics[width=1\textwidth]{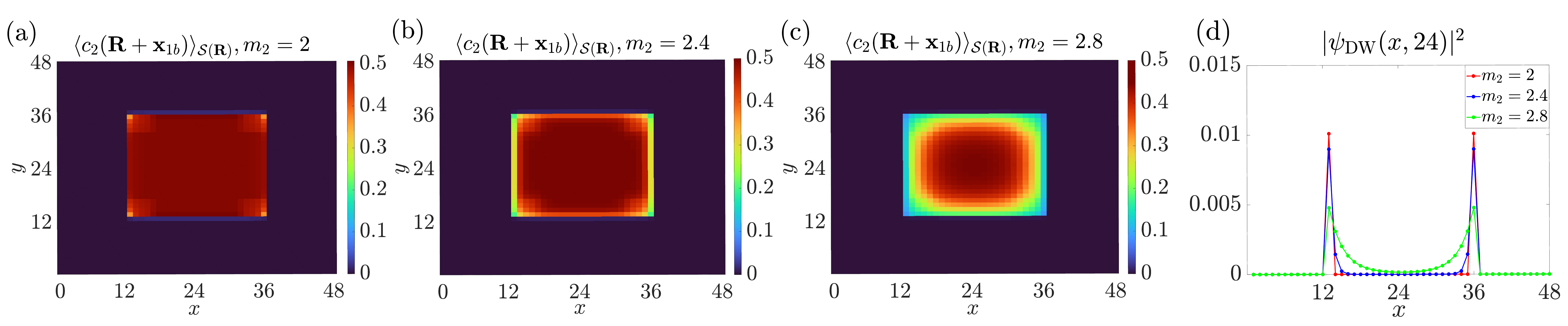}
\caption{
Mesh plot data and probability distribution plots of a domain wall boundary mode for a domain wall configuration identical to Fig.~\ref{fig:domain_pol}(a), with the inner green region being a Chern insulator with $\mc{C}=1$ and the outer blue region being a trivial atomic insulator.
(a)-(c) Plot of the mesh of traced TCMs $\{\vev{c_n(\bR + \bx_{1b})}_{\mc{S}(\bR)}\}$.
The model parameters used for the trivial atomic insulator are $(m_1, m_{H,1}, t_{x,1}, t_{y,1})=(4,0,0.1,0.1)$ and the Chern insulator are $(m_2, m_{H,2}, t_{x,2}, t_{y,2})=(m_2,1,1,1)$ where $m_2=2,2.4,2.8$.
For this choice of $m_{H,2}$, $t_{x,2}$, and $t_{y,2}$, $m_2=3$ is gapless.
As $m_2$ is increased towards this value, the region where the mesh is quantized to $0.5$ reduces in size because of the increase in correlation length, as indicated by each of the plots of the probability distribution of a wavefunction $|\psi_{\trm{DW}}(x,24)|^{2}$ corresponding to a domain wall boundary mode in (d).
$\mc{S}(\mathbf{R})$ was chosen to be a $4\times 4$ window centered around a peak for each traced TCM in the mesh.
}
\label{fig:domain_chern}
\end{figure*}
%%%%%%

%%%%%%
\subsection{Diagnosing crystalline topology in insulators with spatial inhomogeneities}
\label{subsec:domain}
%%%%%%
In this subsection, we will discuss the construction of a mesh of local topological crystalline invariants derived from the partially traced TCMs, and apply the mesh to determining the properties of inhomogeneous insulators in domain wall configurations.
This mesh construction was introduced in Ref.~\onlinecite{MondragonShem2024} where it was used to justify the stability of crystalline topology to symmetry breaking disorder.
For a periodic lattice, any WP $W$ located at $\bR+\bx_W$ can serve as the rotation center for a TCM, with $\mc{X}[c_n(\bR+\bx_W)]$ [c.f., Eq.~\eqref{eq:invariant_positions}] denoting the set of positions invariant under the rotation
(recall the rotation operator with origin at $\bR+\bx_W$ is denoted as $c_n(\bR + \bx_W)$).

The key step is that for {\it every unit cell} on the periodic lattice, one can enumerate the set of TCMs: $\{ \vev{c_n(\bR + \bx_W)}_{\bR'} \}$.
We know from Eq.~\eqref{eq:tcm_decay}, and as illustrated in the previous subsection, that the quantities $\vev{c_n(\bR + \bx_W)}_{\bR'}$ are sharply peaked when $\bR'$ is in the neighborhood of the invariant positions $\mc{X}[c_n(\bR + \bx_W)]$.
One of these invariant positions is the rotation center within the unit cell $\bR$ itself.
For the neighborhood $\mc{S}(\bR)$ of this center we can consider a traced TCM with the support $\mc{S}(\bR)$.
Furthermore, as we scan $\bR$ across the lattice we can define a {\it mesh} of traced TCMs, one for each $\bR$.
Explicitly, as defined in Eq.~\eqref{eq:tcm_traced_def}, we compute the {\it partially traced} TCM,
\bg
\vev{c_n(\bR + \bx_W)}_{\mc{S}(\bR)}
= \sum_{\bR' \in \mc{S}(\bR)} \, \vev{c_n(\bR + \bx_W)}_{\bR'}.
\label{eq:mesh}
\eg
For simplicity, we choose a square neighborhood, and when the linear size of $\mc{S}(\bR)$ is $\xi_\mc{S}$, the sum/partial trace receives contributions only from $\bR'$ such that $|\bR'-\bR| < \xi_\mc{S}$.
Thus, to give meaningful results for insulators, we expect $\xi_\mc{S}$ must be chosen to be comparable to, or larger than, their correlation length $\zeta$.

This construction generates a mesh of traced TCMs $\{ \vev{c_n(\bR + \bx_W)}_{\mc{S}(\bR)} \}$, at the $W$ WP within every unit cell on the position-space lattice.
The TCM mesh provides a set of spatially-resolved crystalline topological data.
That is, each partially traced TCM specified by Eq.~\eqref{eq:mesh} yields a local topological invariant at that position.
For a translationally invariant lattice having periodic boundaries, such a mesh of invariants is uniform and quantized.
However, if translation-symmetry breaking defects such as open boundaries, or domain walls are present, we expect that some of the traced TCMs contained in the mesh will no longer be quantized because the rotational symmetry is not exactly preserved at every rotation center.
Despite this, the exponentially localized nature of the single-particle projector implies that the deviation of each traced TCM from its quantized value in the translationally-invariant limit depends on the relative size of $\mc{S}(\bR)$ with respect to the correlation length $\zeta$.
Indeed, the mesh gives quantitative support to the intuition that locally the system can still appear to have crystal symmetry even if it is broken globally.

To demonstrate the utility of the traced TCM mesh, we consider the model given by Eq.~\eqref{eq:stacked_ssh} and set up a configuration having distinct phases separated by a domain wall.
In Fig.~\ref{fig:domain_pol} we show the results for a domain wall separating two obstructed atomic insulators having different bulk polarizations, while in Fig.~\ref{fig:domain_chern} we show a domain wall separating a Chern insulator and a trivial atomic insulator.
We find that within each region separated by the domain wall, the local values of the mesh of traced TCMs for a specific multiplicity $1$ WP is approximately equal to a quantized value.
In contrast, the traced TCMs along the domain wall are not quantized, possibly because of low-energy modes on the domain wall (see Fig.~\ref{fig:domain_chern}(d) for an illustration of this).
In each of the cases it is clear that the mesh is able to distinguish the spatial structure of the phases with a resolution controlled by the correlation length.

To understand why the partially traced TCMs $\vev{c_n(\bR+\bx_W)}_{\mc{S}(\bR)}$ evaluate to symmetry-protected topological invariants having certain quantized values, let us first understand the quantitative relationship between the fully traced and {\it partially traced} TCMs in a periodic and homogeneous system.
To this end, we focus on the rotation operator $c_2(\bx_W)$ located within the unit cell $\bR=\bb 0$.
As shown in Fig.~\ref{fig:tcm}, the TCM densities for periodic boundaries are exponentially localized around four rotation centers defined by $\mc{X}[c_2(\bx_W)]$, according to Eq.~\eqref{eq:tcm_decay}.
For a homogeneous system with linear dimensions obeying Eq.~\eqref{eq:perfect_constraint}, the four rotation-invariant positions correspond to the same type of WP $W$, and the partially traced TCMs around these centers are all identical.
For instance, consider the intensity of the center peak around $\bx_W$ as shown in Fig.~\ref{fig:tcm}(b).
The intensity can be computed by the partially traced TCM $\vev{c_2(\bx_W)}_{\mc{S}(\bb 0)}$, where the window $\mc{S}(\bb 0)$ is a neighborhood of the center peak.
Since the other three peaks in Fig.~\ref{fig:tcm}(b) have the same intensity, we can conclude that
\bg
\vev{c_2(\bx_W)}_{\mc{S}(\bb 0)}
= \frac{1}{4} \vev{c_2(\bx_W)}_F,
\label{eq:tcm_full_partial}
\eg
up to an exponentially small correction $O(e^{-\xi_{\mc{S}}/\zeta})$ determined by the correlation length $\zeta$ and the linear window size $\xi_{\mc{S}}$.
This relation is implied by Eq.~\eqref{eq:tcm_local_and_global}.

Since one can compute topological invariants like the Chern number, polarization, etc. from the fully traced TCMs, the partially traced TCMs, which we have just argued are proportional to the fully traced ones, also indicate the topological invariants.
Moreover, we find $\vev{c_2(\bx_W)}_{\mc{S}(\bb 0)}$ is approximately quantized as $\Z/4$.
This is because $\vev{c_2(\bx_W)}_F$ is the group theory character for spinless $C_2$ rotation of a given ground state, and takes an integer value, i.e., $\vev{c_2(\bx_W)}_F \in \Z$.

So far, we have focused on the TCMs for $c_2(\bx_W)$ in a homogeneous and periodic system where $c_2(\bx_W)$ is indeed a symmetry of the system.
However, the connection between the partially traced TCMs and topological invariants holds even when open boundaries are used, or if the rotation symmetry is globally broken, but approximately preserved locally.
To illustrate, we can consider the domain wall configurations we introduced in Figs.~\ref{fig:domain_pol} and \ref{fig:domain_chern}.
Although the system is homogeneous away from the domain wall, $c_2(\bR + \bx_W)$ is not preserved for a general unit cell $\bR$ within a homogeneous region.
However, because of the exponential localization of the TCMs [Eq.~\eqref{eq:tcm_decay}], the value of the partially traced TCMs $\vev{c_2(\bR + \bx_W)}_{\mc{S}(\bR)}$ is insensitive to the change of boundary condition (if the correlation length is sufficiently shorter than the linear window size).
Thus, $\vev{c_2(\bR + \bx_W)}_{\mc{S}(\bR)}$ diagnoses topological invariants in a broader class of systems than the fully traced ones.

With the above conclusion, we now understand clearly how the distribution of our partially traced TCMs distinguishes the phases shown in Fig.~\ref{fig:domain_pol}.
In Fig.~\ref{fig:domain_pol}(b), the mesh of $\vev{c_2(\bR+\bx_{1b})}_\mc{S(\bR)}$ is shown.
For the obstructed atomic insulator outside the domain wall, the partially traced TCMs evaluate to $\vev{c_2(\bR+\bx_{1b})}_{\mc{S}(\bR)} \simeq 1$, when $\bR$ is within the homogeneous region, and the size $\xi_\mc{S}$ of the window $\mc{S}(\bR)$ is larger than the correlation length $\zeta$.
These values of partially traced TCMs imply that $\vev{c_2(\bx_{1b})}_F=4$ for the outer obstructed atomic insulator if it were homogeneous and periodic.
By doing a similar analysis for $\vev{c_2(\bR+\bx_{1c,1d})}$, as shown in Figs.~\ref{fig:domain_pol}(c)-(d), we infer that the fully traced TCMs of this insulator would be $\vev{c_2(\bx_{1c,1d})}_F=0$ for a periodic and homogeneous case.
By recalling that $(\bb P)_x = -\frac{e}{8} (\vev{c_2(\bx_{1b})}_F + \vev{c_2(\bx_{1d})}_F)$ mod $e$, and $(\bb P)_y = -\frac{e}{8} (\vev{c_2(\bx_{1c})}_F + \vev{c_2(\bx_{1d})}_F)$ mod $e$ [Table~\ref{table:pol_1}], we conclude that the bulk polarization of this insulator is $\bb P=(e/2)\bb a_1$.
In comparison, the partially traced TCMs for the obstructed insulator inside the domain wall, are $\vev{c_2(\bR + \bx_{1c})}_F \simeq 1$ and $\vev{c_2(\bR + \bx_{1b,1d})}_F \simeq 0$ for $\bR$ in the homogeneous region.
These values imply the bulk polarization $\bb P = \bb (e/2) \bb a_2$.
In this way, the TCM mesh for the $1b,1c,1d$ positions distinguishes atomic insulators with different polarizations.

Finally, we obtain a similar conclusion for a Chern insulating phase.
For a Chern phase with $C_2$ symmetry, the Chern number $\mc{C}$ modulo 2 can be indicated by the traced TCMs $\vev{c_2(\bR + \bx_W)}_{\mc{S}(\bR)}$.
As discussed in Sec.~\ref{subsec:map_chern}, and shown in Table~\ref{table:chern}, $\mc{C}$ mod 2 is determined by the fully traced TCM: $\mc{C}= \frac{1}{2} \vev{c_2(\bx_W)}_F$ mod 2 for any $W \in \{1a,1b,1c,1d\}$.
For the domain wall between a trivial insulator and a Chern insulator with $\mc{C}=1$, we computed the distribution of partially traced TCMs for $\vev{c_2(\bx_{1a})}$, as shown in Fig.~\ref{fig:domain_chern}(a).
Note that we find this distribution is similar to the ones for the $C_2$ rotations centered at other WPs, $c_2(\bx_{1b,1c,1d})$.
The relationship between fully and partially traced TCMs [Eq.~\eqref{eq:tcm_full_partial}] implies that $\mc{C} \simeq 2\vev{c_2(\bR+\bx_W)}_{\mc{S}(\bR)}$ mod 2 for a unit cell $\bR$ within the homogeneous region.
Thus, the nontrivial Chern number of the insulator inside the domain wall is indicated by $\vev{c_2(\bR+\bx_{W})}_{\mc{S}(\bR)} \simeq \frac{1}{2}$ for $W \in \{ 1a, 1b, 1c, 1d \}$, i.e., $\mc{C}=1$ mod 2.

Our analysis using the quantitative relationship between the fully and partially traced TCMs can be extended to other rotation symmetric insulators by figuring out how many times the WPs $W$ are included in $\mc{X}[c_n(\bb R + \bx_W)]$.
For example, for a $C_3$-symmetric insulator obeying Eq.~\eqref{eq:perfect_constraint}, three invariant positions with the same type of WP form $\mc{X}[c_3(\bb R + \bx_W)]$.
In this case, the relationship between the fully and partially traced TCMs changes to $\vev{c_3(\bx_W)}_{\mc{S}(\bb 0)} = \frac{1}{3} \vev{c_3(\bx_W)}_F$.
The quantization of $\vev{c_3(\bx_W)}_{\mc{S}(\bb 0)}$ also follows form the fact that $\vev{c_3(\bx_W)}_F$ is the group theory character for $C_3$ rotation of the ground state.
This means that, for the spinless (spinful) case, $\vev{c_3(\bx_W)}_F$ is always expressed as $\sum_{\ell=0}^2 n_\ell \, e^{\frac{2\pi \ell i}{3}}$ ($\sum_{\ell=0}^2 n_\ell \, e^{\frac{2\pi \ell i + \pi i}{3}}$) for integers $n_\ell$.
Thus, the real and imaginary parts of partially traced TCMs around the center peak are quantized accordingly.
For the lattice types that do not satisfy Eq.~\eqref{eq:perfect_constraint}, a similar analysis can be achieved by applying necessary twisted boundary conditions according to the procedure developed in Sec.~\ref{sec:map_general}.

In summary, as shown in Figs.~\ref{fig:domain_pol} and \ref{fig:domain_chern}, the mesh of traced TCMs clearly distinguishes the different phases of our model within each region separated by the domain wall, thereby demonstrating how TCMs can be utilized to evaluate bulk topological crystalline invariants over a position-space lattice that does not globally preserve translational and crystalline symmetries.
The spatial resolution of this method is limited by the size of the correlation length $\zeta$, but this is expected since a bulk insulating region is only stably established in spatial regions of size larger than $\zeta$.
%%%%%%

%%%%%%
\section{Conclusion}
\label{sec:conclusion}
%%%%%%
In this work, we explored the properties of topological crystalline markers (TCMs) that are locally calculable in position space.
We demonstrated how the TCMs can detect crystalline topology in gapped systems protected by $C_n$-symmetry.
This reformulation of bulk topological invariants and responses using the TCMs encompasses a wide range of insulators, classified under the AZ symmetry classes A, AI, AII, and D.
The basis-independent formulation of topological crystalline invariants presented here, as well as in Ref.~\onlinecite{MondragonShem2024}, shows that bulk invariants can be extracted directly from the ground-state wavefunction using projected symmetry operators, without the need to assume translational invariance.
This approach allows us to characterize the bulk crystalline topology of finite-size lattices with arbitrary numbers of unit cells, even when such lattices do not generate all high-symmetry points in the Brillouin zone.
More crucially, we show that TCMs can spatially resolve bulk crystalline topology, making them a powerful tool for calculating topological crystalline invariants at any point on the crystalline lattice, even in the presence of translation-symmetry breaking.
Furthermore, we demonstrate that TCMs can be universally applied to study both Wannierizable and non-Wannierizable insulating phases, which means the Chern number, bulk polarization, and sector charge can be computed using one formalism.

Our work on the TCMs is closely connected to other works, and can be extended to other systems beyond non-interacting insulators.
For instance, the mapping between the fully traced TCMs and the Wannier orbital configuration discussed in Sec.~\ref{subsec:map_atomic} implies that the fully traced TCMs are directly related to the recently discovered real-space invariants (RSIs)~\cite{Song2020,HerzogArbeitman2023,Herzog-Arbeitman2024}.
In the Supplemental Material~\cite{supple}, we explicitly derive this mapping between the fully traced TCMs and the RSIs for all $C_n$ symmetries in classes A, AI, and AII.
Importantly, although the TCMs and the RSIs can be mapped to each other, we note that both quantities are fundamentally different.
This is because the calculation of RSIs requires momentum-space data obtained from the irreps of the Bloch wavefunctions at the invariant momenta of the BZ.
In contrast to this, the TCMs are built solely from the ground state projector and the crystalline symmetry operators, and their construction does not require an explicit choice of basis to be present.
Additionally, the results presented in this work are similar in spirit to recent approaches of extracting topological invariants directly from the ground state wavefunction~\cite{Fan2023} using expectation values of partial rotation operators~\cite{Zhang2023,Zhang2023a,Manjunath2024,Kobayashi2024}.
This implies at least one route for extending TCMs to interacting systems.

Since the TCMs are expressed solely in terms of the symmetry operators and the ground state projector in any basis, our results can be used to study a wide breadth of systems with broken translation symmetry, including disordered topological phases~\cite{Fu2012,Fulga2014,Diez2015,Song2015}, amorphous systems which weakly preserve the point-group symmetry~\cite{Agarwala2017,Mitchell2018,Agarwala2020,Spring2021,Hannukainen2022,Li2023,Cheng2023,Tao2023,Grushin2023,Corbae2023}, and quasicrystalline lattices~\cite{He2019,Sykes2022,Li2023}.
%%%%%%

\section{Data Availability}
The data that supports the findings of this article is not publicly available. The data is available from the authors upon reasonable request.

%%%%%%
\let\oldaddcontentsline\addcontentsline
\renewcommand{\addcontentsline}[3]{}
%%%%%%
\begin{acknowledgments}
We thank Barry Bradlyn and Penghao Zhu for several useful discussions and guidance during the development of this work.
S.V. thanks Kunal Marwaha, Yuxuan Zhang, Naren Manjunath, Jonah Herzog-Arbeitman, Alexander Cerjan, Pratik Sathe, and Ammar Jahin for insightful discussions on related works and feedback on the results presented in this work.
Y.H. and T.L.H are supported by the US Office of Naval Research (ONR) Multidisciplinary University Research Initiative (MURI) grant N00014-20-1-2325.
Y.H. is also supported by Air Force Office of Scientific Research under award number FA9550-21-1-0131.
S.V. was partly supported by the NSF Graduate Research Fellowship Program under Grant No. DGE-1746047 when this work was initiated and later supported by the University of Illinois.
\end{acknowledgments}
%%%%%%

%%%%%%Refs
%\bibliographystyle{apsrev}
\bibliography{ref.bib}
%%%%%%

\end{document}

\title{\textbf{\ourtitle}}
%%%%%%

%%%%%%Author
\author{Saavanth Velury}
\email{svelury2@illinois.edu}
\affiliation{Department of Physics and Anthony J. Leggett Institute for Condensed Matter Theory, University of Illinois at Urbana-Champaign, Urbana, Illinois 61801-3080, USA}

\author{Yoonseok Hwang}
\affiliation{Department of Physics and Anthony J. Leggett Institute for Condensed Matter Theory, University of Illinois at Urbana-Champaign, Urbana, Illinois 61801-3080, USA}

\author{Taylor L. Hughes}
\affiliation{Department of Physics and Anthony J. Leggett Institute for Condensed Matter Theory, University of Illinois at Urbana-Champaign, Urbana, Illinois 61801-3080, USA}
%%%%%%

\maketitle

%%%%%%
\onecolumngrid
\tableofcontents
%%%%%%

%%%%%%
\section{Review of tight-binding Hamiltonian and symmetry transformations}
\label{sec:tb_hamiltonian_symmetry_supp}
%%%%%%
In this section, we briefly review the properties of the tight-binding Hamiltonian and the action of symmetry operators on the associated single-particle basis. 
%%%%%%

%%%%%%
\tocless{\subsubsection{Tight-binding Hamiltonian}}{\label{subsec:tb_supp}}
%%%%%%
Consider a periodic lattice in $d$-dimensions with $N$ unit cells, and a tight-binding basis (e.g., a set of basis atomic orbitals) which can be denoted by the single-particle kets $\ket{\bb R,\alpha}$ where $\bR$ denotes a (Bravais) lattice vector corresponding to the position of a unit cell, and $\alpha=1,\ldots,N_\trm{int}$ is an index for the internal degree of freedom within the unit cell, denoting an orbital index, sublattice site, etc.
Usually, we express $\bR=\sum\limits_{i=1}^{d}n_i\bb a_i$, where $n_i\in\mathbb{Z}$ for $i\in\{1,\ldots,d\}$ and $\{\bb a_i\}_{i=1}^{d}$ denotes the set of independent (primitive) lattice vectors spanning the lattice.
The tight-binding basis state $\ket{\bR,\alpha}$ represents an electron localized at the position $\bb R+\bx_\alpha$, where $\bx_\alpha$ denotes the sublattice site of the $\alpha^{\text{th}}$ orbital with respect to the unit cell origin.
Note that $\{\ket{\bR,\alpha}\}$ forms an orthonormal basis, i.e. $\langle\bR,\alpha|\bR',\beta\rangle=\delta_{\bb R,\bb R'}\delta_{\alpha,\beta}$.
The single-particle Hamiltonian can be expanded in the tight-binding basis, known as the tight-binding Hamiltonian, as follows,
\bg
H=\sum\limits_{\bb R,\bb R'}\sum\limits_{\alpha,\beta}t_{(\bb R,\alpha),(\bb R',\beta)}\ket{\bR,\alpha}\bra{\bR',\beta},
\label{eq:TB_Hamiltonian_supp}
\eg
where $t_{(\bb R,\alpha),(\bb R',\beta)}$ denotes the hopping amplitude between the $\alpha^{\text{th}}$ and $\beta^{\text{th}}$ orbitals located at $\bb R+\bx_\alpha$ and $\bb R'+\bx_\beta$ respectively.

For a tight-binding Hamiltonian on a periodic lattice with translational symmetry, it is useful to consider the Fourier transform.
Typically, there are two different conventions one can use when considering the Fourier transform of $\ket{\bR,\alpha}$,
\ba
&\ket{\bk,\alpha}\equiv\frac{1}{\sqrt{N}}\sum_\bR e^{i\bk\cdot\bR}\ket{\bR,\alpha}\nn
&\ket{\{\bk,\alpha\}}\equiv\frac{1}{\sqrt{N}}\sum_\bR e^{i\bk\cdot(\bR+\bx_\alpha)}\ket{\bR,\alpha}.
\label{eq:Fourier_Transform_supp}
\ea
The states $\{\ket{\bk,\alpha}\}$ span a periodic basis, because $\ket{\bk+\bb G,\alpha}=\ket{\bk,\alpha}$ where $\bb G$ denotes a reciprocal lattice vector.
On the other hand, the states $\{\ket{\{\bk,\alpha\}}\}$ span a non-periodic basis, because $\ket{\{\bk+\bb G,\alpha\}}\neq\ket{\{\bk,\alpha\}}$ because $\bb G\cdot\bx_\alpha\neq 0 \pmod{2\pi}$. 
However, a basis transformation can be applied to the non-periodic basis states to obtain the periodic basis states,
\bg
\ket{\bk,\alpha}=\sum_\beta [V(\bk)]_{\beta\alpha}\ket{\{\bk,\beta\}},
\label{eq:periodic_non_periodic_supp}
\eg
where $[V(\bk)]_{\beta\alpha}=e^{-i\bk\cdot\bx_\beta}\delta_{\beta\alpha}$ is the sublattice embedding matrix for this unitary transformation.
Note that with the sublattice embedding matrix, one can now implement periodicity on the non-periodic basis states as $\ket{\{\bk+\bb G,\alpha\}}= \sum_\beta V(\bb G)^{-1}_{\beta \alpha} \ket{\{\bk,\beta\}}$, which is called the periodic gauge.
Both the periodic basis states and the non-periodic basis states satisfy orthonormality (i.e., $\langle\bk,\alpha|\bk',\beta\rangle=\langle\{\bk,\alpha\}|\{\bk',\beta\}\rangle=\delta_{\bk,\bk'}\delta_{\alpha,\beta}$). 
Using Eqs.~\eqref{eq:Fourier_Transform_supp} and \eqref{eq:periodic_non_periodic_supp}, the tight-binding Hamiltonian can be expanded in either the periodic or non-periodic basis states,
\ba
H=\sum_\bk \sum_{\alpha,\beta}\ket{\bk,\alpha}H(\bk)_{\alpha\beta}\bra{\bk,\beta}=\sum_\bk \sum_{\alpha,\beta}\ket{\{\bk,\alpha\}}\tilde{H}(\bk)_{\alpha\beta}\bra{\{\bk,\beta\}},
\label{eq:periodic_non_periodic_Hamiltonian_supp}
\ea
where $H(\bk)_{\alpha\beta}\equiv\bra{\bk,\alpha}H(\bk)\ket{\bk,\beta}$ and $\tilde{H}(\bk)_{\alpha\beta}\equiv\bra{\{\bk,\alpha\}}\tilde{H}(\bk)\ket{\{\bk,\beta\}}$.
From Eq.~\eqref{eq:periodic_non_periodic_Hamiltonian_supp}, one has
\ba
&\tilde{H}(\bk)=V(\bk)H(\bk)V(\bk)^{-1}, \nn
&H(\bk+\bb G)=H(\bk), \nn
&\tilde{H}(\bk+\bb G)=V(\bb G)\tilde{H}(\bk)V(\bb G)^{-1},
\label{eq:periodic_gauge_Hamiltonian_supp}
\ea
The matrix elements of the tight-binding Hamiltonian can be explicitly evaluated in either the periodic or non-periodic bases when implementing the Fourier transform given by Eq.~\eqref{eq:Fourier_Transform_supp} on the tight-binding Hamiltonian given by Eq.~\eqref{eq:TB_Hamiltonian_supp},
\ba
&H(\bk)_{\alpha\beta}=\sum_{\bR}t_{(\bb R,\alpha),(\bb 0,\beta)}e^{-i\bk\cdot\bR}, \nn
&\tilde{H}(\bk)_{\alpha\beta}=\sum_{\bR}t_{(\bb R,\alpha),(\bb 0,\beta)}e^{-i\bk\cdot(\bR + \bx_\alpha - \bx_\beta)},
\label{eq:tight_binding_Hamiltonian_matrix_elements_momspace_supp}
\ea
Note that the matrix elements in the non-periodic basis are equivalent to the matrix elements in the periodic basis if \textit{all} the orbitals are located at the unit cell origin (e.g., $\bx_\alpha=\bb 0$ for all $\alpha$). 
$H(\bk)$ and $\tilde{H}(\bk)$ are the Bloch Hamiltonians for the periodic or non-periodic bases respectively.
Diagonalization of the Bloch Hamiltonian yields the following in each basis,
\ba
&\sum_\beta H(\bk)_{\alpha\beta}\ket{u_n(\bk)}_\beta=E_n(\bk)\ket{u_n(\bk)}_\alpha \nn
&\sum_\beta \tilde{H}(\bk)_{\alpha\beta}\ket{\{u_n(\bk)\}}_\beta=E_n(\bk)\ket{\{u_n(\bk)\}}_\alpha
\label{eq:BlochHamiltonian}
\ea
The states $\ket{\{u_n(\bk)\}}$ in the non-periodic basis can be related to the states $\ket{u_n(\bk)}$ via the sublattice embedding matrix, i.e. $\ket{\{u_n(\bk)\}}=V(\bk)\ket{u_n(\bk)}$.
The $\ket{u_n(\bk)}$ states in the periodic basis automatically satisfy the periodic gauge since $\ket{u_n(\bk+\bb G)}=\ket{u_n(\bk)}$, whereas the $\ket{\{u_n(\bk)\}}$ states in the non-periodic basis require the sublattice embedding matrix to satisfy the periodic gauge, i.e., $\ket{\{u_n(\bk+\bb G)\}}=V(\bb G)\ket{\{u_n(\bk)\}}$.
(Thus, in both bases, the Bloch wavefunction $\ket{\psi_n(\bk)} = \sum_{\alpha} \ket{u_n(\bk)}_\alpha \ket{\bk, \alpha} = \sum_{\alpha} \ket{\{u_n(\bk)\}}_\alpha \ket{\{\bk, \alpha\}}$ is periodic such that $\ket{\psi_n(\bk)}=\ket{\psi_n(\bk+\bb G)}$. 
Here, $\ket{u_n(\bk)}_\alpha$ ($\ket{\{u_n(\bk)\}}_\alpha$) denotes the $\alpha^{\rm{th}}$ component of $\ket{u_n(\bk)}$ ($\ket{\{u_n(\bk)\}}$).)

Throughout this Supplemental Material, specifically for Secs.~\ref{sec:TCM_decomp_supp}, and Sec.~\ref{sec:tbc_supp}, we will work in the periodic basis and the Bloch Hamiltonian $H(\bk)$ to present the derivations in the most algebraically simple way possible.
It is straightforward to re-derive the results in the non-periodic basis and the corresponding Bloch Hamiltonian $\tilde{H}(\bk)$ using the basis transformations shown in Eqs.~\eqref{eq:periodic_non_periodic_supp} and \eqref{eq:periodic_gauge_Hamiltonian_supp}.
%%%%%%

%%%%%%
\tocless{\subsubsection{Symmetry transformations}}{\label{subsec:symm_transform_supp}}
%%%%%%
Consider an element $\chi$ of the space group $G$ where $\chi=\{R_\chi|\boldsymbol{\delta}_\chi\}$ which acts on a point in space as follows,
\ba
\chi:\bb r\to R_\chi\bb r+\boldsymbol{\delta}_\chi
\label{eq:symmetry_transform_def_supp}
\ea
where $R_\chi$ is an orthogonal matrix. On the tight-binding basis states, $\chi$ is a unitary operator that acts as follows,
\ba
\chi\ket{\bR,\alpha} = \sum_\beta[U(\chi)]_{\beta \alpha} \ket{\bR_\chi(\alpha), \beta},
\label{eq:symmetry_action_def_supp}
\ea
where $\bR_\chi(\alpha)=R_\chi(\bR + \bx_\alpha)+\boldsymbol{\delta}_\chi-\bx_\beta$ and $U(\chi)$ is a unitary matrix representing $\chi$ in the orbital basis. 
On the periodic basis states, the action of $\chi$ can be determined by combining Eqs.~\eqref{eq:Fourier_Transform_supp} and \eqref{eq:symmetry_action_def_supp},
\ba
& \chi \ket{\bk, \alpha} = \frac{1}{\sqrt{N}} \sum_\bR e^{i\bk \cdot \bR} \chi \ket{\bR, \alpha} = \frac{1}{\sqrt{N}} \sum_\bR \sum_\beta e^{i\bk \cdot \bR}[U(\chi)]_{\beta \alpha} \ket{\bR_\chi(\alpha), \beta}
\nn
& \equiv \frac{1}{\sqrt{N}} \sum_\bR \sum_\beta e^{i\bk\cdot(R_{\chi}^{-1}(\bR+\bx_\beta-\boldsymbol{\delta}_\chi)-\bx_\alpha)}[U(\chi)]_{\beta\alpha}\ket{\bR,\beta}
\nn
& = \sum_\beta e^{i(R_\chi \bk)\cdot(\bx_\beta-R_\chi \bx_\alpha-\boldsymbol{\delta}_\chi)}[U(\chi)]_{\beta\alpha}\left(\frac{1}{\sqrt{N}}\sum_\bR e^{i(R_\chi \bk)\cdot\bR}\ket{\bR,\beta}\right)
\nn
& =\sum_\beta e^{i(R_\chi \bk)\cdot(\bx_\beta-R_\chi \bx_\alpha-\boldsymbol{\delta}_\chi)}[U(\chi)]_{\beta\alpha}\ket{R_\chi \bk,\beta}, \nn
&\equiv \sum_\beta [U_\chi(\bk)]_{\beta\alpha}\ket{R_\chi \bk,\beta}.
\label{eq:symmetry_action_momentum_space_supp}
\ea
In the above, we have introduced the unitary matrix $U_\chi(\bk)$ which is the momentum-space representation of $U_\chi$ and is defined as follows,
\ba
[U_\chi(\bk)]_{\beta\alpha} = e^{i(R_\chi \bk)\cdot(\bx_\beta-R_\chi \bx_\alpha-\boldsymbol{\delta}_\chi)} U(\chi)
\equiv \left[ V(R_\chi \bk)^{\dagger} \left( e^{-i(R_\chi \bk) \cdot \boldsymbol{\delta}_\chi} U(\chi) \right) V(\bk) \right]_{\beta \alpha}.
\label{eq:unitary_matrix_momentum_space_def_supp}
\ea
where $V(\bk)$ is the sublattice embedding matrix (note that it is unitary) defined in Sec.~\ref{subsec:tb_supp}.
If the tight-binding Hamiltonian is symmetric under $\chi$, it is invariant under conjugation by $\chi$, e.g., $H=\chi H\chi^{-1}$.
This implies the Bloch Hamiltonian transforms as follows,
\ba
H(R_\chi \bk)=U_\chi(\bk)H(\bk)U_\chi(\bk)^{\dagger}
\label{eq:Bloch_Hamiltonian_transformation}
\ea

Consider the group element corresponding to a $C_n$ rotation at the unit cell origin, i.e., $c_n(\bb 0)=\{C_n|\bb 0\}$, where $c_n(\bb 0)$ is the rotation operator for a rotation axis located at the unit cell origin (we also assume this to be the $1a$ Wyckoff position (WP), $\bx_{1a}=\bb 0$).
The rotation operator $c_n(\bb 0)$ maps a position $\bb r$ to $c_n(\bb 0)\bb r$.
From Eq.~\eqref{eq:unitary_matrix_momentum_space_def_supp}, we have
\ba
c_n(\bb 0,\bk)=V(C_n\bk)^{\dagger}c_n(\bb 0)V(\bk),
\label{eq:momentum_space_rotation_relation_supp_I}
\ea
Additionally, consider a WP at $\bb r_o$.
The group element
\ba
c_n(\bb r_o)=\{C_n|\bb r_o - c_n(\bb 0)\bb r_o\}=\{E|\bb r_o\}\{C_n|\bb 0\}\{E|-\bb r_o\}\equiv t(\bb r_o)c_n(\bb 0)t(\bb r_o)^{-1}
\ea
is the rotation operator for a rotation axis at $\bb r_o$ and maps a position $\bb r$ to $c_n(\bb r_o)\bb r=c_n(\bb 0)\bb r+\bb r_o - c_n(\bb 0)\bb r_o$.
Note that $\{E|\bx\}\equiv t(\bb x)$ is the group element corresponding to translation by $\bb x$.
From Eq.~\eqref{eq:unitary_matrix_momentum_space_def_supp}, identifying $\boldsymbol{\delta}_{c_n(\bb 0)}=\bb r_o-c_n(\bb 0)\bb r_o$, it follows that
\ba
c_n(\bb r_o,\bk) = e^{-i\bk\cdot((c_n(\bb 0))^{-1}\bb r_o - \bb r_o)}V(C_n\bk)^{\dagger}c_n(\bb 0,\bk)V(\bk) = e^{-i(C_n\bk -\bk)\cdot\bb r_o}V(C_n\bk)^{\dagger}c_n(\bb 0,\bk)V(\bk)
\label{eq:momentum_space_rotation_relation_supp_II}
\ea
This is a generalized version of the result presented in Eq. (15) of the main text, which can be easily obtained assuming that all orbitals within the unit cell lie at the origin (e.g., $\bx_\alpha=\bb 0$ for all $\alpha$).
Then $V(\bk)$ becomes the identity matrix and Eq.~\eqref{eq:momentum_space_rotation_relation_supp_II} is equivalent to Eq. (15) of the main text.
%%%%%%%

%%%%%%
\section{Derivation of topological crystalline marker decomposition in momentum space}
\label{sec:TCM_decomp_supp}
%%%%%%
Here, we will go over the decomposition of the fully traced topological crystalline marker (TCM) for $C_n$ point groups in the momentum-space basis, which is given by Eq. (17) in the main text. This is used to establish the mapping between the fully traced TCMs and the momentum-space irrep multiplicities which is covered in Secs.~\ref{sec:mappings_supp} and \ref{sec:matrices_supp} of this Supplemental Material.
Working in the periodic basis, diagonalizing the Bloch Hamiltonian yields Eq.~\eqref{eq:BlochHamiltonian}.
The ground state projector $P_{GS}$ can be expanded in terms of the Bloch states,
\ba
P_{GS}=\sum_{j=1}^{\nu} \sum_\bk \ket{\psi_j(\bk)}\bra{\psi_j(\bk)},
\label{eq:ground_state_projector_supp}
\ea
where $\nu$ is the number of filled bands.
The Bloch states are expressed in the periodic basis as $\ket{\psi_n(\bk)}=\sum_\alpha \ket{u_n(\bk)}_\alpha \ket{\bk,\alpha}$, and satisfy the orthogonality relation $\langle \psi_m(\bk)|\psi_n(\bk')\rangle=\delta_{mn}\delta_{\bk,\bk'}$ (note that $\langle u_m(\bk)|u_n(\bk')\rangle\neq\delta_{mn}\delta_{\bk,\bk'}$ in general, but $\langle u_m(\bk)|u_n(\bk)\rangle=\delta_{mn}$).
The fully traced TCM can be expanded in momentum-space as follows,
\ba
\vev{c_n(\bb r_o)}_F = \trm{Tr}[c_n(\bb r_o) \, P_{GS}] = \sum_\bk \sum_{m=1}^{N_\trm{int}} \bra{\psi_m(\bk)}c_n(\bb r_o,\bk) \, P_{GS}\ket{\psi_m(\bk)}.
\label{eq:fully_traced_TCM_1_supp}
\ea
Substituting Eq.~\eqref{eq:ground_state_projector_supp} into Eq.~\eqref{eq:fully_traced_TCM_1_supp} yields the following,
\ba
\vev{c_n(\bb r_o)}_F = \sum_{\bk,\bk'} \sum_{m=1}^{N_\trm{int}} \sum_{j=1}^{\nu} \bra{\psi_m(\bk)}c_n(\bb r_o,\bk)\ket{\psi_j(\bk')}\langle \psi_j(\bk')|\psi_m(\bk)\rangle.
\label{eq:fully_traced_TCM_2_supp}
\ea
Given $c_n(\bb r_o,\bk)$ maps $\bk\to C_n\bk$, the non-zero contributions in the fully traced TCMs only occur at the high-symmetry momenta $\bar{\bk}$ where $C_n \bar{\bk}=\bar{\bk}$ modulo $\bb b$, where $\bb b$ is a reciprocal lattice vector.
This means the sum in Eq.~\eqref{eq:fully_traced_TCM_2_supp} is non-zero only for $\bk=\bk'=\bar{\bk}$.
Therefore, Eq.~\eqref{eq:fully_traced_TCM_2_supp} reduces to the following using the orthogonality relation of the Bloch states $\langle \psi_{m}(\bk)|\psi_{n}(\bk')\rangle=\delta_{mn}\delta_{\bk,\bk'}$,
\ba
\vev{c_n(\bb r_o)}_F = \sum_{\bar{\bk}\in\trm{HSM}_n} \sum_{j=1}^{\nu} \bra{\psi_j(\bar{\bk})}c_n(\bb r_o,\bar{\bk})\ket{\psi_j(\bar{\bk})},
\label{eq:fully_traced_TCM_3_supp}
\ea
where $\trm{HSM}_n$ is the set of $C_n$-invariant high-symmetry momenta.
For every $\bar{\bk}\in\trm{HSM}_n$, because $[c_n(\bb r_o,\bar{\bk}),H(\bar{\bk})]=0$, this means $\ket{u_j(\bar{\bk})}$ is a simultaneous eigenstate of both $H(\bar{\bk})$ and $c_n(\bb r_o,\bar{\bk})$.
Eq.~\eqref{eq:fully_traced_TCM_3_supp} is the trace (over the filled bands) of the \textit{sewing matrix} $B_{mn}(\bk)=\bra{\psi_m(\bk)}c_n(\bb r_o,\bk)\ket{\psi_n(\bk)}$ evaluated at the high-symmetry momenta $\bar{\bk}\in\trm{HSM}_n$.
This matrix can be diagonalized to yield the rotation eigenvalues $\{e^{i\phi_{n,p}(\bb r_o,\bar{\bk})}\}_{p=1}^{n}$ and their irrep multiplicities $m(\bar{\bk}_p)$ at each $\bar{\bk}\in\trm{HSM}_n$.
Thus,
\ba
\vev{c_n(\bb r_o)}_F = \sum_{\bar{\bk}\in\trm{HSM}_n} \sum_{p=1}^{n} e^{i\phi_{n,p}(\bb r_o,\bar{\bk})}m(\bar{\bk}_p),
\label{eq:fully_traced_TCM_4_supp}
\ea
where $\phi_{n,p}(\bb r_o,\bar{\bk})=\frac{2\pi}{n}\ell-\bar{\bk}\cdot((c_n(\bb 0))^{-1}\bb r_o - \bb r_o)=\frac{2\pi}{n}\ell-(C_n \bar{\bk}-\bar{\bk})\cdot\bb r_o$ with the orbital angular momentum $\ell=p-1$ for spinless electrons and $\ell=p-1/2$ for spin-1/2 electrons ($p=1,\ldots,n$) as defined in the main text.
The irrep multiplicity $m(\bar{\bk}_p)$ is the number of occupied states at $\bar{\bk}$ for the irrep $\bar{\bk}_p$, and
\ba
\sum_{p=1}^{n}m(\bar{\bk}_p)=\nu.
\label{eq:irrep_mult_condition_supp}
\ea
Eq.~\eqref{eq:irrep_mult_condition_supp} directly applies to $C_2$ and $C_3$ point groups. 
For $C_n$ point groups that contain a $C_{m\leq n}$-invariant subgroup, there exist high-symmetry momenta $\bar{\bk}\neq\Gamma$ which are only invariant under the $C_{m\leq n}$ subgroup, which is true for $C_4$ and $C_6$ point groups.
Therefore, a generalized version of Eq.~\eqref{eq:irrep_mult_condition_supp} is,
\ba
\sum_{p=1}^{m}m(\bar{\bk}_p)=\sum_{p=1}^{n}m(\Gamma_p)=\nu\to\sum_{p=1}^{m}[\bar{\bk}_p]=0
\label{eq:irrep_mult_condition_general_supp}
\ea
where we have used the rotation invariants $[\bar{\bk}_p]$ to simplify the expression for this constraint.
%%%%%%

%%%%%%
\section{Construction of mappings between topological crystalline markers, momentum-space irrep multiplicities, rotation invariants, and wannier orbital irrep multiplicities/real-space invariants (RSI)}
\label{sec:mappings_supp}
%%%%%%
Eq.~\eqref{eq:fully_traced_TCM_4_supp} explicitly demonstrates that the fully traced TCMs can be expressed in terms of the momentum-space irrep multiplicities, thereby demonstrating the existence of a mapping between the fully traced TCMs and the momentum-space irrep multiplicities.
Furthermore, for atomic insulators, the ground state can be expressed in terms of symmetric, exponentially-localized Wannier orbitals, which implies the existence of a mapping between momentum-space irrep multiplicities and Wannier orbital irrep multiplicities.
These mappings can be used to express strong topological invariants, such as the Chern number, and the quantized responses of atomic insulators such as the bulk polarization and sector charge, in terms of one unified set of quantities, which is the set of fully traced TCMs for a given space group.
In this section, we will provide an overview of how these mappings are constructed.

The TCM and the fully traced TCM can be constructed for any group element $\chi=\{R_\chi|\boldsymbol{\delta}_\chi\}\in G$, 
\ba
\vev{\chi}=\sum_\alpha \bra{\bR,\alpha}\chi \, P_{GS}\ket{\bR,\alpha}\to\vev{\chi}_F=\trm{Tr}[\chi \, P_{GS}].
\label{eq:generic_TCM_supp}
\ea
The minimal complete set of fully traced TCMs for a given space group $G$ can be constructed from the generators of the little group $G_{\bb r_o}$ for each \textit{maximal} WP $\bb r_o$.
This includes a fully traced TCM for the identity element (e.g., $\{E|\bb 0\}$) and a fully traced TCM for the inverse element (e.g., $\chi^{-1}=\{(R_\chi)^{-1}|-(R_\chi)^{-1}\boldsymbol{\delta}_\chi \}$).
Focusing on $C_n$ point groups, we introduce a column vector $\bb v_{\trm{TCM}}^{(n)}$, which contains the minimal, complete set of independent fully traced TCMs.
As an example, for a $C_2$ point group, the unit cell contains four WPs ($1a$, $1b$, $1c$, and $1d$), so $\bb v_{\trm{TCM}}^{(2)}$ is given as follows,
\ba
\bb v_{\trm{TCM}}^{(2)} = \begin{pmatrix} \nu, & \vev{c_2(\bx_{1a})}_F, & \vev{c_2(\bx_{1b})}_F, & \vev{c_2(\bx_{1c})}_F, & \vev{c_2(\bx_{1d})}_F \end{pmatrix}^{T}.
\label{eq:TCM_C2_vector_supp}
\ea
where $\nu=\trm{Tr}[P_{GS}]$ is the filling and is equivalent to the TCM for the identity element.
The remaining $\bb v_{\trm{TCM}}^{(3)}$, $\bb v_{\trm{TCM}}^{(4)}$, and $\bb v_{\trm{TCM}}^{(6)}$ for the $C_3$, $C_4$, and $C_6$ point groups can be determined in a similar manner,
\ba
\bb v_{\trm{TCM}}^{(3)}
= (& \nu,  \vev{c_3(\bx_{1a})}_F, \vev{c_3(\bx_{1b})}_F, \vev{c_3(\bx_{1c})}_F, \vev{(c_3(\bx_{1a}))^{-1}}_F, \vev{(c_3(\bx_{1b}))^{-1}}_F, \vev{(c_3(\bx_{1c}))^{-1}}_F )^T, \nn
\bb v_{\trm{TCM}}^{(4)}
= (& \nu, \vev{c_4(\bx_{1a})}_F, \vev{c_4(\bx_{1b})}_F, \vev{(c_4(\bx_{1a}))^{-1}}_F, \vev{(c_4(\bx_{1b}))^{-1}}_F, \vev{c_2(\bx_{1a})}_F, \vev{c_2(\bx_{1b})}_F, \vev{c_2(\bx_{2c})}_F )^T,
\nn
\bb v_{\trm{TCM}}^{(6)}
= (& \nu, \vev{c_6(\bx_{1a})}_F, \vev{(c_6(\bx_{1a}))^{-1}}_F, \vev{c_3(\bx_{1a})}_F, \vev{c_3(\bx_{2b})}_F, \vev{(c_3(\bx_{1a}))^{-1}}_F,
\nn
& \vev{(c_3(\bx_{2b}))^{-1}}_F, \vev{c_2(\bx_{1a})}_F, \vev{c_2(\bx_{3c})}_F )^T.
\label{eq:TCM_C3_C4_C6_vector_supp}
\ea
We now consider the complete set of momentum-space irrep multiplicities for each $C_n$ point group. 
In order to do so, we utilize the momentum-space rotation invariants for each $C_n$ point group as defined in Eq. (21) of the main text.
Constraints arising from filling (as shown in Eq.~\eqref{eq:irrep_mult_condition_general_supp}) and internal symmetries such as time-reversal and/or particle-hole can easily be expressed in terms of the rotation invariants.
Therefore, we introduce a column vector $\bb v_{\trm{MS}}^{(n)}$ which contains the minimal complete set of rotation invariants for each $C_n$ point group,
\ba
&\bb v_{\trm{MS}}^{(2)} = \begin{pmatrix} \nu, & m(\Gamma_1), & [X_1], & [Y_1], & [M_1] \end{pmatrix}^{T}, \nn
&\bb v_{\trm{MS}}^{(3)} = \begin{pmatrix} \nu, & m(\Gamma_1), & m(\Gamma_2), & [K_1], & [K_2], & [K'_1], & [K'_2] \end{pmatrix}^{T}, \nn
&\bb v_{\trm{MS}}^{(4)} = \begin{pmatrix} \nu, & m(\Gamma_1), & m(\Gamma_2), & m(\Gamma_3), & [X_1], & [M_1], & [M_2], & [M_3] \end{pmatrix}^{T}, \nn
&\bb v_{\trm{MS}}^{(6)} = \begin{pmatrix} \nu, & m(\Gamma_1), & m(\Gamma_2), & m(\Gamma_3), & m(\Gamma_4), & m(\Gamma_5), & [K_1], & [K_2], & [M_1] \end{pmatrix}^{T}.
\label{eq:MS_C2_C3_C4_C6_vector_supp}
\ea
Finally, pertaining to atomic insulators, we consider the complete set of Wannier orbital irrep multiplicities over the maximal WPs for each $C_n$ point group.
Similar to rotation invariants, we utilize real-space invariants (RSIs)~\cite{Song2020}, which can be expressed in terms of the Wannier orbital irrep multiplicities for each $C_n$ point group for spinless electrons,
\ba
C_2: \, & \delta_{W} = n_{W}^{(1)} - n_{W}^{(0)} \, \trm{for } W \in \{1a, 1b, 1c, 1d\},
\nn
C_3: \, & \delta_{W}^{(\ell)} = n_{W}^{(\ell)} - n_{W}^{(0)} \, \trm{for } W \in \{1a, 1b, 1c\} \, \trm{and } \ell \in \{1,2\},
\nn
C_4: \, & \delta_{W}^{(\ell)} = n_{W}^{(\ell)} - n_{W}^{(0)} \, \trm{for } W \in \{1a, 1b\} \, \trm{and } \ell \in \{1,2,3\},
\nn
& \delta_{2c} = n_{2c}^{(1)} - n_{2c}^{(0)},
\nn
C_6: \, & \delta_{1a}^{(\ell)} = n_{1a}^{(\ell)} - n_{1a}^{(0)} \, \trm{for } \, \ell \in \{1,2,3,4,5\},
\nn
& \delta_{2b}^{(\ell)} = n_{2b}^{(\ell)} - n_{2b}^{(0)} \, \trm{for } \, \ell \in \{1,2\},
\nn
& \delta_{3c} = n_{3c}^{(1)} - n_{3c}^{(0)}.
\label{eq:RSI_defn_spinless_supp}
\ea
and for spin-1/2 electrons,
\ba
C_2: \, & \delta_{W} = n_{W}^{(3/2)} - n_{W}^{(1/2)} \, \trm{for } W \in \{1a, 1b, 1c, 1d\},
\nn
C_3: \, & \delta_{W}^{(\ell-1/2)} = n_{W}^{(\ell)} - n_{W}^{(1/2)} \, \trm{for } W \in \{1a, 1b, 1c\} \, \trm{and } \ell \in \{3/2,5/2\},
\nn
C_4: \, & \delta_{W}^{(\ell-1/2)} = n_{W}^{(\ell)} - n_{W}^{(1/2)} \, \trm{for } W \in \{1a, 1b\} \, \trm{and } \ell \in \{3/2,5/2,7/2\},
\nn
& \delta_{2c} = n_{2c}^{(3/2)} - n_{2c}^{(1/2)},
\nn
C_6: \, & \delta_{1a}^{(\ell-1/2)} = n_{1a}^{(\ell)} - n_{1a}^{(1/2)} \, \trm{for } \, \ell \in \{3/2,5/2,7/2,9/2,11/2\},
\nn
& \delta_{2b}^{(\ell-1/2)} = n_{2b}^{(\ell)} - n_{2b}^{(1/2)} \, \trm{for } \, \ell \in \{3/2,5/2\},
\nn
& \delta_{3c} = n_{3c}^{(3/2)} - n_{3c}^{(1/2)}.
\label{eq:RSI_defn_spinless_supp}
\ea
Additionally, the filling $\nu$ can be expressed in terms of the Wannier orbital irrep multiplicities,
\ba
\nu=\sum_{W,\ell} M_W n_{W}^{(\ell)}
\label{eq:filling_Wannier_orbital_irrep_mult_supp}
\ea
where $M_W$ is the multiplicity of the maximal WP $W$.
We introduce the column vector $\bb v_{\trm{RSI}}^{(n)}$ which contains the minimal complete set of RSIs for each $C_n$ point group,
\ba
&\bb v_{\trm{RSI}}^{(2)} = \begin{pmatrix} \nu, & \delta_{1a}, & \delta_{1b}, & \delta_{1c}, & \delta_{1d} \end{pmatrix}^{T}, 
\nn
&\bb v_{\trm{RSI}}^{(3)} = \begin{pmatrix} \nu, & \delta_{1a}^{(1)}, & \delta_{1a}^{(2)}, & \delta_{1b}^{(1)}, & \delta_{1b}^{(2)}, & \delta_{1c}^{(1)}, & \delta_{1c}^{(2)} \end{pmatrix}^{T},
\nn
&\bb v_{\trm{RSI}}^{(4)} = \begin{pmatrix} \nu, & \delta_{1a}^{(1)}, & \delta_{1a}^{(2)}, & \delta_{1a}^{(3)}, & \delta_{1b}^{(1)}, & \delta_{1b}^{(2)}, & \delta_{1b}^{(3)}, & \delta_{2c} \end{pmatrix}^{T},
\nn
&\bb v_{\trm{RSI}}^{(6)} = \begin{pmatrix} \nu, & \delta_{1a}^{(1)}, & \delta_{1a}^{(2)}, & \delta_{1a}^{(3)}, & \delta_{1a}^{(4)}, & \delta_{1a}^{(5)}, & \delta_{2b}^{(1)}, & \delta_{2b}^{(2)}, & \delta_{3c} \end{pmatrix}^{T}.
\label{eq:RSI_C2_C3_C4_C6_vector_supp}
\ea
Crucially, $\bb v_{\trm{TCM}}^{(n)}$, $\bb v_{\trm{MS}}^{(n)}$, and $\bb v_{\trm{RSI}}^{(n)}$ all have the same rank, which means that \textit{bijective} mappings exist between each of these quantities,
\ba
\bb v_{\trm{TCM}}^{(n)} = M_{\trm{TCM}\leftarrow\trm{MS}}^{(n)}\bb v_{\trm{MS}}^{(n)} \ \text{and} \ \bb v_{\trm{MS}}^{(n)} = M_{\trm{MS}\leftarrow\trm{RSI}}^{(n)}\bb v_{\trm{RSI}}^{(n)} \to \bb v_{\trm{TCM}}^{(n)} = M_{\trm{TCM}\leftarrow\trm{RSI}}^{(n)}\bb v_{\trm{RSI}}^{(n)},
\label{eq:mappings_supp}
\ea
with $M_{\trm{TCM}\leftarrow\trm{RSI}}^{(n)} = M_{\trm{TCM}\leftarrow\trm{MS}}^{(n)}M_{\trm{MS}\leftarrow\trm{RSI}}^{(n)}$.
$M_{\trm{TCM}\leftarrow\trm{MS}}^{(n)}$ is a matrix for the mapping from the momentum-space rotation invariants to the fully traced TCMs, and similarly, $M_{\trm{MS}\leftarrow\trm{RSI}}^{(n)}$ is the matrix for the mapping from the real-space invariants to the momentum-space rotation invariants.
Note that $M_{\trm{MS}\leftarrow\trm{TCM}}^{(n)}=\left(M_{\trm{TCM}\leftarrow\trm{MS}}^{(n)}\right)^{-1}$ and $M_{\trm{RSI}\leftarrow\trm{MS}}^{(n)}=\left(M_{\trm{MS}\leftarrow\trm{RSI}}^{(n)}\right)^{-1}$.
The entries of $M_{\trm{TCM}\leftarrow\trm{MS}}$ are calculated using Eq.~\eqref{eq:fully_traced_TCM_4_supp}, and the entries of $M_{\trm{MS}\leftarrow\trm{RSI}}$ are determined from elementary band representation tables.
%%%%%%

%%%%%%
\section{Elementary band representations for $C_n$-symmetric atomic insulators}
%%%%%%
In this section, we provide tables of the elementary band representations for spinless and spin-1/2 systems for each $C_n$ symmetry for symmetry class A systems.
Although we do not provide a review of band representation theory in this Supplemental Material, Ref.~\cite{Cano2021} provides an excellent review on the subject, and the elementary band representations for $C_n$ point groups specifically are also derived in Refs.~\cite{Li2020,Takahashi2021}.
{
\renewcommand{\arraystretch}{1.2}
\begin{table}[htpb]
\centering
\caption{
Representation of bands induced from Wannier orbitals for a $C_2$-symmetric lattice.
Each Wannier orbital type can take one of maximal WPs, $W=1a,1b,1c,1d$ with an angular momentum $\ell=0$ or $1$ (spinless) or angular momentum $\ell=1/2$ or $l=3/2$ (spin-1/2).
The 2nd-5th columns denote $C_2$ eigenvalue of bands at high-symmetry points in momentum space.
The 6th-8th columns represent the rotation invariants.
}
\label{table:C2_bandrep_supp}
\setlength{\tabcolsep}{6.5pt}
\begin{tabularx}{0.45\textwidth}{c | c c c c c c c}
\hline \hline
\multirowcell{2}{$(\ell,W)$} & \multicolumn{4}{c}{HSP$_2$} & \multicolumn{3}{c}{Rotation Invariants}
\\ \cline{2-5} \cline{6-8}
& $\Gamma$ & $X$ & $Y$ & $M$ & $[X_1]$ & $[Y_1]$ & $[M_1]$
\\ \hline
$(0,1a)$ & $1$ & $1$ & $1$ & $1$ & 0 & 0 & 0
\\
$(1,1a)$ & $-1$ & $-1$ & $-1$ & $-1$ & 0 & 0 & 0
\\
$(0,1b)$ & $1$ & $-1$ & $1$ & $-1$ & $-1$ & 0 & $-1$
\\
$(1,1b)$ & $-1$ & $1$ & $-1$ & $1$ & $1$ & 0 & $1$
\\
$(0,1c)$ & $1$ & $1$ & $-1$ & $-1$ & 0 & $-1$ & $-1$
\\
$(1,1c)$ & $-1$ & $-1$ & $1$ & $1$ & 0 & $1$ & $1$
\\
$(0,1d)$ & $1$ & $-1$ & $-1$ & $1$ & $-1$ & $-1$ & 0
\\
$(1,1d)$ & $-1$ & $1$ & $1$ & $-1$ & $1$ & $1$ & 0
\\ \hline \hline
\end{tabularx}
\hspace{2em}
\setlength{\tabcolsep}{6pt}
\begin{tabularx}{0.45\textwidth}{c | c c c c c c c}
\hline \hline
\multirowcell{2}{$(\ell,W)$} & \multicolumn{4}{c}{HSP$_2$} & \multicolumn{3}{c}{Rotation Invariants}
\\ \cline{2-5} \cline{6-8}
& $\Gamma$ & $X$ & $Y$ & $M$ & $[X_1]$ & $[Y_1]$ & $[M_1]$
\\ \hline
$(1/2,1a)$ & $i$ & $i$ & $i$ & $i$ & 0 & 0 & 0
\\
$(3/2,1a)$ & $-i$ & $-i$ & $-i$ & $-i$ & 0 & 0 & 0
\\
$(1/2,1b)$ & $i$ & $-i$ & $i$ & $-i$ & $-1$ & 0 & $-1$
\\
$(3/2,1b)$ & $-i$ & $i$ & $-i$ & $i$ & $1$ & 0 & $1$
\\
$(1/2,1c)$ & $i$ & $i$ & $-i$ & $-i$ & 0 & $-1$ & $-1$
\\
$(3/2,1c)$ & $-i$ & $-i$ & $i$ & $i$ & 0 & $1$ & $1$
\\
$(1/2,1d)$ & $i$ & $-i$ & $-i$ & $i$ & $-1$ & $-1$ & 0
\\
$(3/2,1d)$ & $-i$ & $i$ & $i$ & $-i$ & $1$ & $1$ & 0
\\ \hline \hline
\end{tabularx}
\end{table}
}
{
\renewcommand{\arraystretch}{1.2}
\begin{table}[htpb]
\centering
\caption{
Representation of bands induced from Wannier orbitals for a $C_3$-symmetric lattice.
Each Wannier orbital type can take one of maximal WPs, $W=1a,1b,1c$ with an angular momentum $\ell=0,1,2$ (spinless) or angular momentum $\ell=1/2,3/2,5/2$ (spin-1/2).
The 2nd-4th columns denote $C_3$ eigenvalue $\omega=e^{2\pi i/3}$ and $\zeta=e^{\pi i/3}$ (and their complex conjugates, $\bar{\omega}$ and $\bar{\zeta}$ respectively) of bands at high-symmetry points in momentum space.
The 5th-8th columns represent the rotation invariants.
}
\label{table:C3_bandrep_supp}
\setlength{\tabcolsep}{7pt}
\begin{tabularx}{0.45\textwidth}{c | c c c c c c c}
\hline \hline
\multirowcell{2}{$(\ell,W)$} & \multicolumn{3}{c}{HSP$_3$} & \multicolumn{4}{c}{Rotation Invariants}
\\ \cline{2-4} \cline{5-8}
& $\Gamma$ & $K$ & $K'$ & $[K_1]$ & $[K_2]$ & $[K'_1]$ & $[K'_2]$
\\ \hline
$(0,1a)$ & $1$ & $1$ & $1$ & $0$ & $0$ & $0$ & $0$
\\
$(1,1a)$ & $\omega$ & $\omega$ & $\omega$ & $0$ & $0$ & $0$ & $0$
\\
$(2,1a)$ & $\bar{\omega}$ & $\bar{\omega}$ & $\bar{\omega}$ & $0$ & $0$ & $0$ & $0$
\\
$(0,1b)$ & $1$ & $\omega$ & $\bar{\omega}$ & $-1$ & $1$ & $-1$ & $0$
\\
$(1,1b)$ & $\omega$ & $\bar{\omega}$ & $1$ & $0$ & $-1$ & $1$ & $-1$
\\
$(2,1b)$ & $\bar{\omega}$ & $1$ & $\omega$ & $1$ & $0$ & $0$ & $1$
\\
$(0,1c)$ & $1$ & $\bar{\omega}$ & $\omega$ & $-1$ & $0$ & $-1$ & $1$
\\
$(1,1c)$ & $\omega$ & $1$ & $\bar{\omega}$ & $1$ & $-1$ & $0$ & $-1$
\\
$(2,1c)$ & $\bar{\omega}$ & $\omega$ & $1$ & $0$ & $1$ & $1$ & $0$
\\ \hline \hline
\end{tabularx}
\hspace{2em}
\setlength{\tabcolsep}{5.7pt}
\begin{tabularx}{0.45\textwidth}{c | c c c c c c c}
\hline \hline
\multirowcell{2}{$(\ell,W)$} & \multicolumn{3}{c}{HSP$_3$} & \multicolumn{4}{c}{Rotation Invariants}
\\ \cline{2-4} \cline{5-8}
& $\Gamma$ & $K$ & $K'$ & $[K_1]$ & $[K_2]$ & $[K'_1]$ & $[K'_2]$
\\ \hline
$(1/2,1a)$ & $\zeta$ & $\zeta$ & $\zeta$ & $0$ & $0$ & $0$ & $0$
\\
$(3/2,1a)$ & $-1$ & $-1$ & $-1$ & $0$ & $0$ & $0$ & $0$
\\
$(5/2,1a)$ & $\bar{\zeta}$ & $\bar{\zeta}$ & $\bar{\zeta}$ & $0$ & $0$ & $0$ & $0$
\\
$(1/2,1b)$ & $\zeta$ & $-1$ & $\bar{\zeta}$ & $-1$ & $1$ & $-1$ & $0$
\\
$(3/2,1b)$ & $-1$ & $\bar{\zeta}$ & $\zeta$ & $0$ & $-1$ & $1$ & $-1$
\\
$(5/2,1b)$ & $\bar{\zeta}$ & $\zeta$ & $-1$ & $1$ & $0$ & $0$ & $1$
\\
$(1/2,1c)$ & $\zeta$ & $\bar{\zeta}$ & $-1$ & $-1$ & $0$ & $-1$ & $1$
\\
$(3/2,1c)$ & $-1$ & $\zeta$ & $\bar{\zeta}$ & $1$ & $-1$ & $0$ & $-1$
\\
$(5/2,1c)$ & $\bar{\zeta}$ & $-1$ & $\zeta$ & $0$ & $1$ & $1$ & $0$
\\ \hline \hline
\end{tabularx}
\end{table}
}
{
\renewcommand{\arraystretch}{1.2}
\begin{table}[htpb]
\centering
\caption{
Representation of bands induced from Wannier orbitals for a $C_4$-symmetric lattice.
Each Wannier orbital type can take one of maximal WPs, $W=1a,1b$ with an angular momentum $\ell=0,1,2,3$ (spinless) or angular momentum $\ell=1/2,3/2,5/2,7/2$ (spin-1/2) \textit{or} take the $W=2c$ WP with angular momentum $\ell=0$ or $\ell=1$ (spinless) or angular momentum $\ell=1/2$ or $\ell=3/2$ (spin-1/2).
The 2nd-4th columns denote $C_4$ eigenvalues ($\eta=e^{i\pi/4}$ and its complex conjugate $\bar{\eta}$) or $C_2$ eigenvalues of bands at high-symmetry points in momentum space.
Note that $X$ (and $X'$) are $C_2$-invariant HSPs contained in $\trm{HSP}_2$, whereas $\Gamma$ and $M$ are $C_4$-invariant HSPs contained in $\trm{HSP}_4$.
The 5th-8th columns represent the rotation invariants.
}
\label{table:C4_bandrep_supp}
\setlength{\tabcolsep}{2.75pt}
\begin{tabularx}{0.45\textwidth}{c | c c c c c c c}
\hline \hline
\multirowcell{2}{$(\ell,W)$} & \multicolumn{3}{c}{HSP$_4$/HSP$_2$} & \multicolumn{4}{c}{Rotation Invariants}
\\ \cline{2-4} \cline{5-8}
& $\Gamma$ & $X$ & $M$ & $[X_1]$ & $[M_1]$ & $[M_2]$ & $[M_3]$
\\ \hline
$(0,1a)$ & $1$ & $1$ & $1$ & $0$ & $0$ & $0$ & $0$
\\
$(1,1a)$ & $i$ & $-1$ & $i$ & $0$ & $0$ & $0$ & $0$
\\
$(2,1a)$ & $-1$ & $1$ & $-1$ & $0$ & $0$ & $0$ & $0$
\\
$(3,1a)$ & $-i$ & $-1$ & $-i$ & $0$ & $0$ & $0$ & $0$
\\
$(0,1b)$ & $1$ & $-1$ & $-1$ & $-1$ & $-1$ & $0$ & $1$
\\
$(1,1b)$ & $i$ & $1$ & $-i$ & $1$ & $0$ & $-1$ & $0$
\\
$(2,1b)$ & $-1$ & $-1$ & $1$ & $-1$ & $1$ & $0$ & $-1$
\\
$(3,1b)$ & $-i$ & $1$ & $i$ & $1$ & $0$ & $1$ & $0$
\\
$(0,2c)$ & $\{1,-1\}$ & $\{1,-1\}$ & $\{i,-i\}$ & $-1$ & $-1$ & $1$ & $-1$
\\
$(1,2c)$ & $\{i,-i\}$ & $\{1,-1\}$ & $\{1,-1\}$ & $1$ & $1$ & $-1$ & $1$
\\ \hline \hline
\end{tabularx}
\hspace{2em}
\setlength{\tabcolsep}{2.25pt}
\begin{tabularx}{0.45\textwidth}{c | c c c c c c c}
\hline \hline
\multirowcell{2}{$(\ell,W)$} & \multicolumn{3}{c}{HSP$_4$/HSP$_2$} & \multicolumn{4}{c}{Rotation Invariants}
\\ \cline{2-4} \cline{5-8}
& $\Gamma$ & $X$ & $M$ & $[X_1]$ & $[M_1]$ & $[M_2]$ & $[M_3]$
\\ \hline
$(1/2,1a)$ & $\eta$ & $i$ & $\eta$ & $0$ & $0$ & $0$ & $0$
\\
$(3/2,1a)$ & $-\bar{\eta}$ & $-i$ & $-\bar{\eta}$ & $0$ & $0$ & $0$ & $0$
\\
$(5/2,1a)$ & $-\eta$ & $i$ & $-\eta$ & $0$ & $0$ & $0$ & $0$
\\
$(7/2,1a)$ & $\bar{\eta}$ & $-i$ & $\bar{\eta}$ & $0$ & $0$ & $0$ & $0$
\\
$(1/2,1b)$ & $\eta$ & $-i$ & $-\eta$ & $-1$ & $-1$ & $0$ & $1$
\\
$(3/2,1b)$ & $-\bar{\eta}$ & $i$ & $\bar{\eta}$ & $1$ & $0$ & $-1$ & $0$
\\
$(5/2,1b)$ & $-\eta$ & $-i$ & $\eta$ & $-1$ & $1$ & $0$ & $-1$
\\
$(7/2,1b)$ & $\bar{\eta}$ & $i$ & $-\bar{\eta}$ & $1$ & $0$ & $1$ & $0$
\\
$(1/2,2c)$ & $\{\eta,-\eta\}$ & $\{i,-i\}$ & $\{-\bar{\eta},\bar{\eta}\}$ & $-1$ & $-1$ & $1$ & $-1$
\\
$(3/2,2c)$ & $\{-\bar{\eta},\bar{\eta}\}$ & $\{i,-i\}$ & $\{\eta,-\eta\}$ & $1$ & $1$ & $-1$ & $1$
\\ \hline \hline
\end{tabularx}
\end{table}
}
{
\renewcommand{\arraystretch}{1.2}
\begin{table}[htpb]
\centering
\caption{
Representation of bands induced from Wannier orbitals on a $C_6$-symmetric lattice.
Each Wannier orbital type can take one of maximal WPs, $W=1a$ with angular momentum $\ell=0,1,2,3,4,5$ (spinless) or angular momentum $\ell=1/2,3/2,5/2,7/2,11/2$ (spin-1/2), \textit{or} take the $W=2b$ WP with angular momentum $\ell=0,1,2$ (spinless) or angular momentum $\ell=1/2,3/2,5/2$, \textit{or} take the $W=3c$ WP with angular momentum $\ell=0,1$ (spinless) or angular momentum $\ell=1/2,3/2$ (spin-1/2).
The 2nd-4th columns denote $C_6$ eigenvalues or $C_3$ eigenvalues ($\gamma=e^{\pi i/6}$, $\zeta=e^{\pi i/3}$, $\omega=e^{2\pi i/3}$ and their complex conjugates $\bar{\gamma}$, $\bar{\zeta}$, and $\bar{\omega}$ respectively) or $C_2$ eigenvalues of bands at high-symmetry points in momentum space.
Note that $K$ (and $K'$) are $C_3$ invariant HSPs contained in $\trm{HSP}_3$ and $M$ (and $M'$, $M''$) are $C_2$ invariants HSPs contained in $\trm{HSP}_2$, whereas $\Gamma$ is a $C_6$ invariant HSP contained in $\trm{HSP}_6$.
The 5th-7th columns represent the rotation invariants.
}
\label{table:C6_bandrep_supp_1}
\begin{tabularx}{0.54\textwidth}{l*{7}{>{\centering\arraybackslash}p{4em}}}
\hline \hline
\multirowcell{2}{$(\ell,W)$} & \multicolumn{3}{c}{HSP$_6$/HSP$_3$/HSP$_2$} & \multicolumn{3}{c}{Rotation Invariants}
\\ \cline{2-4} \cline{5-7}
& $\Gamma$ & $K$ & $M$ & $[K_1]$ & $[K_2]$ & $[M_1]$
\\ \hline
$(0,1a)$ & $1$ & $1$ & $1$ & $0$ & $0$ & $0$ 
\\
$(1,1a)$ & $\zeta$ & $\omega$ & $-1$ & $0$ & $0$ & $0$
\\
$(2,1a)$ & $\omega$ & $\bar{\omega}$ & $1$ & $0$ & $0$ & $0$
\\
$(3,1a)$ & $-1$ & $1$ & $-1$ & $0$ & $0$ & $0$
\\
$(4,1a)$ & $\bar{\omega}$ & $\omega$ & $1$ & $0$ & $0$ & $0$
\\
$(5,1a)$ & $\bar{\zeta}$ & $\bar{\omega}$ & $-1$ & $0$ & $0$ & $0$
\\
$(0,2b)$ & $\{1,-1\}$ & $\{\omega,\bar{\omega}\}$ & $\{1,-1\}$ & $-2$ & $1$ & $0$
\\
$(1,2b)$ & $\{\zeta,\bar{\omega}\}$ & $\{\bar{\omega},1\}$ & $\{1,-1\}$ & $1$ & $-2$ & $0$
\\
$(2,2b)$ & $\{\omega,\bar{\zeta}\}$ & $\{1,\omega\}$ & $\{1,-1\}$ & $1$ & $1$ & $0$
\\
$(0,3c)$ & $\{1,\omega,\bar{\omega}\}$ & $\{1,\omega,\bar{\omega}\}$ & $\{1,-1,-1\}$ & $0$ & $0$ & $-2$
\\
$(1,3c)$ & $\{\zeta,-1,\bar{\zeta}\}$ & $\{\zeta,-1,\bar{\zeta}\}$ & $\{1,1,-1\}$ & $0$ & $0$ & $2$
\\ \hline \hline
\end{tabularx}
\end{table}
}
{
\renewcommand{\arraystretch}{1.2}
\begin{table}[htpb]
\centering
\label{table:C6_bandrep_supp_2}
\begin{tabularx}{0.625\textwidth}{l*{7}{>{\centering\arraybackslash}p{4.5em}}}
\hline \hline
\multirowcell{2}{$(\ell,W)$} & \multicolumn{3}{c}{HSP$_6$/HSP$_3$/HSP$_2$} & \multicolumn{3}{c}{Rotation Invariants}
\\ \cline{2-4} \cline{5-7}
& $\Gamma$ & $K$ & $M$ & $[K_1]$ & $[K_2]$ & $[M_1]$
\\ \hline
$(1/2,1a)$ & $\gamma$ & $\zeta$ & $i$ & $0$ & $0$ & $0$
\\
$(3/2,1a)$ & $i$ & $-1$ & $-i$ & $0$ & $0$ & $0$
\\
$(5/2,1a)$ & $-\bar{\gamma}$ & $\bar{\zeta}$ & $i$ & $0$ & $0$ & $0$
\\
$(7/2,1a)$ & $-\gamma$ & $\zeta$ & $-i$ & $0$ & $0$ & $0$
\\
$(9/2,1a)$ & $-i$ & $-1$ & $i$ & $0$ & $0$ & $0$
\\
$(11/2,1a)$ & $\bar{\gamma}$ & $\bar{\zeta}$ & $-i$ & $0$ & $0$ & $0$
\\
$(1/2,2b)$ & $\{\gamma,-\gamma\}$ & $\{-1,\bar{\zeta}\}$ & $\{i,-i\}$ & $-2$ & $1$ & $0$
\\
$(3/2,2b)$ & $\{i,-i\}$ & $\{\bar{\zeta},\zeta\}$ & $\{i,-i\}$ & $1$ & $-2$ & $0$
\\
$(5/2,2b)$ & $\{-\bar{\gamma},\bar{\gamma}\}$ & $\{\zeta,-1\}$ & $\{i,-i\}$ & $1$ & $1$ & $0$
\\
$(1/2,3c)$ & $\{\gamma,-\bar{\gamma},-i\}$ & $\{\zeta,-1,\bar{\zeta}\}$ & $\{i,-i,-i\}$ & $0$ & $0$ & $-2$
\\
$(3/2,3c)$ & $\{i,-\gamma,\bar{\gamma}\}$ & $\{\zeta,-1,\bar{\zeta}\}$ & $\{i,i,-i\}$ & $0$ & $0$ & $2$
\\ \hline \hline
\end{tabularx}
\end{table}
}
%

%%%%%%
\section{Comments on class D superconductors and pairing symmetry}
\label{sec:pairingsymm}
%%%%%%
The mappings between the fully traced TCMs and the rotation invariants provided in Sec.~\ref{sec:matrices_supp} are for symmetry classes A, AI, AII, and D of the Altland-Zirnbauer (AZ) classification.
However, class D topological crystalline superconductors (TCS), unlike class A, AI, and AII insulators, are expressed in terms of Bogoliubov de-Gennes (BdG) Hamiltonians.
The BdG Hamiltonian is typically given as follows,
\bg
H_{\rm BdG}(\bk)=\begin{pmatrix} H(\bk) & \Delta(\bk) \\ \Delta^{\dagger}(\bk) & -H^*(-\bk) \end{pmatrix},
\label{eq:BdGHamiltonian}
\eg
where $H(\bk)$ is the Hamiltonian encoding the band structure in the normal phase, and the BdG Hamiltonian $H_{\rm BdG}(\bk)$ encodes the band structure for the superconducting phase, and $\Delta(\bk)$ denotes the gap function.
Since class D TCSs must obey particle-hole symmetry, the BdG Hamiltonian in Eq.~\eqref{eq:BdGHamiltonian} satisfies $\Xi H_{\rm BdG}(\bk) \Xi^{-1} = -H_{\rm BdG}(-\bk)$ with $\Delta^{\dagger}(-\bk)=-\Delta^*(\bk)$, where $\Xi=\tau_1 K$ is the particle-hole operator, $\tau_1$ is the Pauli matrix in the Nambu basis, and $K$ is the complex conjugation operator.
Consider a spatial symmetry $\chi=\{R_{\chi}|\boldsymbol{\delta}_{\chi}\}\in G$, the momentum-space representation of which is given by $U_{\chi}(\bk)$.
In order for the BdG Hamiltonian to be symmetric under $\chi$, the following relations must be satisfied,
\ba
&U_{\chi}(\bk)H(\bk)U_{\chi}^{\dagger}(\bk)=H(R_{\chi}\bk), 
\nn
&U_{\chi}(\bk)\Delta(\bk)U_{\chi}^{T}(-\bk)=\Theta(\chi)\Delta(R_{\chi}\bk),
\label{eq:BdGspatialsymm}
\ea
where a $U(1)$ phase factor $\Theta(\chi)$ takes one-dimensional representation of wallpaper or space group.
This allows for $U_{\trm{BdG},\chi}(\bk)H_{\trm{BdG}}(\bk)U_{\trm{BdG},\chi}^{\dagger}(\bk)=H_{\trm{BdG}}(R_{\chi}\bk)$ where $U_{\trm{BdG},\chi}(\bk)$ is the momentum-space representation of $\chi$ in the Nambu basis.
Here, $U_{\trm{BdG},\chi}(\bk)$ is defined as follows,
\bg
U_{\trm{BdG},\chi}(\bk)=\begin{pmatrix} U_{\chi}(\bk) & 0 \\ 0 & \Theta(\chi)U_{\chi}^*(-\bk) \end{pmatrix}
\label{eq:BdGspatialsymmII}
\eg
which implies the following,
\bg
\Xi\chi\Xi^{-1}=\Theta^*(\chi)\chi\leftrightarrow \tau_1 \, U_{\trm{BdG},\chi}^*(\bk) \, \tau_1=\Theta^*(\chi)U_{\trm{BdG},\chi}(-\bk)
\label{eq:BdGspatialsymmIII}
\eg
According to Eqs.~\eqref{eq:BdGspatialsymm}-\eqref{eq:BdGspatialsymmIII}, $\Theta(\chi)$ determines the \textit{pairing symmetry}, which is the symmetry associated with the gap function $\Delta(\bk)$.
Both $\chi$ and $\Theta(\chi)$ determine the symmetries of the BdG Hamiltonian.
For this work, we focus only on the trivial pairing symmetry for $C_n$ rotations, where $\Theta(c_n(\bb r_o))=1$ and $[\Xi,c_n(\bb r_o)]=0$.
Thus, in Sec.~\ref{sec:matrices_supp}, all class D results obtained are for TCSs with pairing symmetries that allow for commuting particle-hole and rotation operators.

Different values of $\Theta(\chi)$ can change the pairing symmetry of the BdG Hamiltonian, but it is not always possible to apply symmetry indicators to properly diagnose the bulk topological properties of the TCS for certain values of $\Theta(\chi)$~\cite{Geier2020,Ono2020b}.
However, if the bulk topology is symmetry-indicated for a particular non-trivial $\Theta(\chi)$, then it is possible to also express the bulk topology in terms of the fully traced TCMs.
This can lead to an issue where the spatial symmetry $\chi$ and the particle-hole operator $\Xi$ no longer commute as per Eq.~\eqref{eq:BdGspatialsymmIII} (although this can be partially resolved through a redefinition of the unitary operator for $\chi$, e.g., see Ref.~\onlinecite{Peng2022}.)
This implies that a broader characterization of the bulk topology in terms of the fully traced TCMs exists for class D TCSs for different pairing symmetries. For each pairing symmetry that exhibits symmetry-indicated bulk topology, the mapping between the fully traced TCMs and the momentum-space data (e.g., rotation invariants for $C_n$ rotations) can be modified by properly incorporating the U(1) phase factor $\Theta(\chi)$, which remains to be studied. 
%%%%%%

%%%%%%
\section{Matrices for mappings for each symmetry class and $C_n$ symmetry}
\label{sec:matrices_supp}
%%%%%%
In this section, we derive the matrices $M_{\trm{TCM}\leftarrow\trm{MS}}^{(n)}$, $M_{\trm{MS}\leftarrow\trm{RSI}}^{(n)}$, and consequently, $M_{\trm{TCM}\leftarrow\trm{RSI}}^{(n)}$ for every $C_n$ symmetry and symmetry class (classes A, AI, AII, and D). 
Then, using these matrices, we explain how to determine the bulk Chern number, bulk polarization, and sector charge in terms of the fully traced TCMs.
The definitions of $\bb v_{\trm{TCM}}^{(n)}$, $\bb v_{\trm{MS}}^{(n)}$, and $\bb v_{\trm{RSI}}^{(n)}$ are given by Eqs.~\eqref{eq:TCM_C2_vector_supp}-\eqref{eq:MS_C2_C3_C4_C6_vector_supp}, and \eqref{eq:RSI_C2_C3_C4_C6_vector_supp}. 
The results derived in this section only hold for lattices that satisfy the constraint given as,
\ba
& C_2: \, (N_1, N_2) = (0, 0) \pmod 2
\nn
& C_4: \, N = 0 \pmod 2
\nn
& C_{n=3,6}: \, N = 0 \pmod n.
\label{eq:perfect_constraint_supp}
\ea
Note that the matrices $M_{\trm{MS}\leftarrow\trm{RSI}}^{(n)}$ only apply to symmetry classes A, AI, and AII, since atomic insulators fall under these symmetry classes.
In this section, note that the expressions for the bulk polarization are stated as ``$\bb P = p_1\bb a_1 + p_2\bb a_2 \pmod{e}$", which means $p_1 \pmod{e}$ and $p_2 \pmod{e}$.
%%%%%%

%%%%%%
\tocless{\subsubsection{$C_2$ symmetry}}{\label{subsec:c2_map}}
%%%%%%
For a $C_2$-symmetric system, the filling $\nu$ constrains the rotation invariants, the $m(\Gamma_p)$ irrep multiplicities, and the Wannier orbital irrep multiplicities to satisfy the following equation,
\ba
\nu = \sum_{p=1}^{2}m(\Gamma_p)=\sum_W \sum_{\ell}n_{W}^{(\ell)}\to\sum_{p=1}^{2}[X_p]=\sum_{p=1}^{2}[Y_p]=\sum_{p=1}^{2}[M_p]=0
\label{eq:C2_filling_constraint_supp}
\ea
Using Eq.~\eqref{eq:fully_traced_TCM_4_supp} and Table~\ref{table:C2_bandrep_supp}, we can determine the $M_{\trm{TCM}\leftarrow\trm{MS}}^{(2)}$ and $M_{\trm{MS}\leftarrow\trm{RSI}}^{(2)}$ matrices for symmetry classes A and AI pertaining to spinless electrons,
\ba
&\bb v_{\trm{TCM}}^{(2)}=M_{\trm{TCM}\leftarrow\trm{MS}}^{(2)}\bb v_{\trm{MS}}^{(2)}\to \begin{pmatrix} \nu \\\\ \vev{c_2(\bx_{1a})}_F \\\\ \vev{c_2(\bx_{1b})}_F \\\\ \vev{c_2(\bx_{1c})}_F \\\\ \vev{c_2(\bx_{1d})}_F \end{pmatrix} =\begin{pmatrix} 1 & 0 & 0 & 0 & 0 \\\\ -4 & 8 & 2 & 2 & 2 \\\\ 0 & 0 & -2 & 2 & -2 \\\\ 0 & 0 & 2 & -2 & -2 \\\\ 0 & 0 & -2 & -2 & 2 \end{pmatrix}\begin{pmatrix} \nu \\\\ m(\Gamma_1) \\\\ [X_1] \\\\ [Y_1] \\\\ [M_1] \end{pmatrix},
\nnnn
&\bb v_{\trm{MS}}^{(2)}=M_{\trm{MS}\leftarrow\trm{RSI}}^{(2)}\bb v_{\trm{RSI}}^{(2)}\to \begin{pmatrix} \nu \\\\ m(\Gamma_1) \\\\ [X_1] \\\\ [Y_1] \\\\ [M_1] \end{pmatrix}=\begin{pmatrix} 1 & 0 & 0 & 0 & 0 \\\\ 1/2 & -1/2 & -1/2 & -1/2 & -1/2 \\\\ 0 & 0 & 1 & 0 & 1 \\\\ 0 & 0 & 0 & 1 & 1 \\\\ 0 & 0 & 1 & 1 & 0 \end{pmatrix}\begin{pmatrix} \nu \\\\ \delta_{1a} \\\\ \delta_{1b} \\\\ \delta_{1c} \\\\ \delta_{1d} \end{pmatrix}.
\label{eq:C2_Matrices_A_AI_supp_I}
\ea
From Eq.~\eqref{eq:C2_Matrices_A_AI_supp_I}, the matrix $M_{\trm{MS}\leftarrow\trm{TCM}}^{(2)}=(M_{\trm{TCM}\leftarrow\trm{MS}}^{(2)})^{-1}$ can be computed and the following relations can be determined,
\ba
&2\nu+\frac{1}{2}\vev{c_2(\bx_{1a})}_F=4m(\Gamma_1)+[X_1]+[Y_1]+[M_1],
\nn
&\frac{1}{2}\vev{c_2(\bx_{1b})}_F=-[X_1]+[Y_1]-[M_1],
\nn
&\frac{1}{2}\vev{c_2(\bx_{1c})}_F=[X_1]-[Y_1]-[M_1],
\nn
&\frac{1}{2}\vev{c_2(\bx_{1d})}_F=-[X_1]-[Y_1]+[M_1],
\label{eq:C2_Matrices_A_AI_supp_II}
\ea
which leads to the relation 
\ba
\mc{C}=[X_1]+[Y_1]+[M_1]=\frac{1}{2}\vev{c_2(\bx_W)}_F \pmod{2} \ \text{for} \ W \in \{1a,1b,1c,1d\}
\label{eq:C2_Matrices_A_AI_supp_III}
\ea
for the Chern number for symmetry classes A and D pertaining to spinless electrons. 
The matrix $M_{\trm{RSI}\leftarrow\trm{TCM}}^{(2)}=(M_{\trm{TCM}\leftarrow\trm{RSI}}^{(2)})^{-1}$ can also be computed and the following relations can be determined,
\ba
\delta_{W}=-\frac{1}{4}\vev{c_2(\bx_{W})}_F \ \text{for} \ W \in \{1a,1b,1c,1d\}.
\label{eq:C2_Matrices_A_AI_supp_IV}
\ea
Note that for atomic insulators, the Chern number must be zero, which enforces a constraint between the rotation invariants as $[X_1]=-([Y_1]+[M_1])\pmod{2}$.
In terms of RSIs and TCMs, the bulk polarization and the sector charge can be expressed as follows (for symmetry classes A and AI pertaining to spinless electrons),
\ba
\bb P
=& \frac{e}{2}[(\delta_{1b}+\delta_{1d})\bb a_1+(\delta_{1c}+\delta_{1d})\bb a_2] \pmod{e}
\nn
=& -\frac{e}{8}[(\vev{c_2(\bx_{1b})}_F+\vev{c_2(\bx_{1d})}_F)\bb a_1+(\vev{c_2(\bx_{1c})}_F+\vev{c_2(\bx_{1d})}_F)\bb a_2] \pmod{e},
\nn
Q_{\trm{sector}}
=& -\frac{e}{2}\delta_{1d}=-\frac{e}{8}\vev{c_2(\bx_{1d})}_F \pmod{e}.
\label{eq:C2_Matrices_A_AI_supp_V}
\ea
For symmetry class A pertaining to spin-1/2 electrons, the matrix $M_{\trm{MS}\leftarrow\trm{RSI}}^{(2)}$ is the same as determined by Table~\ref{table:C2_bandrep_supp}, but the matrix $M_{\trm{TCM}\leftarrow\trm{MS}}^{(2)}$ is
\ba
\bb v_{\trm{TCM}}^{(2)} = M_{\trm{TCM} \leftarrow \trm{MS}}^{(2)} \bb v_{\trm{MS}}^{(2)} \to
\begin{pmatrix} \nu \\\\ \vev{c_2(\bx_{1a})}_F \\\\ \vev{c_2(\bx_{1b})}_F \\\\ \vev{c_2(\bx_{1c})}_F \\\\ \vev{c_2(\bx_{1d})}_F
\end{pmatrix}
= \begin{pmatrix}
1 & 0 & 0 & 0 & 0 \\\\
-4i & 8i & 2i & 2i & 2i \\\\
0 & 0 & -2i & 2i & -2i \\\\
0 & 0 & 2i & -2i & -2i \\\\ 0 & 0 & -2i & -2i & 2i
\end{pmatrix}
\begin{pmatrix}
\nu \\\\ \delta_{1a} \\\\ \delta_{1b} \\\\ \delta_{1c} \\\\ \delta_{1d}
\end{pmatrix},
\label{eq:C2_Matrices_A_spin_half_supp_I}
\ea
which implies the following relations
\ba
&2i\nu-\frac{i}{2}\vev{c_2(\bx_{1a})}_F=4m(\Gamma_1)+[X_1]+[Y_1]+[M_1],
\nn
&-\frac{i}{2}\vev{c_2(\bx_{1b})}_F=-[X_1]+[Y_1]-[M_1],
\nn
&-\frac{i}{2}\vev{c_2(\bx_{1c})}_F=[X_1]-[Y_1]-[M_1],
\nn
&-\frac{i}{2}\vev{c_2(\bx_{1d})}_F=-[X_1]-[Y_1]+[M_1].
\label{eq:C2_Matrices_A_spin_half_supp_II}
\ea
This means the Chern number for symmetry classes A and D pertaining to spin-1/2 electrons is
\ba
\mc{C}=-\frac{i}{2}\vev{c_2(\bx_W)}_F \pmod{2}.
\label{eq:C2_Matrices_A_D_supp_I}
\ea
Computing $M_{\trm{RSI}\leftarrow\trm{TCM}}^{(2)}=(M_{\trm{TCM}\leftarrow\trm{RSI}}^{(2)})^{-1}$ from Eq.~\eqref{eq:C2_Matrices_A_spin_half_supp_I} yields the following relation,
\ba
\delta_W = \frac{i}{4}\vev{c_2(\bx_W)}_F \ \text{for} \ W \in \{1a,1b,1c,1d\}
\label{eq:C2_Matrices_A_spin_half_supp_III}
\ea
Thus, the bulk polarization and the sector charge can be expressed as follows for symmetry class A pertaining to spin-1/2 electrons,
\ba
&\bb P=\frac{ie}{8}[(\vev{c_2(\bx_{1b})}_F+\vev{c_2(\bx_{1d})}_F)\bb a_1+(\vev{c_2(\bx_{1c})}_F+\vev{c_2(\bx_{1d})}_F)\bb a_2] \pmod{e},
\nn
&Q_{\trm{sector}}=\frac{ie}{8}\vev{c_2(\bx_{1d})}_F \pmod{e}.
\label{eq:C2_Matrices_A_spin_half_supp_IV}
\ea
For symmetry class AII, time-reversal symmetry imposes the constraint that $[X_p]=[Y_p]=[M_p]=0$ for $p\in\{1,2\}$.
This also implies that
\ba
\delta_{W}=0 \quad \textrm{and}
\quad \vev{c_2(\bx_{W})}_F = 0 \quad \textrm{for} \quad
W \in \{1a,1b,1c,1d\},
\label{eq:C2_Matrices_AII_supp_II}
\ea
as a consequence of Kramers' theorem, which means $\bb P = \bb 0 \pmod{e}$ and $Q_{\trm{sector}}=0 \pmod{e}$ for symmetry class AII.
%%%%%%

%%%%%%
\tocless{\subsubsection{$C_3$ symmetry}}{\label{subsec:c3_map}}
%%%%%%
For a $C_3$-symmetric system, the filling $\nu$ constrains the rotation invariants, the $m(\Gamma_p)$ irrep multiplicities, and the Wannier orbital irrep multiplicities to satisfy the following equation,
\ba
\nu=\sum_{p=1}^{3}m(\Gamma_p)=\sum_W \sum_{\ell}n_W^{(\ell)}\to\sum_{p=1}^{3}[K_p]=\sum_{p=1}^{3}[K'_p]=0
\label{eq:C3_filling_constraint_supp}
\ea
Using Eq.~\eqref{eq:fully_traced_TCM_4_supp} and Table~\ref{table:C3_bandrep_supp}, we can determine the $M_{\trm{TCM}\leftarrow\trm{MS}}^{(3)}$ and $M_{\trm{MS}\leftarrow\trm{RSI}}^{(3)}$ matrices for symmetry class A pertaining to spinless electrons,
\ba
&\begin{pmatrix} \nu \\\\ \vev{c_3(\bx_{1a})}_F \\\\ \vev{c_3(\bx_{1b})}_F \\\\ \vev{c_3(\bx_{1c})}_F \\\\ \vev{(c_3(\bx_{1a}))^{-1}}_F \\\\ \vev{(c_3(\bx_{1b}))^{-1}}_F \\\\ \vev{(c_3(\bx_{1c}))^{-1}}_F \end{pmatrix}=\begin{pmatrix} 1 & 0 & 0 & 0 & 0 & 0 & 0 \\\\ 3\bar{\omega} & 3(1-\bar{\omega}) & 3(\omega-\bar{\omega}) & 1-\bar{\omega} & \omega-\bar{\omega} & 1-\bar{\omega} & \omega-\bar{\omega} \\\\ 0 & 0 & 0 & -(\omega-\bar{\omega}) & -(\omega-1) & \omega-1 & -(1-\bar{\omega}) \\\\ 0 & 0 & 0 & \omega-1 & -(1-\bar{\omega}) & -(\omega-\bar{\omega}) & -(\omega-1) \\\\ 3\omega & -3(\omega-1) & -3(\omega-\bar{\omega}) & -(\omega-1) & -(\omega-\bar{\omega}) & -(\omega-1) & -(\omega-\bar{\omega}) \\\\ 0 & 0 & 0 & \omega-\bar{\omega} & 1-\bar{\omega} & -(1-\bar{\omega}) & \omega-1 \\\\ 0 & 0 & 0 & -(1-\bar{\omega}) & \omega-1 & \omega-\bar{\omega} & 1-\bar{\omega} \end{pmatrix}\begin{pmatrix} \nu \\\\ m(\Gamma_1) \\\\ m(\Gamma_2) \\\\ [K_1] \\\\ [K_2] \\\\ [K'_1] \\\\ [K'_2] \end{pmatrix},
\nnnn
&\begin{pmatrix} \nu \\\\ m(\Gamma_1) \\\\ m(\Gamma_2) \\\\ [K_1] \\\\ [K_2] \\\\ [K'_1] \\\\ [K'_2] \end{pmatrix}=\begin{pmatrix} 1 & 0 & 0 & 0 & 0 & 0 & 0 \\\\ 1/3 & -1/3 & -1/3 & -1/3 & -1/3 & -1/3 & -1/3 \\\\ 1/3 & 2/3 & -1/3 & 2/3 & -1/3 & 2/3 & -1/3 \\\\ 0 & 0 & 0 & 0 & 1 & 1 & 0 \\\\ 0 & 0 & 0 & -1 & 0 & -1 & 1 \\\\ 0 & 0 & 0 & 1 & 0 & 0 & 1 \\\\ 0 & 0 & 0 & -1 & 1 & -1 & 0 \end{pmatrix}\begin{pmatrix} \nu \\\\ \delta_{1a}^{(1)} \\\\ \delta_{1a}^{(2)} \\\\ \delta_{1b}^{(1)} \\\\ \delta_{1b}^{(2)} \\\\ \delta_{1c}^{(1)} \\\\ \delta_{1c}^{(2)} \end{pmatrix},
\label{eq:C3_Matrices_A_supp_I}
\ea
where $\omega=e^{2\pi i/3}$ and $\bar{\omega}=e^{-2\pi i/3}$.
Note that the entries of the rows of $M_{\trm{TCM}\leftarrow\trm{MS}}^{(3)}$ corresponding to $\vev{(c_3(\bx_{W}))^{-1}}_F$ are complex conjugates of the entries of the rows corresponding to $\vev{c_3(\bx_{W})}_F$, for $W\in\{1a,1b,1c\}$ (e.g., $(\vev{c_3(\bx_{W})}_F)^*=\vev{(c_3(\bx_{W}))^{-1}}_F$).
From Eq.~\eqref{eq:C3_Matrices_A_supp_I} and computing $M_{\trm{MS}\leftarrow\trm{TCM}}^{(3)}=(M_{\trm{TCM}\leftarrow\trm{MS}}^{(3)})^{-1}$, the following relations can be determined,
\ba
&(\omega-\bar{\omega})^{-1}(\vev{c_3(\bx_{1a})}_F-\vev{(c_3(\bx_{1a}))^{-1}}_F)=-3\nu+3m(\Gamma_1)+6m(\Gamma_2)+[K_1]+2[K_2]+[K'_1]+2[K'_2],
\nn
&(\omega-\bar{\omega})^{-1}(\vev{c_3(\bx_{1b})}_F-\vev{(c_3(\bx_{1b}))^{-1}}_F )=-2[K_1]-[K_2]+[K'_1]-[K'_2],
\nn
&(\omega-\bar{\omega})^{-1}(\vev{c_3(\bx_{1c})}_F-\vev{(c_3(\bx_{1c}))^{-1}}_F)=[K_1]-[K_2]-2[K'_1]-[K'_2],
\label{eq:C3_Matrices_A_supp_II}
\ea
which leads to the relation
\ba
\mc{C}=[K_1]+2[K_2]+[K'_1]+2[K'_2]=\frac{2}{\sqrt{3}}\text{Im}[\vev{c_3(\bx_W)}_F] \pmod{3} \ \text{for} \ W \in \{1a,1b,1c\}
\label{eq:C3_Matrices_A_supp_III}
\ea
for the Chern number for spinless electron systems in symmetry class A.
The matrix $M_{\trm{RSI}\leftarrow\trm{TCM}}^{(3)}=(M_{\trm{TCM}\leftarrow\trm{RSI}}^{(3)})^{-1}$ can also be computed and the following relations can be determined,
\ba
&\delta_{W}^{(1)}=\frac{i}{3\sqrt{3}}\omega\vev{c_3(\bx_{W})}_F+\frac{1}{9}(\omega-1)(\vev{c_3(\bx_{W})}_F)^*,
\nn
&\delta_{W}^{(2)}=\frac{1}{9}(\omega-1)\vev{c_3(\bx_{W})}_F+\frac{i}{3\sqrt{3}}\omega(\vev{c_3(\bx_{W})}_F)^*,
\label{eq:C3_Matrices_A_supp_IV}
\ea
for $W\in\{1a,1b,1c\}$.
Note that for atomic insulators, the Chern number must be zero, which enforces a constraint between the rotation invariants as $[K_1]+[K'_1]=[K_2]+[K'_2]\pmod{3}$.
In terms of RSIs and TCMs, the bulk polarization and the sector charge can be expressed as follows (for symmetry class A pertaining to spinless electrons),
\ba
&\bb P = \frac{e}{3}(\delta_{1b}^{(1)}+\delta_{1b}^{(2)}-\delta_{1c}^{(1)}-\delta_{1c}^{(2)})(\bb a_1 + \bb a_2) = -\frac{2e}{3}(\delta_{1b}^{(1)}+\delta_{1b}^{(2)}-\delta_{1c}^{(1)}-\delta_{1c}^{(2)})(\bb a_1 + \bb a_2)
\nn
& = \frac{4e}{9}\text{Re}[\vev{c_3(\bx_{1b})}_F-\vev{c_3(\bx_{1c})}_F](\bb a_1 + \bb a_2) \pmod{e},
\nn
&Q_{\trm{sector}} = \frac{e}{3}(\delta_{1a}^{(1)}+\delta_{1a}^{(2)}) = -\frac{2e}{3}(\delta_{1a}^{(1)}+\delta_{1a}^{(2)}) = \frac{4e}{9}\trm{Re}[\vev{c_3(\bx_{1a})}_F] \pmod{e}.
\label{eq:C3_Matrices_A_supp_V}
\ea
For symmetry class AI systems, time-reversal symmetry introduces the following constraints on the rotation invariants, $[K_2]=-([K'_1]+[K'_2])$ and $[K'_2]=-([K_1]+[K_2])$, which implies $[K_1]=[K'_1]$.
Furthermore, time-reversal symmetry also imposes a constraint that $m(\Gamma_2)=m(\Gamma_3)$, which implies $\nu=m(\Gamma_1)+2m(\Gamma_2)$, as well as $\delta_{W}^{(1)}=\delta_{W}^{(2)}\equiv\delta_{W}$ which also implies $\vev{c_3(\bx_{W})}_F=\vev{(c_3(\bx_{W}))^{-1}}_F=(\vev{c_3(\bx_{W})}_F)^*$ for all $W\in\{1a,1b,1c\}$.
Therefore, to evaluate the matrices for symmetry class AI, we must incorporate the additional constraints imposed by the symmetries and modify $\bb v_{\trm{TCM}}^{(3)}$, $\bb v_{\trm{MS}}^{(3)}$, and $\bb v_{\trm{RSI}}^{(3)}$ to include only the independent set of quantities for each,
\ba
&\bb v_{\trm{TCM}}^{(3)}=\begin{pmatrix} \nu, & \vev{c_3(\bx_{1a})}_F, & \vev{c_3(\bx_{1b})}_F, & \vev{c_3(\bx_{1c})}_F \end{pmatrix}^{T},
\nn
&\bb v_{\trm{MS}}^{(3)}=\begin{pmatrix} \nu, & m(\Gamma_1), & [K_1], & [K_2] \end{pmatrix}^{T},
\nn
&\bb v_{\trm{RSI}}^{(3)}=\begin{pmatrix} \nu, & \delta_{1a}, & \delta_{1b}, & \delta_{1c} \end{pmatrix}^{T}.
\label{eq:C3_Matrices_AI_supp_I}
\ea
Thus, $M_{\trm{TCM}\leftarrow\trm{MS}}^{(3)}$ and $M_{\trm{MS}\leftarrow\trm{RSI}}^{(3)}$ for symmetry class AI are given as follows,
\ba
&\begin{pmatrix} \nu \\\\ \vev{c_3(\bx_{1a})}_F \\\\ \vev{c_3(\bx_{1b})}_F \\\\ \vev{c_3(\bx_{1c})}_F \end{pmatrix}=\begin{pmatrix} 1 & 0 & 0 & 0 \\\\ -3/2 & 9/2 & 3 & 0 \\\\ 0 & 0 & 0 & 3 \\\\ 0 & 0 & -3 & -3 \end{pmatrix}\begin{pmatrix} \nu \\\\ m(\Gamma_1) \\\\ [K_1] \\\\ [K_2] \end{pmatrix},
\nnnn
&\begin{pmatrix} \nu \\\\ m(\Gamma_1) \\\\ [K_1] \\\\ [K_2] \end{pmatrix}=\begin{pmatrix} 1 & 0 & 0 & 0 \\\\ 1/3 & -2/3 & -2/3 & -2/3 \\\\ 0 & 0 & 1 & 1 \\\\ 0 & 0 & -1 & 0 \end{pmatrix}\begin{pmatrix} \nu \\\\ \delta_{1a} \\\\ \delta_{1b} \\\\ \delta_{1c} \end{pmatrix}.
\label{eq:C3_Matrices_AI_supp_II}
\ea
Computing $M_{\trm{RSI}\leftarrow\trm{TCM}}^{(3)}=(M_{\trm{TCM}\leftarrow\trm{RSI}}^{(3)})^{-1}$ yields the following relation,
\ba
\delta_W=-\frac{1}{3}\vev{c_3(\bx_W)}_F \ \text{for} \ W \in \{1a,1b,1c\}
\label{eq:C3_Matrices_AI_supp_III}
\ea
which means the bulk polarization and the sector charge for symmetry class AI are,
\ba
&\bb P = \frac{2e}{3}(\delta_{1b}-\delta_{1c})(\bb a_1 + \bb a_2) = -\frac{e}{3}(\delta_{1b}-\delta_{1c})(\bb a_1 + \bb a_2) = \frac{e}{9}(\vev{c_3(\bx_{1b})}_F-\vev{c_3(\bx_{1c})}_F)(\bb a_1 + \bb a_2) \pmod{e},
\nn
& Q_{\trm{sector}} = \frac{2e}{3}\delta_{1a} = -\frac{e}{3}\delta_{1a} = \frac{e}{9}\vev{c_3(\bx_{1a})}_F \pmod{e}.
\label{eq:C3_Matrices_AI_supp_IV}
\ea
Taking the original definitions of $\bb v_{\trm{TCM}}^{(3)}$, $\bb v_{\trm{MS}}^{(3)}$, and $\bb v_{\trm{RSI}}^{(3)}$ given by Eqs.~\eqref{eq:TCM_C3_C4_C6_vector_supp}, \eqref{eq:MS_C2_C3_C4_C6_vector_supp}, and \eqref{eq:RSI_C2_C3_C4_C6_vector_supp}, one can construct the matrix $M_{\trm{TCM}\leftarrow\trm{MS}}^{(3)}$ once again using Eq.~\eqref{eq:fully_traced_TCM_4_supp} and Table~\ref{table:C3_bandrep_supp} for symmetry class A systems pertaining to spin-1/2 electrons,
\ba
\begin{pmatrix} \nu \\\\ \vev{c_3(\bx_{1a})}_F \\\\ \vev{c_3(\bx_{1b})}_F \\\\ \vev{c_3(\bx_{1c})}_F \\\\ \vev{(c_3(\bx_{1a}))^{-1}}_F \\\\ \vev{(c_3(\bx_{1b}))^{-1}}_F \\\\ \vev{(c_3(\bx_{1c}))^{-1}}_F \end{pmatrix}=\begin{pmatrix} 1 & 0 & 0 & 0 & 0 & 0 & 0 \\\\ 3\bar{\zeta} & 3(\zeta-\bar{\zeta}) & -3(1+\bar{\zeta}) & \zeta-\bar{\zeta} & -(1+\bar{\zeta}) & \zeta-\bar{\zeta} & -(1+\bar{\zeta}) \\\\ 0 & 0 & 0 & 1+\bar{\zeta} & 1+\zeta & -(1+\zeta) & -(\zeta-\bar{\zeta}) \\\\ 0 & 0 & 0 & -(1+\zeta) & -(\zeta-\bar{\zeta}) & 1+\bar{\zeta} & 1+\zeta \\\\ 3\zeta & -3(\zeta-\bar{\zeta}) & -3(1+\zeta) & -(\zeta-\bar{\zeta}) & -(1+\zeta) & -(\zeta-\bar{\zeta}) & -(1+\zeta) \\\\ 0 & 0 & 0 & 1+\zeta & 1+\bar{\zeta} & -(1+\bar{\zeta}) & \zeta-\bar{\zeta} \\\\ 0 & 0 & 0 & -(1+\bar{\zeta}) & \zeta-\bar{\zeta} & 1+\zeta & 1+\bar{\zeta} \end{pmatrix}\begin{pmatrix} \nu \\\\ m(\Gamma_1) \\\\ m(\Gamma_2) \\\\ [K_1] \\\\ [K_2] \\\\ [K'_1] \\\\ [K'_2] \end{pmatrix},
\label{eq:C3_Matrices_A_spin_half_supp_I}
\ea
where $\zeta=e^{\pi i/3}$ and $\bar{\zeta}=e^{-\pi i/3}$, and $M_{\trm{MS}\leftarrow\trm{RSI}}^{(3)}$ is identical to Eq.~\eqref{eq:C3_Matrices_A_supp_I}. 
From Eq.~\eqref{eq:C3_Matrices_A_spin_half_supp_I}, the following relations can be determined,
\ba
&(\bar{\zeta}-\zeta)^{-1}(\vev{c_3(\bx_{1a})}_F-\vev{(c_3(\bx_{1a}))^{-1}}_F)=3\nu-6m(\Gamma_1)-3m(\Gamma_2)-2[K_1]-[K_2]-2[K'_1]-[K'_2],
\nn
&(\bar{\zeta}-\zeta)^{-1}(\vev{c_3(\bx_{1b})}_F-\vev{(c_3(\bx_{1b}))^{-1}}_F)=[K_1]-[K_2]+[K'_1]+2[K'_2],
\nn
&(\bar{\zeta}-\zeta)^{-1}(\vev{c_3(\bx_{1c})}_F-\vev{(c_3(\bx_{1c})^{-1}}_F)=[K_1]+2[K_2]+[K'_1]-[K'_2],
\label{eq:C3_Matrices_A_spin_half_supp_II}
\ea
which leads to the relation
\ba
\mc{C}=[K_1]+2[K_2]+[K'_1]+2[K'_2]=-\frac{2}{\sqrt{3}}\text{Im}[\vev{c_3(\bx_W)}_F] \pmod{3} \ \text{for} \ W \in \{1a,1b,1c\}
\label{eq:C3_Matrices_A_spin_half_supp_III}
\ea
for the Chern number for spin-1/2 electron systems in symmetry class A. 
The matrix $M_{\trm{RSI}\leftarrow\trm{TCM}}^{(3)}=(M_{\trm{TCM}\leftarrow\trm{RSI}}^{(3)})^{-1}$ can also be computed and the following relations can be determined,
\ba
&\delta_{W}^{(1)}=\frac{1}{9}(\omega-1)\vev{c_3(\bx_{W})}_F+\frac{1}{9}(\bar{\omega}-1)(\vev{c_3(\bx_{W})}_F)^*,
\nn
&\delta_{W}^{(2)}=\frac{i}{3\sqrt{3}}(\vev{c_3(\bx_{W})}_F-(\vev{c_3(\bx_{W})}_F)^*),
\label{eq:C3_Matrices_A_spin_half_supp_IV}
\ea
for $W\in\{1a,1b,1c\}$.
In terms of the TCMs, the bulk polarization and the sector charge can be expressed as follows (for symmetry class A pertaining to spin-1/2 electrons),
\ba
&\bb P = \frac{e}{3}(\delta_{1b}^{(1)}+\delta_{1b}^{(2)}-\delta_{1c}^{(1)}-\delta_{1c}^{(2)})(\bb a_1 + \bb a_2) = \frac{e}{3}(-2\delta_{1b}^{(1)}+\delta_{1b}^{(1)}+2\delta_{1c}^{(1)}-\delta_{1c}^{(2)})(\bb a_1 + \bb a_2)
\nn
& = \frac{2e}{9}\text{Re}[\vev{c_3(\bx_{1b})}_F-\vev{c_3(\bx_{1c})}_F](\bb a_1 + \bb a_2) \pmod{e},
\nn
&Q_{\trm{sector}} = \frac{e}{3}(\delta_{1a}^{(1)}+\delta_{1a}^{(2)}) = \frac{e}{3}(-2\delta_{1a}^{(1)}+\delta_{1a}^{(2)}) = \frac{2e}{9}\trm{Re}[\vev{c_3(\bx_{1a})}_F] \pmod{e}.
\label{eq:C3_Matrices_A_spin_half_supp_V}
\ea
For symmetry class AII systems, time-reversal symmetry introduces the following constraints on the rotation invariants, $[K_1]=-([K'_1]+[K'_2])$ and $[K'_1]=-([K_1]+[K_2])$, which implies $[K_2]=[K'_2]$.
Furthermore, time-reversal symmetry also imposes a constraint $m(\Gamma_1)=m(\Gamma_3)$, which implies $\nu=2m(\Gamma_1)+m(\Gamma_2)$, as well as $\delta_W^{(2)}=0$ which therefore implies $\delta_W^{(1)}\equiv\delta_W$.
As expected, $\vev{c_3(\bx_{W})}_F=\vev{(c_3(\bx_{W}))^{-1}}_F=(\vev{c_3(\bx_{W})}_F)^*$ for all $W\in\{1a,1b,1c\}$. 
To evaluate the matrices for symmetry class AII, we will use the modified definition of $\bb v_{\trm{TCM}}^{(3)}$, $\bb v_{\trm{MS}}^{(3)}$ and $\bb v_{\trm{RSI}}^{(3)}$ given by Eq.~\eqref{eq:C3_Matrices_AI_supp_I}. 
Thus, $M_{\trm{TCM}\leftarrow\trm{MS}}^{(3)}$ and $M_{\trm{MS}\leftarrow\trm{RSI}}^{(3)}$ for symmetry class AII are given as follows,
\ba
&\begin{pmatrix} \nu \\\\ \vev{c_3(\bx_{1a})}_F \\\\ \vev{c_3(\bx_{1b})}_F \\\\ \vev{c_3(\bx_{1c})}_F \end{pmatrix}=\begin{pmatrix} 1 & 0 & 0 & 0 \\\\ -3 & 9 & 0 & -3 \\\\ 0 & 0 & 3 & 3 \\\\ 0 & 0 & -3 & 0 \end{pmatrix}\begin{pmatrix} \nu \\\\ m(\Gamma_1) \\\\ [K_1] \\\\ [K_2] \end{pmatrix},
\nnnn
&\begin{pmatrix} \nu \\\\ m(\Gamma_1) \\\\ [K_1] \\\\ [K_2] \end{pmatrix}=\begin{pmatrix} 1 & 0 & 0 & 0 \\\\ 1/3 & -1/3 & -1/3 & -1/3 \\\\ 0 & 0 & 0 & 1 \\\\ 0 & 0 & -1 & -1 \end{pmatrix}\begin{pmatrix} \nu \\\\ \delta_{1a} \\\\ \delta_{1b} \\\\ \delta_{1c} \end{pmatrix}.
\label{eq:C3_Matrices_AII_supp_I}
\ea
Computing $M_{\trm{RSI}\leftarrow\trm{TCM}}^{(3)}=(M_{\trm{TCM}\leftarrow\trm{RSI}}^{(3)})^{-1}$ yields the same relations between the RSIs and TCMs given by Eq.~\eqref{eq:C3_Matrices_AI_supp_III}.
Because $\delta_{W}^{(2)}=0$ and $\delta_{W}\equiv\delta_{W}^{(1)}$, this implies that the bulk polarization and the sector charge for symmetry class AII are,
\ba
&\bb P = \frac{e}{3}(\delta_{1b}-\delta_{1c})(\bb a_1 + \bb a_2) = -\frac{2e}{3}(\delta_{1b}-\delta_{1c})(\bb a_1 + \bb a_2) = \frac{2e}{9}(\vev{c_3(\bx_{1b})}_F-\vev{c_3(\bx_{1c})}_F)(\bb a_1 + \bb a_2) \pmod{e},
\nn
&Q_{\trm{sector}} = \frac{e}{3}(\delta_{1a}^{(1)}+\delta_{1a}^{(2)}) = -\frac{2e}{3}(\delta_{1a}^{(1)}+\delta_{1a}^{(2)}) = \frac{2e}{9}\vev{c_3(\bx_{1a})}_F \pmod{e}.
\label{eq:C3_Matrices_AII_supp_II}
\ea
Finally, for symmetry class D systems, we consider both spinless and spin-1/2 electron systems. 
For both systems, particle-hole symmetry will introduce constraints on the rotation invariants.
Therefore, similar to the approach taken for symmetry classes AI and AII, to evaluate the matrices for symmetry class D, we once again incorporate the additional constraints imposed by the symmetries and modify $\bb v_{\trm{TCM}}^{(3)}$ and $\bb v_{\trm{MS}}^{(3)}$ (we do not consider $\bb v_{\trm{RSI}}^{(3)}$ because symmetry class D does not support atomic insulators) to include only the independent set of quantities for each,
\ba
&\bb v_{\trm{TCM}}^{(3)}=\begin{pmatrix} \nu, & \vev{c_3(\bx_{1a})}_F, & \vev{c_3(\bx_{1b})}_F, & \vev{c_3(\bx_{1c})}_F, & \vev{(c_3(\bx_{1a}))^{-1}}_F \end{pmatrix}^{T},
\nn
&\bb v_{\trm{MS}}^{(3)}=\begin{pmatrix} \nu, & m(\Gamma_1), & m(\Gamma_2), & [K_1], & [K_2] \end{pmatrix}^{T}.
\label{eq:C3_Matrices_D_supp_I}
\ea
For spinless systems, particle-hole symmetry introduces the following constraints on the rotation invariants, $[K_2]=[K'_1]+[K'_2]$ and $[K'_2]=[K_1]+[K_2]$ which implies $[K_1]=-[K'_1]$.
Furthermore, using Eq.~\eqref{eq:fully_traced_TCM_4_supp}, it follows that $\vev{c_3(\bx_W)}_F=-\vev{(c_3(\bx_W))^{-1}}_F=-(\vev{c_3(\bx_W)}_F)^*$ \textit{only for} $W\in\{1b,1c\}$.
Thus, $M_{\trm{TCM}\leftarrow\trm{MS}}^{(3)}$ for symmetry class D for spinless electrons is given as follows,
\ba
\begin{pmatrix} \nu \\\\ \vev{c_3(\bx_{1a})}_F \\\\ \vev{c_3(\bx_{1b})}_F \\\\ \vev{c_3(\bx_{1c})}_F \\\\ \vev{(c_3(\bx_{1a}))^{-1}}_F \end{pmatrix}=\begin{pmatrix} 1 & 0 & 0 & 0 & 0 \\\\ 3\bar{\omega} & 3(1-\bar{\omega}) & 3(\omega-\bar{\omega}) & \omega-\bar{\omega} & 2(\omega-\bar{\omega}) \\\\ 0 & 0 & 0 & -2(\omega-\bar{\omega}) & -(\omega-\bar{\omega}) \\\\ 0 & 0 & 0 & \omega-\bar{\omega} & -(\omega-\bar{\omega}) \\\\ 3\omega & -3(\omega-1) & -3(\omega-\bar{\omega}) & -(\omega-\bar{\omega}) & -2(\omega-\bar{\omega}) \end{pmatrix}\begin{pmatrix} \nu \\\\ m(\Gamma_1) \\\\ m(\Gamma_2) \\\\ [K_1] \\\\ [K_2] \end{pmatrix}
\label{eq:C3_Matrices_D_supp_II}
\ea
From Eq.~\eqref{eq:C3_Matrices_D_supp_II}, the same relations as Eq.~\eqref{eq:C3_Matrices_A_supp_II} can be obtained, with the additional relations $[K'_2]=[K_1]+[K_2]$, $[K'_1]=-[K_1]$, and $\vev{c_3(\bx_{W})}_F=-\vev{(c_3(\bx_{W}))^{-1}}_F=-(\vev{c_3(\bx_{W})}_F)^*$ for $W\in\{1b,1c\}$. 
Thus, the Chern number for symmetry class D pertaining to spinless electrons is equivalent to the Chern number for symmetry class for spinless electrons given by Eq.~\eqref{eq:C3_Matrices_A_supp_III} (note that because $\trm{Re}[\vev{c_3(\bx_{W})}_F]=0$ for $W\in\{1b,1c\}$, the expression for the Chern number can be further simplified).
Similarly, for spin-1/2 systems, particle-hole symmetry introduces the following constraints on the rotation invariants, $[K_1]=[K'_1]+[K'_2]$ and $[K'_1]=[K_1]+[K_2]$ which implies $[K_2]=-[K'_2]$.
The same constraint of $\vev{c_3(\bx_{W})}_F=-\vev{(c_3(\bx_{W}))^{-1}}_F=-(\vev{c_3(\bx_{W})}_F)^*$ also applies for $W\in\{1b,1c\}$.
Thus, $M_{\trm{TCM}\leftarrow\trm{MS}}^{(3)}$ for symmetry class D is given as follows,
\ba
\begin{pmatrix} \nu \\\\ \vev{c_3(\bx_{1a})}_F \\\\ \vev{c_3(\bx_{1b})}_F \\\\ \vev{c_3(\bx_{1c})}_F \\\\ \vev{(c_3(\bx_{1a}))^{-1}}_F \end{pmatrix}=\begin{pmatrix} 1 & 0 & 0 & 0 & 0 \\\\ 3\bar{\zeta} & 3(\zeta-\bar{\zeta}) & -3(1+\bar{\zeta}) & 2(\zeta-\bar{\zeta}) & \zeta-\bar{\zeta} \\\\ 0 & 0 & 0 & -(\zeta-\bar{\zeta}) & \zeta-\bar{\zeta} \\\\ 0 & 0 & 0 & -(\zeta-\bar{\zeta}) & -2(\zeta-\bar{\zeta}) \\\\ 3\zeta & -3(\zeta-\bar{\zeta}) & -3(1+\zeta) & -2(\zeta-\bar{\zeta}) & -(\zeta-\bar{\zeta}) \end{pmatrix}\begin{pmatrix} \nu \\\\ m(\Gamma_1) \\\\ m(\Gamma_2) \\\\ [K_1] \\\\ [K_2] \end{pmatrix}
\label{eq:C3_Matrices_D_supp_III}
\ea
From Eq.~\eqref{eq:C3_Matrices_D_supp_III}, the same relations as Eq.~\eqref{eq:C3_Matrices_A_spin_half_supp_II} can be obtained, with the additional relations $[K'_1]=[K_1]+[K_2]$, $[K'_2]=-[K_2]$, and $\vev{c_3(\bx_{W})}_F=-\vev{(c_3(\bx_{W}))^{-1}}_F=-(\vev{c_3(\bx_{W})}_F)^*$ for $W\in\{1b,1c\}$.
Thus, the Chern number for symmetry class D for spin-1/2 electrons is equivalent to the Chern number for symmetry class A for spin-1/2 electrons given by Eq.~\eqref{eq:C3_Matrices_A_spin_half_supp_III} (note that because $\trm{Re}[\vev{c_3(\bx_{W})}_F]=0$ for $W\in\{1b,1c\}$, the expression for the Chern number can be further simplified).
%%%%%%

%%%%%%
\tocless{\subsubsection{$C_4$ symmetry}}{\label{subsec:c4_map}}
%%%%%%
For a $C_4$-symmetric system, the filling $\nu$ constrains the rotation invariants, the $m(\Gamma_p)$ irrep multiplicities, and the Wannier orbital irrep multiplicities to satisfy the following equation,
\ba
\nu=\sum_{p=1}^{4}m(\Gamma_p)=\sum_W \sum_\ell M_W n_{W}^{(\ell)}\to\sum_{p=1}^{2}[X_p]=\sum_{p=1}^{4}[M_p]=0
\label{eq:C4_filling_constraint_supp}
\ea
where $M_{1a}=M_{1b}=1$ and $M_{2c}=2$ are the corresponding Wannier orbital multiplicities at WPs $1a$, $1b$, and $2c$ respectively.
Using Eq.~\eqref{eq:fully_traced_TCM_4_supp} and Table~\ref{table:C4_bandrep_supp}, we can determine the $M_{\trm{TCM}\leftarrow\trm{MS}}^{(4)}$ and $M_{\trm{MS}\leftarrow\trm{RSI}}^{(4)}$ matrices for symmetry class A pertaining to spinless electrons,
\ba
&\begin{pmatrix} \nu \\\\ \vev{c_4(\bx_{1a})}_F \\\\ \vev{c_4(\bx_{1b})}_F \\\\ \vev{(c_4(\bx_{1a}))^{-1}}_F \\\\ \vev{(c_4(\bx_{1b}))^{-1}}_F \\\\ \vev{c_2(\bx_{1a})}_F \\\\ \vev{c_2(\bx_{1b})}_F \\\\ \vev{c_2(\bx_{2c})}_F \end{pmatrix}=\begin{pmatrix} 1 & 0 & 0 & 0 & 0 & 0 & 0 & 0 \\\\ -2i & 2(1+i) & 4i & -2(1-i) & 0 & 1+i & 2i & -(1-i) \\\\ 0 & 0 & 0 & 0 & 0 & -(1+i) & -2i & 1-i \\\\ 2i & 2(1-i) & -4i & -2(1+i) & 0 & 1-i & -2i & -(1+i) \\\\ 0 & 0 & 0 & 0 & 0 & -(1-i) & 2i & 1+i \\\\ -4 & 8 & 0 & 8 & 4 & 2 & 0 & 2 \\\\ 0 & 0 & 0 & 0 & -4 & 2 & 0 & 2 \\\\ 0 & 0 & 0 & 0 & 0 & -2 & 0 & -2 \end{pmatrix}\begin{pmatrix} \nu \\\\ m(\Gamma_1) \\\\ m(\Gamma_2) \\\\ m(\Gamma_3) \\\\ [X_1] \\\\ [M_1] \\\\ [M_2] \\\\ [M_3] \end{pmatrix},
\nnnn
&\begin{pmatrix} \nu \\\\ m(\Gamma_1) \\\\ m(\Gamma_2) \\\\ m(\Gamma_3) \\\\ [X_1] \\\\ [M_1] \\\\ [M_2] \\\\ [M_3] \end{pmatrix}=\begin{pmatrix} 1 & 0 & 0 & 0 & 0 & 0 & 0 & 0 \\\\ 1/4 & -1/4 & -1/4 & -1/4 & -1/4 & -1/4 & -1/4 & -1/2 \\\\ 1/4 & 3/4 & -1/4 & -1/4 & 3/4 & -1/4 & -1/4 & 1/2 \\\\ 1/4 & -1/4 & 3/4 & -1/4 & -1/4 & 3/4 & -1/4 & -1/2 \\\\ 0 & 0 & 0 & 0 & 1 & -1 & 1 & 1 \\\\ 0 & 0 & 0 & 0 & 0 & 1 & 0 & 1 \\\\ 0 & 0 & 0 & 0 & -1 & 0 & 1 & -1 \\\\ 0 & 0 & 0 & 0 & 0 & -1 & 0 & 1 \end{pmatrix}\begin{pmatrix} \nu \\\\ \delta_{1a}^{(1)} \\\\ \delta_{1a}^{(2)} \\\\ \delta_{1a}^{(3)} \\\\ \delta_{1b}^{(1)} \\\\ \delta_{1b}^{(2)} \\\\ \delta_{1b}^{(3)} \\\\ \delta_{2c} \end{pmatrix}.
\label{eq:C4_Matrices_A_supp_I}
\ea
Note that the entries of the rows of $M_{\trm{TCM}\leftarrow\trm{MS}}^{(4)}$ corresponding to $\vev{(c_4(\bx_{W}))^{-1}}_F$ are complex conjugates of the entries of the rows corresponding to $\vev{c_4(\bx_{W})}_F$, for $W\in\{1a,1b\}$ (e.g., $(\vev{c_4(\bx_{W})}_F)^*=\vev{(c_4(\bx_{W}))^{-1}}_F$).
From Eq.~\eqref{eq:C4_Matrices_A_supp_I} and computing $M_{\trm{MS}\leftarrow\trm{TCM}}^{(4)}=(M_{\trm{TCM}\leftarrow\trm{MS}}^{(4)})^{-1}$, the following relations can be determined,
\ba
&2m(\Gamma_1)-2m(\Gamma_3)+[M_1]-[M_3]=\trm{Re}[\vev{c_4(\bx_{1a})}_F],
\nn
&-2\nu+2m(\Gamma_1)+4m(\Gamma_2)+2m(\Gamma_3)+[M_1]+2[M_2]+[M_3]=\trm{Im}[\vev{c_4(\bx_{1a})}_F],
\nn
&-2\nu+4m(\Gamma_1)+4m(\Gamma_3)+2[X_1]+[M_1]+[M_3]=\frac{1}{2}\vev{c_2(\bx_{1a})}_F,
\nn
&-[M_1]+[M_3]=\trm{Re}[\vev{c_4(\bx_{1b})}_F],
\nn
&-[M_1]-2[M_2]-[M_3]=\trm{Im}[\vev{c_4(\bx_{1b})}_F],
\nn
&-2[X_1]+[M_1]+[M_3]=\frac{1}{2}\vev{c_2(\bx_{1b})}_F,
\label{eq:C4_Matrices_A_supp_II}
\ea
which implies the following \textit{multiple} possible relations for the Chern number for symmetry class A pertaining to spinless electrons,
\ba
\mc{C}
=& [M_1]+2[M_2]+3[M_3]+2[X_1]
= \pm{\sqrt{2}}\trm{Re}[e^{i\pi/4}\vev{c_4(\bx_{W})}_F]+\frac{1}{2}\vev{c_2(\bx_{W})}_F \pmod{4} \ \text{for} \ W \in \{1a,1b\},
\label{eq:C4_Matrices_A_supp_III}
\\
\mc{C}
=& [M_1]+2[M_2]+3[M_3]+2[X_1]
= \pm{\sqrt{2}} \trm{Re} [e^{-i\pi/4}\vev{c_4(\bx_{W})}_F]-\frac{1}{2}\vev{c_2(\bx_{W})}_F \pmod{4} \ \text{for} \ W \in \{1a,1b\}.
\label{eq:C4_Matrices_A_supp_IV}
\ea
The matrix $M_{\trm{TCM}\leftarrow\trm{RSI}}^{(4)}$ can also be computed and the following relations can be determined,
\ba
&i\delta_{W}^{(1)}-\delta_{W}^{(2)}-i\delta_{W}^{(3)}=\frac{1}{2}\vev{c_4(\bx_W)}_F \ \text{for} \ W \in \{1a,1b\},
\nn
&\delta_{W}^{(1)}-\delta_{W}^{(2)}+\delta_{W}^{(3)}=-\frac{1}{4}\vev{c_2(\bx_W)}_F \ \text{for} \ W \in \{1a,1b\},
\nn
&\delta_{2c}=-\frac{1}{4}\vev{c_2(\bx_{2c})}_F.
\label{eq:C4_Matrices_A_supp_V}
\ea
Note that for atomic insulators, the Chern number must be zero, which enforces an additional constraint between the rotation invariants, $[M_3]=[M_1]+2[M_2]+2[X_1]\pmod{4}$.
In terms of RSIs and TCMs, the bulk polarization for symmetry class A pertaining to spinless electrons can be expressed in three possible ways,
\ba
\bb P
=& \frac{e}{2}(\delta_{1b}^{(1)}+\delta_{1b}^{(2)} + \delta_{1b}^{(3)}+\delta_{2c})(\bb a_1 + \bb a_2)
\nn
=& \frac{e}{8}(s_{1}\vev{c_2(\bx_{1b})}_F + s_{2}\vev{c_2(\bx_{1c})}_F)(\bb a_1 + \bb a_2) \pmod{e} \ \text{for} \ s_{1,2} \in \{-1,1\}
\label{eq:C4_Matrices_A_supp_VI}
\\
\bb P
=& \frac{e}{8}(s_{1} \vev{c_2(\bx_{2c})}_F + 2s_{2} \trm{Re} [\vev{c_4(\bx_{1b})}_F] + 2s_{3} \trm{Im} [\vev{c_4(\bx_{1b})}_F]) (\bb a_1 + \bb a_2) \pmod{e} \ \text{for} \ s_{1,2,3} \in \{-1,1\}
\label{eq:C4_Matrices_A_supp_VII}
\\
\bb P
=& \frac{e}{8}(4s_1 \trm{Re}[\vev{c_4(\bx_{1b})}_F] + s_2\vev{c_2(\bx_{1b})}_F + s_3\vev{c_2(\bx_{2c})}_F) (\bb a_1 + \bb a_2) \pmod{e} \ \text{for} \ s_{1,2,3} \in \{-1,+1\}.
\label{eq:C4_Matrices_A_supp_VIII}
\ea
The sector charge for symmetry class A pertaining to spinless electrons can also be expressed in in terms of the TCMs as,
\ba
Q_{\trm{sector}}=\frac{e}{4}(\delta_{1b}^{(1)}+\delta_{1b}^{(2)}+\delta_{1b}^{(3)})=\frac{e}{16}(-4\trm{Re}[\vev{c_{4}(\bx_{1b})}_F]-\vev{c_2(\bx_{1b})}_F) \pmod{e}.
\label{eq:C4_Matrices_A_supp_IX}
\ea
It should be noted that the sector charge expression is \textit{not unique} owing to the mod $e$ constraint; as such, there are multiple other possible expressions for the sector charge which are all equivalent.

For symmetry class AI systems, time-reversal symmetry introduces the constraints $[M_3]=-([M_1]+2[M_2])$ and $m(\Gamma_2)=m(\Gamma_4)$, implying $\nu=m(\Gamma_1)+2m(\Gamma_2)+m(\Gamma_3)$.
Additionally, time-reversal symmetry imposes the constraint $\delta_{W}^{(1)}=\delta_{W}^{(3)}$ and $\vev{c_4(\bx_{W})}_F=\vev{(c_4(\bx_{W}))^{-1}}_F=(\vev{c_4(\bx_{W})}_F)^*$ for all $W\in\{1a,1b\}$. 
Therefore, to evaluate the matrices for symmetry class AI, we must incorporate the additional constraints imposed by the symmetries and modify $\bb v_{\trm{TCM}}^{(4)}$, $\bb v_{\trm{MS}}^{(4)}$, and $\bb v_{\trm{RSI}}^{(4)}$ to include only the independent set of quantities for each,
\ba
&\bb v_{\trm{TCM}}^{(4)} = \begin{pmatrix} \nu, & \vev{c_4(\bx_{1a})}_F, & \vev{c_4(\bx_{1b})}_F, & \vev{c_2(\bx_{1a})}_F, & \vev{c_2(\bx_{1b})}_F, & \vev{c_2(\bx_{2c})}_F \end{pmatrix}^{T},
\nn
&\bb v_{\trm{MS}}^{(4)} = \begin{pmatrix} \nu, & m(\Gamma_1), & m(\Gamma_2), & [X_1], & [M_1], & [M_2] \end{pmatrix}^{T},
\nn
&\bb v_{\trm{RSI}}^{(4)} = \begin{pmatrix} \nu, & \delta_{1a}^{(1)}, & \delta_{1a}^{(2)}, & \delta_{1b}^{(1)}, & \delta_{1b}^{(2)}, & \delta_{2c} \end{pmatrix}^{T}.
\label{eq:C4_Matrices_AI_supp_I}
\ea 
Thus, $M_{\trm{TCM}\leftarrow\trm{MS}}^{(4)}$ and $M_{\trm{MS}\leftarrow\trm{RSI}}^{(4)}$ for symmetry class AI are given as follows,
\ba
&\begin{pmatrix} \nu \\\\ \vev{c_4(\bx_{1a})}_F \\\\ \vev{c_4(\bx_{1b})}_F \\\\ \vev{c_2(\bx_{1a})}_F \\\\ \vev{c_2(\bx_{1b})}_F \\\\ \vev{c_2(\bx_{2c})}_F \end{pmatrix} = \begin{pmatrix} 1 & 0 & 0 & 0 & 0 & 0 \\\\ -2 & 4 & 4 & 0 & 2 & 2 \\\\ 0 & 0 & 0 & 0 & -2 & -2 \\\\ 4 & 0 & -16 & 4 & 0 & -4 \\\\ 0 & 0 & 0 & -4 & 0 & -4 \\\\ 0 & 0 & 0 & 0 & 0 & 4 \end{pmatrix}\begin{pmatrix} \nu \\\\ m(\Gamma_1) \\\\ m(\Gamma_2) \\\\ [X_1] \\\\ [M_1] \\\\ [M_2] \end{pmatrix},
\nnnn
&\begin{pmatrix} \nu \\\\ m(\Gamma_1) \\\\ m(\Gamma_2) \\\\ [X_1] \\\\ [M_1] \\\\ [M_2] \end{pmatrix} = \begin{pmatrix} 1 & 0 & 0 & 0 & 0 & 0 \\\\ 1/4 & -1/2 & -1/4 & -1/2 & -1/4 & -1/2 \\\\ 1/4 & 1/2 & -1/4 & 1/2 & -1/4 & 1/2 \\\\ 0 & 0 & 0 & 2 & -1 & 1 \\\\ 0 & 0 & 0 & 0 & 1 & 1 \\\\ 0 & 0 & 0 & 0 & 0 & -1 \end{pmatrix}\begin{pmatrix} \nu \\\\ \delta_{1a}^{(1)} \\\\ \delta_{1a}^{(2)} \\\\ \delta_{1b}^{(1)} \\\\ \delta_{1b}^{(2)} \\\\ \delta_{2c} \end{pmatrix}.
\label{eq:C4_Matrices_AI_supp_II}
\ea
Computing $M_{\trm{TCM}\leftarrow\trm{RSI}}^{(4)}$ yields the following relations,
\ba
&\delta_{W}^{(2)}=-\frac{1}{2}\vev{c_4(\bx_W)}_F \ \text{for} \  W \in \{1a,1b\},
\nn
&2\delta_{W}^{(1)}-\delta_{W}^{(2)}=-\frac{1}{4}\vev{c_2(\bx_W)}_F \ \text{for} \ W \in \{1a,1b\},
\nn
&\delta_{2c}=-\frac{1}{4}\vev{c_2(\bx_{2c})}_F,
\label{eq:C4_Matrices_AI_supp_III}
\ea
which means the bulk polarization is the same as Eq.~\eqref{eq:C4_Matrices_A_supp_VI} or can be expressed as,
\ba
\bb P = \frac{e}{2}(2\delta_{1b}^{(1)} + \delta_{1b}^{(2)} + \delta_{2c})(\bb a_1 + \bb a_2) = \frac{e}{8}(s_{1}\vev{c_2(\bx_{2c})}_F + 2s_{2}\vev{c_4(\bx_{1b})}_F) \pmod{e} \ \text{for} \ s_{1,2} \in \{-1,1\},
\label{eq:C4_Matrices_AI_supp_IV}
\ea
for symmetry class AI.
The sector charge for symmetry class AI can be expressed as,
\ba
Q_{\trm{sector}}=\frac{e}{4}(2\delta_{1b}^{(1)}+\delta_{1b}^{(2)})=-\frac{e}{16}(4\vev{c_4(\bx_{1b})}_F+\vev{c_2(\bx_{1b})}_F) \pmod{e}.
\label{eq:C4_Matrices_AI_supp_V}
\ea

Returning to the original definitions of $\bb v_{\trm{TCM}}^{(4)}$, $\bb v_{\trm{MS}}^{(4)}$, and $\bb v_{\trm{RSI}}^{(4)}$ given by Eqs.~\eqref{eq:TCM_C3_C4_C6_vector_supp}, \eqref{eq:MS_C2_C3_C4_C6_vector_supp}, and \eqref{eq:RSI_C2_C3_C4_C6_vector_supp}, one can construct the matrix $M_{\trm{TCM}\leftarrow\trm{MS}}^{(4)}$ once again using Eq.~\eqref{eq:fully_traced_TCM_4_supp} and Table~\ref{table:C4_bandrep_supp} for symmetry class A systems pertaining to spin-1/2 electrons,
\ba
\begin{pmatrix} \nu \\\\ \vev{c_4(\bx_{1a})}_F \\\\ \vev{c_4(\bx_{1b})}_F \\\\ \vev{(c_4(\bx_{1a}))^{-1}}_F \\\\ \vev{(c_4(\bx_{1b}))^{-1}}_F \\\\ \vev{c_2(\bx_{1a})}_F \\\\ \vev{c_2(\bx_{1b})}_F \\\\ \vev{c_2(\bx_{2c})}_F \end{pmatrix} = \begin{pmatrix} 1 & 0 & 0 & 0 & 0 & 0 & 0 & 0 \\\\ 2\bar{\eta} & 2(\eta-\bar{\eta}) & - 4\bar{\eta} & -2(\eta+\bar{\eta}) & 0 & \eta-\bar{\eta} & -2\bar{\eta} & -(\eta+\bar{\eta}) \\\\ 0 & 0 & 0 & 0 & 0 & -(\eta-\bar{\eta}) & 2\bar{\eta} & \eta+\bar{\eta} \\\\ 2\eta & -2(\eta-\bar{\eta}) & -4\eta & -2(\eta+\bar{\eta}) & 0 & -(\eta-\bar{\eta}) & -2\eta & -(\eta+\bar{\eta}) \\\\ 0 & 0 & 0 & 0 & 0 & \eta-\bar{\eta} & 2\eta & \eta+\bar{\eta} \\\\ -4i & 8i & 0 & 8i & 4i & 2i & 0 & 2i \\\\ 0 & 0 & 0 & 0 & -4i & 2i & 0 & 2i \\\\ 0 & 0 & 0 & 0 & 0 & -2i & 0 & -2i \end{pmatrix}\begin{pmatrix} \nu \\\\ m(\Gamma_1) \\\\ m(\Gamma_2) \\\\ m(\Gamma_3) \\\\ [X_1] \\\\ [M_1] \\\\ [M_2] \\\\ [M_3] \end{pmatrix},
\label{eq:C4_Matrices_A_spin_half_supp_I}
\ea
where $\eta=e^{\pi i/4}$ and $\bar{\eta}=e^{-\pi i/4}$, and $M_{\trm{MS}\leftarrow\trm{RSI}}^{(4)}$ is identical to Eq.~\eqref{eq:C4_Matrices_A_supp_I}. 
From Eq.~\eqref{eq:C4_Matrices_A_spin_half_supp_I}, the following relations can be determined,
\ba
&\nu-2m(\Gamma_2)-2m(\Gamma_3)-([M_2]+[M_3])=\frac{1}{\sqrt{2}}\trm{Re}[\vev{c_4(\bx_{1a})}_F],
\nn
&-\nu+2m(\Gamma_1)+2m(\Gamma_2)+[M_1]+[M_2]=\frac{1}{\sqrt{2}}\trm{Im}[\vev{c_4(\bx_{1a})}_F],
\nn
&-2\nu+4m(\Gamma_1)+4m(\Gamma_3)+2[X_1]+[M_1]+[M_3]=\frac{1}{2i}\vev{c_2(\bx_{1a})}_F,
\nn
&[M_2]+[M_3]=\frac{1}{\sqrt{2}}\trm{Re}[\vev{c_4(\bx_{1b})}_F],
\nn
&-([M_1]+[M_2])=\frac{1}{\sqrt{2}}\trm{Im}[\vev{c_4(\bx_{1b})}_F],
\nn
&-2[X_1]+[M_1]+[M_3]=\frac{1}{2i}\vev{c_2(\bx_{1b})}_F,
\nn
&-([M_1]+[M_3])=\frac{1}{2i}\vev{c_2(\bx_{2c})}_F
\label{eq:C4_Matrices_A_spin_half_supp_II}
\ea
which implies the following \textit{multiple} possible relations for the Chern number for symmetry class A pertaining to spin-1/2 electrons,
\ba
\mc{C}
=& [M_1]+2[M_2]+3[M_3]+2[X_1]
= \pm{\sqrt{2}}\trm{Re}[\vev{c_4(\bx_{W})}_F]-\frac{i}{2}\vev{c_2(\bx_{W})}_F\pmod{4} \ \text{for} \ W \in \{1a,1b\},
\label{eq:C4_Matrices_A_spin_half_supp_III}
\\
\mc{C}
=& [M_1]+2[M_2]+3[M_3]+2[X_1]
= \pm{\sqrt{2}}\trm{Im}[\vev{c_4(\bx_{W})}_F] + \frac{i}{2} \vev{c_2(\bx_{W})}_F \pmod{4} \ \text{for} \ W \in \{1a,1b\}.
\label{eq:C4_Matrices_A_spin_half_supp_IV}
\ea
Computing the matrix $M_{\trm{TCM}\leftarrow\trm{RSI}}^{(4)}$ yields the following relations,
\ba
&-\bar{\eta}\delta_{W}^{(1)}-\eta\delta_{W}^{(2)}+\bar{\eta}\delta_{W}^{(3)}=\frac{1}{2}\vev{c_4(\bx_{W})}_F \ \text{for} \ W \in \{1a,1b\},
\nn
&i(\delta_{W}^{(1)}-\delta_{W}^{(2)}+\delta_{W}^{(3)})=-\frac{1}{4}\vev{c_2(\bx_W)}_F \ \text{for} \ W \in \{1a,1b\},
\nn
&i\delta_{2c}=-\frac{1}{4}\vev{c_2(\bx_{2c})}_F
\label{eq:C4_Matrices_A_spin_half_supp_V}
\ea
which means the bulk polarization can be expressed in four possible ways,
\ba
\bb P
=& \frac{e}{2}(\delta_{1b}^{(1)}+\delta_{1b}^{(2)}+\delta_{1b}^{(3)}+\delta_{2c})(\bb a_1 + \bb a_2)
\nn
&= \frac{ie}{8}(s_{1}\vev{c_2(\bx_{1b})}_F + s_{2}\vev{c_2(\bx_{2c})}_F)(\bb a_1 + \bb a_2)\pmod{e} \ \text{for} \ s_{1,2} \in \{-1,1\},
\label{eq:C4_Matrices_A_spin_half_supp_VI}
\\
\bb P
=& \frac{e}{8}(is_{1}\vev{c_2(\bx_{2c})}_F + 2\sqrt{2}s_{2}\trm{Re}[\vev{c_4(\bx_{1b})}_F])(\bb a_1 + \bb a_2)\pmod{e} \ \text{for} \ s_{1,2} \in \{-1,1\},
\label{eq:C4_Matrices_A_spin_half_supp_VII}
\\
\bb P
=& \frac{e}{8}(is_{1}\vev{c_2(\bx_{2c})}_F + 2\sqrt{2}s_{2}\trm{Im}[\vev{c_4(\bx_{1b})}_F])(\bb a_1 + \bb a_2)\pmod{e} \ \text{for} \ s_{1,2} \in \{-1,1\},
\label{eq:C4_Matrices_A_spin_half_supp_VIII}
\\
\bb P
=& \frac{e}{8}(4s_1\trm{Re}[e^{-\frac{i\pi}{4}}\vev{c_4(\bx_{1b})}_F]+i(s_2\vev{c_2(\bx_{1b})}_F+s_3\vev{c_2(\bx_{2c})}_F))(\bb a_1 + \bb a_2)\pmod{e} \ \text{for} \ s_{1,2,3} \in \{-1,+1\},
\label{eq:C4_Matrices_A_spin_half_supp_IX}
\ea
for symmetry class A pertaining to spin-1/2 electrons.
The sector charge for symmetry class A pertaining to spin-1/2 electrons can be expressed as,
\ba
Q_{\trm{sector}}=\frac{e}{4}(\delta_{1b}^{(1)}+\delta_{1b}^{(2)}+\delta_{1b}^{(3)})=\frac{e}{16}(-4\trm{Re}[e^{-\frac{i\pi}{4}}\vev{c_4(\bx_{1b})}_F]+i\vev{c_2(\bx_{1b})}_F) \pmod{e}.
\label{eq:C4_Matrices_A_spin_half_supp_X}
\ea
Once again, we note that the above expression for the sector charge is \textit{not unique} owing to the$\pmod{e}$ constraint. 
Thus, there are several other expressions for the sector charge that are equivalent.

For symmetry class AII, time-reversal symmetry introduces the constraints $[M_1]=-[M_2]=-[M_3]=[M_4]$, $[X_1]=[X_2]=0$, and $\nu=2(m(\Gamma_1)+m(\Gamma_2))$.
Furthermore, one also has $\delta_{W}^{(1)}=\delta_{W}^{(2)}\equiv\delta_{W}$, $\delta_{W}^{(3)}=0$, and $\delta_{2c}=0$. 
Additionally, time-reversal symmetry also enforces the constraints $\vev{c_4(\bx_W)}_F=\vev{(c_4(\bx_W))^{-1}}_F=(\vev{c_4(\bx_W)}_F)^*$ for $W\in\{1a,1b\}$ and $\vev{c_2(\bx_{W})}_F=0$ for $W\in\{1a,1b,2c\}$.
To evaluate the matrices for the symmetry class AII, we must incorporate the additional constraints imposed by the symmetries and modify $\bb v_{\trm{TCM}}^{(4)}$, $\bb v_{\trm{MS}}^{(4)}$, and $\bb v_{\trm{RSI}}^{(4)}$ to include only the independent sets of quantities for each,
\ba
&\bb v_{\trm{TCM}}^{(4)} = \begin{pmatrix} \nu, & \vev{c_4(\bx_{1a})}_F, & \vev{c_4(\bx_{1b})}_F \end{pmatrix}^{T}, 
\nn
&\bb v_{\trm{MS}}^{(4)} = \begin{pmatrix} \nu, & m(\Gamma_1), & [M_1] \end{pmatrix}^{T},
\nn
&\bb v_{\trm{RSI}}^{(4)} = \begin{pmatrix} \nu, & \delta_{1a}, & \delta_{1b} \end{pmatrix}^{T}.
\label{eq:C4_Matrices_AII_supp_I}
\ea
Thus, $M_{\trm{TCM}\leftarrow\trm{MS}}^{(4)}$ and $M_{\trm{MS}\leftarrow\trm{RSI}}^{(4)}$ for symmetry class AII are given as follows,
\ba
&\begin{pmatrix} \nu \\\\ \vev{c_4(\bx_{1a})}_F \\\\ \vev{c_4(\bx_{1b})}_F \end{pmatrix}=\begin{pmatrix} 1 & 0 & 0 \\\\ (\eta+\bar{\eta}) & -4(\eta+\bar{\eta}) & 2(\eta+\bar{\eta}) \\\\ 0 & 0 & -2(\eta+\bar{\eta}) \end{pmatrix}\begin{pmatrix} \nu \\\\ m(\Gamma_1) \\\\ [M_1] \end{pmatrix},
\nnnn
&\begin{pmatrix} \nu \\\\ m(\Gamma_1) \\\\ [M_1] \end{pmatrix}=\begin{pmatrix} 1 & 0 & 0 \\\\ 1/4 & -1/2 & -1/2 \\\\ 0 & 0 & 1 \end{pmatrix}\begin{pmatrix} \nu \\\\ \delta_{1a} \\\\ \delta_{1b} \end{pmatrix}.
\label{eq:C4_Matrices_AII_supp_II}
\ea
Computing $M_{\trm{RSI}\leftarrow\trm{MS}}^{(4)}=(M_{\trm{MS}\leftarrow\trm{RSI}}^{(4)})^{-1}$ yields the following relation,
\ba
\delta_{W}=-\frac{1}{2\sqrt{2}}\vev{c_4(\bx_{W})}_F \ \text{for} \ W \in \{1a,1b\},
\label{eq:C4_Matrices_AII_supp_III}
\ea
which means the bulk polarization for symmetry class AII is simply $\bb P = \bb 0 \pmod{e}$.
Meanwhile, the sector charge for symmetry class AII can be expressed as,
\ba
Q_{\trm{sector}}=\frac{e}{2}\delta_{1b}=-\frac{e}{4\sqrt{2}}\vev{c_4(\bx_{1b})}_F \pmod{e}.
\label{eq:C4_Matrices_AII_supp_IV}
\ea

Finally, for symmetry class D systems, we consider both spinless and spin-1/2 electron systems.
For spinless systems, particle-hole symmetry introduces the constraint on the rotation invariants $[M_2]=-[M_4]$ which implies $[M_1]=-[M_3]$.
Furthermore, using Eq.~\eqref{eq:fully_traced_TCM_4_supp}, it follows that $\vev{c(\bx_{2c})}_F=0$.
Therefore, similar to the approaches taken for symmetry classes AI and AII, for symmetry class D pertaining to spinless electrons, we modify $\bb v_{\trm{TCM}}^{(4)}$ and $\bb v_{\trm{MS}}^{(4)}$ (we do not consider $\bb v_{\trm{RSI}}^{(4)}$ since atomic insulators do not exist in symmetry class D) to include only the independent set of quantities for each,
\ba
&\bb v_{\trm{TCM}}^{(4)}= \begin{pmatrix} \nu, & \vev{c_4(\bx_{1a})}_F, & \vev{c_4(\bx_{1b})}_F, & \vev{(c_4(\bx_{1a}))^{-1}}_F, & \vev{(c_4(\bx_{1b}))^{-1}}_F, & \vev{c_2(\bx_{1a})}_F, & \vev{c_2(\bx_{1b})}_F \end{pmatrix}^{T},
\nn
&\bb v_{\trm{MS}}^{(4)}= \begin{pmatrix} \nu, & m(\Gamma_1), & m(\Gamma_2), & m(\Gamma_3), & [X_1], & [M_1], & [M_2] \end{pmatrix}^{T}.
\label{eq:C4_Matrices_D_supp_I}
\ea
Thus, $M_{\trm{TCM}\leftarrow\trm{MS}}^{(4)}$ for symmetry class D for spinless electrons is given as follows,
\ba
\begin{pmatrix} \nu \\\\ \vev{c_4(\bx_{1a})}_F \\\\ \vev{c_4(\bx_{1b})}_F \\\\ \vev{(c_4(\bx_{1a}))^{-1}}_F \\\\ \vev{(c_4(\bx_{1b}))^{-1}}_F \\\\ \vev{c_2(\bx_{1a})}_F \\\\ \vev{c_2(\bx_{1b})}_F \end{pmatrix}=\begin{pmatrix} 1 & 0 & 0 & 0 & 0 & 0 & 0 \\\\ -2i & 2(1+i) & 4i & -2(1-i) & 0 & 2 & 2i \\\\ 0 & 0 & 0 & 0 & 0 & -2 & -2i \\\\ 2i & 2(1-i) & -4i & -2(1+i) & 0 & 2 & -2i \\\\ 0 & 0 & 0 & 0 & 0 & -2 & 2i \\\\ -4 & 8 & 0 & 8 & 4 & 0 & 0 \\\\ 0 & 0 & 0 & 0 & -4 & 0 & 0 \end{pmatrix}\begin{pmatrix} \nu \\\\ m(\Gamma_1) \\\\ m(\Gamma_2) \\\\ m(\Gamma_3) \\\\ [X_1] \\\\ [M_1] \\\\ [M_2] \end{pmatrix}
\label{eq:C4_Matrices_D_supp_II}
\ea
From Eq.~\eqref{eq:C4_Matrices_D_supp_II}, the same relations can be obtained as Eq.~\eqref{eq:C4_Matrices_A_supp_II}, but with the additional relations $[M_2]=-[M_4]$, $[M_1]=-[M_3]$, and $\vev{c_2(\bx_{2c})}_F=0$.
Therefore, this yields the same relations for the Chern number derived for symmetry class A for spinless electrons given by Eqs.~\eqref{eq:C4_Matrices_A_supp_III}-\eqref{eq:C4_Matrices_A_supp_IV}.
Similarly, for spin-1/2 systems, particle-hole symmetry introduces the constraints $[M_1]=-[M_4]$, $[M_2]=-[M_3]$.
Additionally, particle-hole symmetry imposes the constraint $\vev{c_4(\bx_{1b})}_F = -\vev{(c_4(\bx_{1b}))^{-1}}_F = -(\vev{c_4(\bx_{1b})}_F)^*$.
Hence, for symmetry class D pertaining to spin-1/2 electrons, we once again modify $\bb v_{\trm{TCM}}^{(4)}$ and $\bb v_{\trm{MS}}^{(4)}$ to include only the independent set of quantities for each,
\ba
&\bb v_{\trm{TCM}}^{(4)}= \begin{pmatrix} \nu, & \vev{c_4(\bx_{1a})}_F, & \vev{c_4(\bx_{1b})}_F, & \vev{(c_4(\bx_{1a}))^{-1}}_F, & \vev{c_2(\bx_{1a})}_F, & \vev{c_2(\bx_{1b})}_F, & \vev{c_2(\bx_{2c})}_F \end{pmatrix}^{T},
\nn
&\bb v_{\trm{MS}}^{(4)}= \begin{pmatrix} \nu, & m(\Gamma_1), & m(\Gamma_2), & m(\Gamma_3), & [X_1], & [M_1], & [M_2] \end{pmatrix}^{T}.
\label{eq:C4_Matrices_D_supp_III}
\ea
Thus, $M_{\trm{TCM}\leftarrow\trm{MS}}^{(4)}$ for symmetry class D for spin-1/2 electrons is given as follows,
\ba
\begin{pmatrix} \nu \\\\ \vev{c_4(\bx_{1a})}_F \\\\ \vev{c_4(\bx_{1b})}_F \\\\ \vev{(c_4(\bx_{1a}))^{-1}}_F \\\\ \vev{c_2(\bx_{1a})}_F \\\\ \vev{c_2(\bx_{1b})}_F \\\\ \vev{c_2(\bx_{2c})}_F \end{pmatrix}=\begin{pmatrix} 1 & 0 & 0 & 0 & 0 & 0 & 0 \\\\ 2\bar{\eta} & 2(\eta-\bar{\eta}) & -4\bar{\eta} & -2(\eta+\bar{\eta}) & 0 & \eta-\bar{\eta} & \eta-\bar{\eta} \\\\ 0 & 0 & 0 & 0 & 0 & -(\eta-\bar{\eta}) & -(\eta-\bar{\eta}) \\\\ 2\eta & -2(\eta-\bar{\eta}) & -4\eta & -2(\eta+\bar{\eta}) & 0 & -(\eta-\bar{\eta}) & -(\eta-\bar{\eta}) \\\\ -4i & 8i & 0 & 8i & 4i & 2i & -2i \\\\ 0 & 0 & 0 & 0 & -4i & 2i & -2i \\\\ 0 & 0 & 0 & 0 & 0 & -2i & 2i \end{pmatrix}\begin{pmatrix} \nu \\\\ m(\Gamma_1) \\\\ m(\Gamma_2) \\\\ m(\Gamma_3) \\\\ [X_1] \\\\ [M_1] \\\\ [M_2] \end{pmatrix}
\label{eq:C4_Matrices_D_supp_IV}
\ea
From Eq.~\eqref{eq:C4_Matrices_D_supp_IV}, the same relations can be obtained as Eq.~\eqref{eq:C4_Matrices_A_spin_half_supp_II}, but with the additional relations $[M_2]=-[M_3]$, $[M_1]=-[M_4]$, and $\vev{c_4(\bx_{1b})}_F=-\vev{(c_4(\bx_{1b}))^{-1}}_F=-(\vev{c_4(\bx_{1b})}_F)^*$.
This yields the same relations for the Chern number derived for symmetry class A for spin-1/2 electrons given by Eqs.~\eqref{eq:C4_Matrices_A_spin_half_supp_III}-\eqref{eq:C4_Matrices_A_spin_half_supp_IV} (note that $\trm{Re}[\vev{c_4(\bx_{1b})}_F]=0$ so the Chern number can also be further simplified to $\mc{C}=-\frac{i}{2}\vev{c_2(\bx_{W})}_F \pmod{4}$ for $W\in\{1a,1b\}$).
%%%%%%

%%%%%%
\tocless{\subsubsection{$C_6$ symmetry}}{\label{subsec:c6_map}}
%%%%%%
For a $C_6$-symmetric system, the filling $\nu$ constrains the rotation invariants, the $m(\Gamma_p)$ irrep multiplicities, and the Wannier orbital irrep multiplicities to satisfy the following equation,
\ba
\nu=\sum_{p=1}^{6}m(\Gamma_p)=\sum_{W}\sum_{\ell} M_{W}n_{W}^{(\ell)}\to\sum_{p=1}^{3}[K_p]=\sum_{p=1}^{2}[M_p]=0
\label{eq:C6_filling_constraint_supp}
\ea
where $M_{1a}=1$, $M_{2b}=2$, and $M_{3c}=3$ are the corresponding Wannier orbital multiplicities at WPs $1a$, $2b$, and $3c$ respectively.
Using Eq.~\eqref{eq:fully_traced_TCM_4_supp} and Table~\ref{table:C6_bandrep_supp_1}, we can determine the $M_{\trm{TCM}\leftarrow\trm{MS}}^{(6)}$ and $M_{\trm{MS}\leftarrow\trm{RSI}}^{(6)}$ matrices for symmetry class A pertaining to spinless electrons,
\ba
&\begin{pmatrix} \nu \\\\ \vev{c_6(\bx_{1a})}_F \\\\ \vev{(c_6(\bx_{1a}))^{-1}}_F \\\\ \vev{c_3(\bx_{1a})}_F \\\\ \vev{c_3(\bx_{2b})}_F \\\\ \vev{(c_3(\bx_{1a}))^{-1}}_F \\\\ \vev{(c_3(\bx_{2b}))^{-1}}_F \\\\ \vev{c_2(\bx_{1a})}_F \\\\ \vev{c_2(\bx_{3c})}_F \end{pmatrix} = \begin{pmatrix} 1 & 0 & 0 & 0 & 0 & 0 & 0 & 0 & 0 \\\\ \bar{\zeta} & 1-\bar{\zeta} & \zeta-\bar{\zeta} & \omega-\bar{\zeta} & -(1+\bar{\zeta}) & \bar{\omega}-\bar{\zeta} & 0 & 0 & 0 \\\\ \zeta & 1-\zeta & -(\zeta-\bar{\zeta}) & \bar{\omega}-\zeta & -(1+\zeta) & \omega-\zeta & 0 & 0 & 0 \\\\ 3\bar{\omega} & 3(1-\bar{\omega}) & 3(\omega-\bar{\omega}) & 0 & 3(1-\bar{\omega}) & 3(\omega-\bar{\omega}) & 2(1-\bar{\omega}) & 2(\omega-\bar{\omega}) & 0 \\\\ 0 & 0 & 0 & 0 & 0 & 0 & -(1-\bar{\omega}) & -(\omega-\bar{\omega}) & 0 \\\\ 3\omega & -3(\omega-1) & -3(\omega-\bar{\omega}) & 0 & -3(\omega-1) & -3(\omega-\bar{\omega}) & -2(\omega-1) & -2(\omega-\bar{\omega}) & 0 \\\\ 0 & 0 & 0 & 0 & 0 & 0 & \omega-1 & \omega-\bar{\omega} & 0 \\\\ -4 & 8 & 0 & 8 & 0 & 8 & 0 & 0 & 6 \\\\ 0 & 0 & 0 & 0 & 0 & 0 & 0 & 0 & -2 \end{pmatrix}\begin{pmatrix} \nu \\\\ m(\Gamma_1) \\\\ m(\Gamma_2) \\\\ m(\Gamma_3) \\\\ m(\Gamma_4) \\\\ m(\Gamma_5) \\\\ [K_1] \\\\ [K_2] \\\\ [M_1] \end{pmatrix},
\nnnn
&\begin{pmatrix} \nu \\\\ m(\Gamma_1) \\\\ m(\Gamma_2) \\\\ m(\Gamma_3) \\\\ m(\Gamma_4) \\\\ m(\Gamma_5) \\\\ [K_1] \\\\ [K_2] \\\\ [M_1] \end{pmatrix}=\begin{pmatrix} 1 & 0 & 0 & 0 & 0 & 0 & 0 & 0 & 0 \\\\ 1/6 & -1/6 & -1/6 & -1/6 & -1/6 & -1/6 & -1/3 & -1/3 & -1/2 \\\\ 1/6 & 5/6 & -1/6 & -1/6 & -1/6 & -1/6 & 2/3 & -1/3 & 1/2 \\\\ 1/6 & -1/6 & 5/6 & -1/6 & -1/6 & -1/6 & -1/3 & 2/3 & -1/2 \\\\ 1/6 & -1/6 & -1/6 & 5/6 & -1/6 & -1/6 & -1/3 & -1/3 & 1/2 \\\\ 1/6 & -1/6 & -1/6 & -1/6 & 5/6 & -1/6 & 2/3 & -1/3 & -1/2 \\\\ 0 & 0 & 0 & 0 & 0 & 0 & 1 & 1 & 0 \\\\ 0 & 0 & 0 & 0 & 0 & 0 & -2 & 1 & 0 \\\\ 0 & 0 & 0 & 0 & 0 & 0 & 0 & 0 & 2 \end{pmatrix}\begin{pmatrix} \nu \\\\ \delta_{1a}^{(1)} \\\\ \delta_{1a}^{(2)} \\\\ \delta_{1a}^{(3)} \\\\ \delta_{1a}^{(4)} \\\\ \delta_{1a}^{(5)} \\\\ \delta_{2b}^{(1)} \\\\ \delta_{2b}^{(2)} \\\\ \delta_{3c}\end{pmatrix}.
\label{eq:C6_Matrices_A_supp_I}
\ea
where $\zeta=e^{\pi i/3}$, $\omega=e^{2\pi i/3}$, $\bar{\zeta}=e^{-\pi i/3}$, and $\bar{\omega}=e^{-2\pi i/3}$. 
Note that the entries of the rows of $M_{\trm{TCM}\leftarrow\trm{MS}}^{(6)}$ corresponding to $\vev{(c_6(\bx_{1a}))^{-1}}_F$ are complex conjugates of the entries of the rows corresponding to $\vev{c_6(\bx_{1a})}_F$, and the same holds true for $\vev{(c_3(\bx_{1a}))^{-1}}_F$ and $\vev{(c_3(\bx_{2b})^{-1}}_F$ with respect to $\vev{c_3(\bx_{1a})}_F$ and $\vev{c_3(\bx_{2b})}_F$ respectively (e.g., $(\vev{c_6(\bx_{1a})}_F)^*=\vev{(c_6(\bx_{1a}))^{-1}}_F$ and $(\vev{c_3(\bx_{W})})^*=\vev{(c_3(\bx_{W}))^{-1}}_F$ for $W\in\{1a,2b\}$).
From Eq.~\eqref{eq:C6_Matrices_A_supp_I} and computing $M_{\trm{MS}\leftarrow\trm{TCM}}^{(6)}=(M_{\trm{TCM}\leftarrow\trm{MS}}^{(6)})^{-1}$, the following relations can be determined,
\bg
6(m(\Gamma_1)-2m(\Gamma_2)+2m(\Gamma_3)-m(\Gamma_4))-4[K_1]-8[K_2]+9[M_1]
\nn
=\frac{2i}{\sqrt{3}}(\vev{c_3(\bx_{1a})}_F-\vev{(c_3(\bx_{1a}))^{-1}}_F)+\frac{3}{2}\vev{c_2(\bx_{1a})}_F
\\
2[K_1]+4[K_2]-3[M_1]=\frac{2i}{\sqrt{3}}(\vev{c_3(\bx_{2b})}_F-\vev{(c_3(\bx_{2b}))^{-1}}_F)+\frac{3}{2}\vev{c_2(\bx_{3c})}_F
\label{eq:C6_Matrices_A_supp_II}
\eg
which leads to the following \textit{multiple} relations for the Chern number for spinless electron systems in symmetry class A,
\ba
\mc{C}
=& 2[K_1]-2[K_2]+3[M_1]
= -\frac{4}{\sqrt{3}} \trm{Im}[\vev{c_3(\bx_{1a})}_F] + \frac{3}{2} \vev{c_2(\bx_{1a})}_F \pmod{6},
\label{eq:C6_Matrices_A_supp_III}
\\
\mc{C}
=& 2[K_1]-2[K_2]+3[M_1]
= -\frac{4}{\sqrt{3}} \trm{Im}[\vev{c_3(\bx_{2b})}_F] + \frac{3}{2} \vev{c_2(\bx_{3c})}_F \pmod{6},
\label{eq:C6_Matrices_A_supp_III}
\ea
Computing $M_{\trm{TCM}\leftarrow\trm{RSI}}^{(6)}$ yields the following relations,
\ba
&\zeta\delta_{1a}^{(1)}+\omega\delta_{1a}^{(2)}-\delta_{1a}^{(3)}+\bar{\omega}\delta_{1a}^{(4)}+\bar{\zeta}\delta_{1a}^{(5)}=\vev{c_6(\bx_{1a})}_F,
\nn
&\omega\delta_{1a}^{(1)}+\bar{\omega}\delta_{1a}^{(2)}+\delta_{1a}^{(3)}+\omega\delta_{1a}^{(4)}+\bar{\omega}\delta_{1a}^{(5)}=\frac{1}{3}\vev{c_3(\bx_{1a})}_F,
\nn
&\omega\delta_{2b}^{(1)}+\bar{\omega}\delta_{2b}^{(2)}=\frac{1}{3}\vev{c_3(\bx_{2b})}_F,
\nn
&\delta_{1a}^{(1)}-\delta_{1a}^{(2)}+\delta_{1a}^{(3)}-\delta_{1a}^{(4)}+\delta_{1a}^{(5)}=-\frac{1}{4}\vev{c_2(\bx_{1a})}_F,
\nn
&\delta_{3c}=-\frac{1}{4}\vev{c_2(\bx_{3c})}_F.
\label{eq:C6_Matrices_A_supp_IV}
\ea
Note that for atomic insulators, the Chern number must be zero, which enforces an additional constraint on the rotation invariants, $[K_2]=[K_1]+\frac{3}{2}[M_1]\pmod{6}$.
The sector charge for symmetry class A systems pertaining to spinless electrons expressed in terms of RSIs and TCMs is given as,
\ba
Q_{\trm{sector}} = \frac{e}{6}\sum_{\ell=1}^{5}\delta_{1a}^{(\ell)} = -\frac{e}{72}(24\trm{Re}[\vev{c_6(\bx_{1a})}_F]+8\trm{Re}\vev{c_3(\bx_{1a})}_F+3\vev{c_2(\bx_{1a})}_F) \pmod{e}.
\label{eq:C6_Matrices_A_supp_V}
\ea
Owing to the$\pmod{e}$ constraint on the sector charge, this expression is non-unique - there are multiple other possible ways of expressing the sector charge in terms of the TCMs that are also equivalent.

For symmetry class AI systems, time-reversal symmetry imposes the constraints $m(\Gamma_2)=m(\Gamma_6)$, $m(\Gamma_3)=m(\Gamma_5)$, and $[K_1]=-2[K_2]$ which implies $\nu=m(\Gamma_1)+2(m(\Gamma_2)+m(\Gamma_3))+m(\Gamma_4)$. 
Additionally, one also has $\delta_{1a}^{(1)}=\delta_{1a}^{(5)}$, $\delta_{1a}^{(2)}=\delta_{1a}^{(4)}$, $\delta_{2b}^{(1)}=\delta_{2b}^{(2)}\equiv\delta_{2b}$, $\vev{c_6(\bx_{1a})}_F=\vev{(c_6(\bx_{1a}))^{-1}}_F=(\vev{c_6(\bx_{1a})}_F)^*$, and $\vev{c_3(\bx_W)}_F=\vev{(c_3(\bx_W))^{-1}}_F=(\vev{c_3(\bx_W)}_F)^*$ for $W\in\{1a,2b\}$.
Therefore, to evaluate the matrices for symmetry class AI, we must incorporate the additional constraints imposed by the symmetries and modify $\bb v_{\trm{TCM}}^{(6)}$, $\bb v_{\trm{MS}}^{(6)}$, and $\bb v_{\trm{RSI}}^{(6)}$ to include only the independent set of quantities for each,
\ba
&\bb v_{\trm{TCM}}^{(6)}=\begin{pmatrix} \nu, & \vev{c_6(\bx_{1a})}_F, & \vev{c_3(\bx_{1a})}_F, & \vev{c_3(\bx_{2b})}_F, & \vev{c_2(\bx_{1a})}_F, & \vev{c_2(\bx_{3c})}_F \end{pmatrix}^{T},
\nn
&\bb v_{\trm{MS}}^{(6)}=\begin{pmatrix} \nu, & m(\Gamma_1), & m(\Gamma_2), & m(\Gamma_3), & [K_1], & [M_1] \end{pmatrix}^{T},
\nn
&\bb v_{\trm{RSI}}^{(6)}=\begin{pmatrix} \nu, & \delta_{1a}^{(1)}, & \delta_{1a}^{(2)}, & \delta_{1a}^{(3)}, & \delta_{2b}, & \delta_{3c} \end{pmatrix}^{T}
\label{eq:C6_Matrices_AI_supp_I}
\ea
Thus, $M_{\trm{TCM}\leftarrow\trm{MS}}^{(6)}$ and $M_{\trm{MS}\leftarrow\trm{RSI}}^{(6)}$ for symmetry class AI are given as follows,
\ba
&\begin{pmatrix} \nu \\\\ \vev{c_6(\bx_{1a})}_F \\\\ \vev{c_3(\bx_{1a})}_F \\\\ \vev{c_3(\bx_{2b})}_F \\\\ \vev{c_2(\bx_{1a})}_F \\\\ \vev{c_2(\bx_{3c})}_F \end{pmatrix}=\begin{pmatrix} 1 & 0 & 0 & 0 & 0 & 0 \\\\ -1 & 2 & 3 & 1 & 0 & 0 \\\\ 3 & 0 & -9 & -9 & 3 & 0 \\\\ 0 & 0 & 0 & 0 & -3/2 & 0 \\\\ -4 & 8 & 0 & 16 & 0 & 6 \\\\ 0 & 0 & 0 & 0 & 0 & -2 \end{pmatrix}\begin{pmatrix} \nu \\\\ m(\Gamma_1) \\\\ m(\Gamma_2) \\\\ m(\Gamma_3) \\\\ [K_1] \\\\ [M_1] \end{pmatrix},
\nnnn
&\begin{pmatrix} \nu \\\\ m(\Gamma_1) \\\\ m(\Gamma_2) \\\\ m(\Gamma_3) \\\\ [K_1] \\\\ [M_1] \end{pmatrix}\begin{pmatrix} 1 & 0 & 0 & 0 & 0 & 0 \\\\ 1/6 & -1/3 & -1/3 & -1/6 & -2/3 & -1/2 \\\\ 1/6 & 2/3 & -1/3 & -1/6 & 1/3 & 1/2 \\\\ 1/6 & -1/3 & 2/3 & -1/6 & 1/3 & -1/2 \\\\ 0 & 0 & 0 & 0 & 2 & 0 \\\\ 0 & 0 & 0 & 0 & 0 & 2 \end{pmatrix}\begin{pmatrix} \nu \\\\ \delta_{1a}^{(1)} \\\\ \delta_{1a}^{(2)} \\\\ \delta_{1a}^{(3)} \\\\ \delta_{2b} \\\\ \delta_{3c} \end{pmatrix}
\label{eq:C6_Matrices_AI_supp_II}
\ea
Computing $M_{\trm{TCM}\leftarrow\trm{RSI}}^{(6)}$ yields the following relations,
\ba
&\delta_{1a}^{(1)}-\delta_{1a}^{(2)}-\delta_{1a}^{(3)}=\vev{c_6(\bx_{1a})}_F,
\nn
&-\delta_{1a}^{(1)}-\delta_{1a}^{(2)}+\delta_{1a}^{(3)}=\frac{1}{3}\vev{c_3(\bx_{1a})}_F,
\nn
&\delta_{2b}=-\frac{1}{3}\vev{c_3(\bx_{2b})}_F,
\nn
&2\delta_{1a}^{(1)}-2\delta_{1a}^{(2)}+\delta_{1a}^{(3)}=-\frac{1}{4}\vev{c_2(\bx_{1a})}_F,
\nn
&\delta_{3c}=-\frac{1}{4}\vev{c_2(\bx_{3c})}_F.
\label{eq:C6_Matrices_AI_supp_III}
\ea
The sector charge for symmetry class AI is therefore given as,
\ba
Q_{\trm{sector}}=\frac{e}{6}(2\delta_{1a}^{(1)}+2\delta_{1a}^{(2)}+\delta_{1a}^{(3)})=-\frac{e}{72}(24\vev{c_6(\bx_{1a})}_F+8\vev{c_3(\bx_{1a})}_F+3\vev{c_2(\bx_{1a})}_F) \pmod{e}.
\label{eq:C6_Matrices_AI_supp_IV}
\ea
Returning to the original definition of $\bb v_{\trm{TCM}}^{(6)}$, $\bb v_{\trm{MS}}^{(6)}$, and $\bb v_{\trm{RSI}}^{(6)}$ given by Eqs.~\eqref{eq:TCM_C3_C4_C6_vector_supp}, \eqref{eq:MS_C2_C3_C4_C6_vector_supp}, and \eqref{eq:RSI_C2_C3_C4_C6_vector_supp}, one can construct the matrix $M_{\trm{TCM}\leftarrow\trm{MS}}^{(6)}$ once again using Eq.~\eqref{eq:fully_traced_TCM_4_supp} and Table~\ref{table:C6_bandrep_supp_2} for symmetry class A systems pertaining to spin-1/2 electrons,
\ba
\begin{pmatrix} \nu \\\\ \vev{c_6(\bx_{1a})}_F \\\\ \vev{(c_6(\bx_{1a}))^{-1}}_F \\\\ \vev{c_3(\bx_{1a})}_F \\\\ \vev{c_3(\bx_{2b})}_F \\\\ \vev{(c_3(\bx_{1a}))^{-1}}_F \\\\ \vev{(c_3(\bx_{2b}))^{-1}}_F \\\\ \vev{c_2(\bx_{1a})}_F \\\\ \vev{c_2(\bx_{3c})}_F \end{pmatrix}=\begin{pmatrix} 1 & 0 & 0 & 0 & 0 & 0 & 0 & 0 & 0 \\\\ \bar{\gamma} & \gamma-\bar{\gamma} & i-\bar{\gamma} & -2\bar{\gamma} & -(\gamma+\bar{\gamma}) & -(i+\bar{\gamma}) & 0 & 0 & 0 \\\\ \gamma & -(\gamma-\bar{\gamma}) & -(i+\gamma) & -2\gamma & -(\gamma+\bar{\gamma}) & i-\gamma & 0 & 0 & 0 \\\\ 3\bar{\zeta} & 3(\zeta-\bar{\zeta}) & -3(1+\bar{\zeta}) & 0 & 3(\zeta-\bar{\zeta}) & -3(1+\bar{\zeta}) & 2(\zeta-\bar{\zeta}) & -2(1+\bar{\zeta}) & 0 \\\\ 0 & 0 & 0 & 0 & 0 & 0 & -(\zeta-\bar{\zeta}) & 1+\bar{\zeta} & 0 \\\\ 3\zeta & -3(\zeta-\bar{\zeta}) & -3(1+\zeta) & 0 & -3(\zeta-\bar{\zeta}) & -3(1+\zeta) & -2(\zeta-\bar{\zeta}) & -2(1+\zeta) & 0 \\\\ 0 & 0 & 0 & 0 & 0 & 0 & \zeta-\bar{\zeta} & 1+\zeta & 0 \\\\ -4i & 8i & 0 & 8i & 0 & 8i & 0 & 0 & 6i \\\\ 0 & 0 & 0 & 0 & 0 & 0 & 0 & 0 & -2i \end{pmatrix}\begin{pmatrix} \nu \\\\ m(\Gamma_1) \\\\ m(\Gamma_2) \\\\ m(\Gamma_3) \\\\ m(\Gamma_4) \\\\ m(\Gamma_5) \\\\ [K_1] \\\\ [K_2] \\\\ [M_1] \end{pmatrix}
\label{eq:C6_Matrices_A_spin_half_supp_I}
\ea
where $\gamma=e^{\pi i/6}$, $\zeta=e^{\pi i/3}$, $\bar{\gamma}=e^{-\pi i/6}$, and $\bar{\zeta}=e^{-\pi i/3}$, and $M_{\trm{MS}\leftarrow\trm{RSI}}^{(6)}$ is identical to Eq.~\eqref{eq:C6_Matrices_A_supp_I}. 
From Eq.~\eqref{eq:C6_Matrices_A_spin_half_supp_I}, the following relations can be determined,
\bg
6(m(\Gamma_2)-2m(\Gamma_3)+2m(\Gamma_4)-m(\Gamma_5))+8[K_1]+4[K_2]-9[M_1]
\nn
= -\frac{2i}{\sqrt{3}}(\vev{c_3(\bx_{1a})}_F-\vev{(c_3(\bx_{1a}))^{-1}}_F)+\frac{3i}{2}\vev{c_2(\bx_{1a})},
\nn
-4[K_1]-2[K_2]+3[M_1]
= -\frac{2i}{\sqrt{3}}(\vev{c_3(\bx_{2b})}_F-\vev{(c_3(\bx_{2b}))^{-1}}_F)+\frac{3i}{2}\vev{c_2(\bx_{3c})}_F,
\label{eq:C6_Matrices_A_spin_half_supp_II}
\eg
which leads to the following \textit{multiple} relations for the Chern number for spin-1/2 electron systems in symmetry class A,
\ba
\mc{C}
=& 2[K_1]-2[K_2]+3[M_1]
=\frac{4}{\sqrt{3}}\trm{Im}[\vev{c_3(\bx_{1a})}_F]+\frac{3i}{2}\vev{c_2(\bx_{1a})}_F \pmod{6},
\label{eq:C6_Matrices_A_spin_half_supp_III}
\\
\mc{C}
=& 2[K_1]-2[K_2]+3[M_1]
= \frac{4}{\sqrt{3}}\trm{Im}[\vev{c_3(\bx_{2b})}_F] + \frac{3i}{2}\vev{c_2(\bx_{3c})}_F \pmod{6},
\label{eq:C6_Matrices_A_spin_half_supp_IV}
\ea
Computing $M_{\trm{TCM}\leftarrow\trm{RSI}}^{(6)}$ yields the following relations,
\ba
&i\delta_{1a}^{(1)}-\bar{\gamma}\delta_{1a}^{(2)}-\gamma\delta_{1a}^{(3)}-i\delta_{1a}^{(4)}+\bar{\gamma}\delta_{1a}^{(5)}=\vev{c_6(\bx_{1a})}_F,
\nn
&-\delta_{1a}^{(1)}+\bar{\zeta}\delta_{1a}^{(2)}+\zeta\delta_{1a}^{(3)}-\delta_{1a}^{(4)}+\bar{\zeta}\delta_{1a}^{(5)}=\frac{1}{3}\vev{c_3(\bx_{1a})}_F,
\nn
&-\delta_{2b}^{(1)}+\bar{\zeta}\delta_{2b}^{(2)}=\frac{1}{3}\vev{c_3(\bx_{2b})}_F,
\nn
&\delta_{1a}^{(1)}-\delta_{1a}^{(2)}+\delta_{1a}^{(3)}-\delta_{1a}^{(4)}+\delta_{1a}^{(5)}=\frac{i}{4}\vev{c_2(\bx_{1a})}_F,
\nn
&\delta_{3c}=\frac{i}{4}\vev{c_2(\bx_{3c})}_F
\label{eq:C6_Matrices_A_spin_half_supp_V}
\ea
The sector charge for symmetry class A pertaining to spin-half electrons is therefore given as,
\bg
Q_{\trm{sector}}
= \frac{e}{6} \sum_{\ell=1}^{5} \delta_{1a}^{(\ell)}
= \frac{e}{6} \left( -5\delta_{1a}^{(1)}+\sum_{\ell=2}^{5}\delta_{1a}^{(\ell)} \right)
\nn
= \frac{e}{72} (-24\trm{Im}[\vev{c_6(\bx_{1a})}_F] + 8\trm{Re}[\vev{c_3(\bx_{1a})}_F] - 3i\vev{c_2(\bx_{1a})}_F) \pmod{e}.
\label{eq:C6_Matrices_A_spin_half_supp_VI}
\eg
Once again, we note that the expression for the sector charge is not unique owing to the$\pmod{e}$ constraint, so there are other equivalent expressions for the sector charge.

For symmetry class AII systems, time-reversal symmetry imposes the constraints $m(\Gamma_1)=m(\Gamma_6)$, $m(\Gamma_2)=m(\Gamma_5)$, $m(\Gamma_3)=m(\Gamma_4)$, which implies $\nu=2(m(\Gamma_1)+m(\Gamma_2)+m(\Gamma_3))$, in addition to $[K_2]=-2[K_1]$, and $[M_1]=[M_2]=0$.
Furthermore, this also imposes the constraints $\delta_{1a}^{(1)}=\delta_{1a}^{(4)}$, $\delta_{1a}^{(2)}=\delta_{1a}^{(3)}$, $\delta_{1a}^{(5)}=0$, $\delta_{2b}^{(1)}\equiv\delta_{2b}$, and $\delta_{2b}^{(2)}=\delta_{3c}=0$, in addition to $\vev{c_6(\bx_{1a})}_F=\vev{(c_6(\bx_{1a}))^{-1}}_F=(\vev{c_6(\bx_{1a})}_F)^*$, $\vev{c_3(\bx_{W})}_F=\vev{(c_3(\bx_{W}))^{-1}}_F=(\vev{c_3(\bx_{W})}_F)^*$ for $W\in\{1a,2b\}$, and $\vev{c_2(\bx_{W})}_F=0$ for $W\in\{1a,2b,3c\}$.
To evaluate the matrices for symmetry class AII, we must incorporate the additional constraints imposed by the symmetries and modify $\bb v_{\trm{TCM}}^{(6)}$, $\bb v_{\trm{MS}}^{(6)}$, and $\bb v_{\trm{RSI}}^{(6)}$ to include only the independent sets of quantities for each,
\ba
&\bb v_{\trm{TCM}}^{(6)}=\begin{pmatrix} \nu, & \vev{c_6(\bx_{1a})}_F, & \vev{c_3(\bx_{1a})}_F, & \vev{c_3(\bx_{2b})}_F \end{pmatrix}^{T},
\nn
&\bb v_{\trm{MS}}^{(6)}=\begin{pmatrix} \nu, & m(\Gamma_1), & m(\Gamma_2), & [K_1] \end{pmatrix}^{T},
\nn
&\bb v_{\trm{RSI}}^{(6)}=\begin{pmatrix} \nu, & \delta_{1a}^{(1)}, & \delta_{1a}^{(2)}, & \delta_{2b} \end{pmatrix}^{T}.
\label{eq:C6_Matrices_AII_supp_I}
\ea
Thus, $M_{\trm{TCM}\leftarrow\trm{MS}}^{(6)}$ and $M_{\trm{MS}\leftarrow\trm{RSI}}^{(6)}$ for symmetry class AII are given as follows,
\ba
&\begin{pmatrix} \nu \\\\ \vev{c_6(\bx_{1a})}_F \\\\ \vev{c_3(\bx_{1a})}_F \\\\ \vev{c_3(\bx_{2b})}_F \end{pmatrix}=\begin{pmatrix} 1 & 0 & 0 & 0 \\\\ -\frac{1}{2}(\gamma+\bar{\gamma}) & 2(\gamma+\bar{\gamma}) & \gamma+\bar{\gamma} & 0 \\\\ \frac{3}{2}(\zeta+\bar{\zeta}) & 0 & -9 & 6 \\\\ 0 & 0 & 0 & -3 \end{pmatrix}\begin{pmatrix} \nu \\\\ m(\Gamma_1) \\\\ m(\Gamma_2) \\\\ [K_1] \end{pmatrix},
\nnnn
&\begin{pmatrix} \nu \\\\ m(\Gamma_1) \\\\ m(\Gamma_2) \\\\ [K_1] \end{pmatrix}=\begin{pmatrix} 1 & 0 & 0 & 0 \\\\ 1/6 & -1/3 & -1/3 & -1/3 \\\\ 1/6 & 2/3 & -1/3 & 2/3 \\\\ 0 & 0 & 0 & 1 \end{pmatrix}\begin{pmatrix} \nu \\\\ \delta_{1a}^{(1)} \\\\ \delta_{1a}^{(2)} \\\\ \delta_{2b} \end{pmatrix}.
\label{eq:C6_Matrices_AII_supp_II}
\ea
Computing $M_{\trm{TCM}\leftarrow\trm{RSI}}^{(6)}$ yields the following relations,
\ba
&\delta_{1a}^{(2)}=-\frac{1}{\sqrt{3}}\vev{c_6(\bx_{1a})}_F,
\nn
&2\delta_{1a}^{(1)}-\delta_{1a}^{(2)}=-\frac{1}{3}\vev{c_3(\bx_{1a})}_F,
\nn
&\delta_{2b}=-\frac{1}{3}\vev{c_3(\bx_{2b})}_F.
\label{eq:C6_Matrices_AII_supp_III}
\ea
The sector charge for symmetry class AII systems is therefore given by,
\ba
Q_{\trm{sector}}=\frac{e}{3}(\delta_{1a}^{(1)}+\delta_{1a}^{(2)})=\frac{e}{3}(-2\delta_{1a}^{(1)}+\delta_{1a}^{(2)})=\frac{e}{9}\vev{c_3(\bx_{1a})}_F \pmod{e}.
\label{eq:C6_Matrices_AII_supp_IV}
\ea

Finally, for symmetry class D systems, we consider both spinless and spin-1/2 electron systems.
For spinless systems, particle-hole symmetry introduces the constraint on the rotation invariants $[K_1]=0$ in addition to $\vev{c_3(\bx_{2b})}_F = -\vev{(c_3(\bx_{2b}))^{-1}}_F = -(\vev{c_3(\bx_{2b})}_F)^*$.
Therefore, similar to approaches for the taken for symmetry classes AI and AII, for symmetry class D pertaining to spinless electrons, we modify $\bb v_{\trm{TCM}}^{(6)}$ and $\bb v_{\trm{MS}}^{(6)}$ (we do not consider $\bb v_{\trm{RSI}}^{(6)}$ since atomic insulators do not exist in symmetry class D) to include only the independent quantities for each,
\ba
&\bb v_{\trm{TCM}}^{(6)}=\begin{pmatrix} \nu, & \vev{c_6(\bx_{1a})}_F, & \vev{(c_6(\bx_{1a}))^{-1}}_F, & \vev{c_3(\bx_{1a})}_F, & \vev{c_3(\bx_{2b})}_F, & \vev{(c_3(\bx_{1a}))^{-1}}_F, & \vev{c_2(\bx_{1a})}_F, & \vev{c_2(\bx_{3c})}_F \end{pmatrix}^{T},
\nn
&\bb v_{\trm{MS}}^{(6)}=\begin{pmatrix} \nu, & m(\Gamma_1), & m(\Gamma_2), & m(\Gamma_3), & m(\Gamma_4), & m(\Gamma_5), & [K_2], & [M_1] \end{pmatrix}^{T},
\label{eq:C6_Matrices_D_supp_I}
\ea
Therefore, $M_{\trm{TCM}\leftarrow\trm{MS}}^{(6)}$ for symmetry class D pertaining to spinless electrons is,
\ba
\begin{pmatrix} \nu \\\\ \vev{c_6(\bx_{1a})}_F \\\\ \vev{(c_6(\bx_{1a}))^{-1}}_F \\\\ \vev{c_3(\bx_{1a})}_F \\\\ \vev{c_3(\bx_{2b})}_F \\\\ \vev{(c_3(\bx_{1a}))^{-1}}_F \\\\ \vev{c_2(\bx_{1a})}_F \\\\ \vev{c_2(\bx_{3c})}_F \end{pmatrix} = \begin{pmatrix} 1 & 0 & 0 & 0 & 0 & 0 & 0 & 0 \\\\ \bar{\zeta} & 1-\bar{\zeta} & \zeta-\bar{\zeta} & \omega-\bar{\zeta} & -(1+\bar{\zeta}) & \bar{\omega}-\bar{\zeta} & 0 & 0 \\\\ \zeta & 1-\zeta & -(\zeta-\bar{\zeta}) & \bar{\omega}-\zeta & -(1+\zeta) & \omega-\zeta & 0 & 0 \\\\ 3\bar{\omega} & 3(1-\bar{\omega}) & 3(\omega-\bar{\omega}) & 0 & 3(1-\bar{\omega}) & 3(\omega-\bar{\omega}) & 2(\omega-\bar{\omega}) & 0 \\\\ 0 & 0 & 0 & 0 & 0 & 0 & -(\omega-\bar{\omega}) & 0 \\\\ 3\omega & -3(\omega-1) & -3(\omega-\bar{\omega}) & 0 & -3(\omega-1) & -3(\omega-\bar{\omega}) & -2(\omega-\bar{\omega}) & 0 \\\\ -4 & 8 & 0 & 8 & 0 & 8 & 0 & 6 \\\\ 0 & 0 & 0 & 0 & 0 & 0 & 0 & -2 \end{pmatrix}\begin{pmatrix} \nu \\\\ m(\Gamma_1) \\\\ m(\Gamma_2) \\\\ m(\Gamma_3) \\\\ m(\Gamma_4) \\\\ m(\Gamma_5) \\\\ [K_2] \\\\ [M_1] \end{pmatrix}
\label{eq:C6_Matrices_D_supp_II}
\ea
From Eq.~\eqref{eq:C6_Matrices_D_supp_II}, it follows that the same relations as Eqs.~\eqref{eq:C6_Matrices_A_supp_II} and \eqref{eq:C6_Matrices_A_supp_III} can be obtained with $[K_1]=0$ and $\vev{c_3(\bx_{2b})}_F = -\vev{(c_3(\bx_{2b}))^{-1}}_F = -(\vev{c_3(\bx_{2b})}_F)^*$.
This means the Chern number for symmetry class D is equivalent to the Chern number for symmetry class A for spinless electrons.
Similarly, for spin-1/2 systems, particle-hole symmetry imposes the constraints $[K_2]=0$ and the same constraint $\vev{c_3(\bx_{2b})}_F=-\vev{(c_3(\bx_{2b}))^{-1}}_F=-(\vev{c_3(\bx_{2b})}_F)^*$.
Once again, we modify $\bb v_{\trm{MS}}^{(6)}$ to incorporate the constraint imposed by the symmetry, but retain the definition of $\bb v_{\trm{TCM}}^{(6)}$ given by Eq.~\eqref{eq:C6_Matrices_D_supp_I},
\ba
\bb v_{\trm{MS}}^{(6)}=\begin{pmatrix} \nu, & m(\Gamma_1), & m(\Gamma_2), & m(\Gamma_3), & m(\Gamma_4), & m(\Gamma_5), & [K_1], & [M_1] \end{pmatrix}^{T},
\label{eq:C6_Matrices_D_supp_III}
\ea
which means $M_{\trm{TCM}\leftarrow\trm{MS}}^{(6)}$ for symmetry class D pertaining to spin-1/2 electrons is,
\ba
\begin{pmatrix} \nu \\\\ \vev{c_6(\bx_{1a})}_F \\\\ \vev{(c_6(\bx_{1a}))^{-1}}_F \\\\ \vev{c_3(\bx_{1a})}_F \\\\ \vev{c_3(\bx_{2b})}_F \\\\ \vev{(c_3(\bx_{1a}))^{-1}}_F \\\\ \vev{c_2(\bx_{1a})}_F \\\\ \vev{c_2(\bx_{3c})}_F \end{pmatrix}=\begin{pmatrix} 1 & 0 & 0 & 0 & 0 & 0 & 0 & 0 \\\\ \bar{\gamma} & \gamma-\bar{\gamma} & i-\bar{\gamma} & -2\bar{\gamma} & -(\gamma+\bar{\gamma}) & -(i+\bar{\gamma}) & 0 & 0 \\\\ \gamma & -(\gamma-\bar{\gamma}) & -(i+\gamma) & -2\gamma & -(\gamma+\bar{\gamma}) & i-\gamma & 0 & 0 \\\\ 3\bar{\zeta} & 3(\zeta-\bar{\zeta}) & -3(1+\bar{\zeta}) & 0 & 3(\zeta-\bar{\zeta}) & -3(1+\bar{\zeta}) & 2(\zeta-\bar{\zeta}) & 0 \\\\ 0 & 0 & 0 & 0 & 0 & 0 & -(\zeta-\bar{\zeta}) & 0 \\\\ 3\zeta & -3(\zeta-\bar{\zeta}) & -3(1+\zeta) & 0 & -3(\zeta-\bar{\zeta}) & -3(1+\zeta) & -2(\zeta-\bar{\zeta}) & 0 \\\\ -4i & 8i & 0 & 8i & 0 & 8i & 0 & 6i \\\\ 0 & 0 & 0 & 0 & 0 & 0 & 0 & -2i \end{pmatrix}\begin{pmatrix} \nu \\\\ m(\Gamma_1) \\\\ m(\Gamma_2) \\\\ m(\Gamma_3) \\\\ m(\Gamma_4) \\\\ m(\Gamma_5) \\\\ [K_1] \\\\ [M_1] \end{pmatrix}.
\label{eq:C6_Matrices_D_supp_IV}
\ea
From Eq.~\eqref{eq:C6_Matrices_D_supp_IV}, one can obtain the same relations as Eq.~\eqref{eq:C6_Matrices_A_spin_half_supp_II}, but with $[K_2]=0$.
It follows that the Chern number for symmetry class D is equivalent to the Chern number for symmetry class A for spin-1/2 electrons.
%%%%%%

%%%%%%
\section{Derivation of momentum quantization condition under twisted boundary conditions}
\label{sec:tbc_supp}
%%%%%%
In this section, we derive the momentum quantization condition presented in the main text,
\ba
\bb k = \left(\frac{n_1}{N_1} - \frac{\theta_1}{2\pi N_1}\right)\bb b_1 + \left(\frac{n_2}{N_2} - \frac{\theta_2}{2\pi N_2}\right)\bb b_2.
\label{eq:momentum_quantization_tbc_supp}
\ea
where $n_{i}\in\{0,\ldots,N_{i}-1\}$ for $i\in\{1,2\}$ (in this section, we are working with the convention $\bb b_i\cdot\bb a_j=2\pi\delta_{ij}$).
This applies to a lattice with dimensions $N_1\times N_2$ and twisted boundary conditions parameterized by $(\theta_1,\theta_2)$, which can be understood as two fluxes introduced along two distinct co-cycles of the torus in the $\bb a_1$ and $\bb a_2$ directions respectively.
The twisted boundary condition are defined as follows,
\ba
\ket{\bR + N_i \bb a_i,\alpha}=e^{-i\theta_i}\ket{\bR,\alpha} \ \text{for} \ i \in \{1,2\}
\label{eq:tbc_def_supp}
\ea
Let us extend $n_1,n_2$ to be any integer, i.e., $n_1,n_2\in\mathbb{Z}$, introduce $l_1,l_2\in\mathbb{Z}$, and denote $C$ as the normalization constant for the Fourier transform.
The Fourier transform in the periodic basis is given as,
\ba
\ket{\bk,\alpha}\equiv\frac{1}{\sqrt{N}}\sum_\bk e^{2\pi i\bk\cdot\bR}\ket{\bR,\alpha}.
\label{eq:Fourier_Transform_supp_II}
\ea
Re-expressing Eq.~\eqref{eq:Fourier_Transform_supp_II} in the following manner,
\ba
&\ket{\bk,\alpha}=C\sum_{\bR,l_1,l_2}e^{2\pi i\bk\cdot(\bR+l_1 N_1\bb a_1+l_2 N_2\bb a_2)}\ket{\bR+l_1 N_1\bb a_1+l_2 N_2\bb a_2,\alpha},
\nn
&=C\sum_{\bR,l_1,l_2}e^{2\pi i\bk\cdot(\bR+l_1 N_1\bb a_1+l_2 N_2\bb a_2)-i(l_1\theta_1 + l_2\theta_2)}\ket{\bR,\alpha},
\nn
&=C\left(\sum_{\bR}e^{2\pi i\bk\cdot\bR}\ket{\bR,\alpha}\right)\left(\sum_{l_1,l_2}e^{2\pi i\left(\bk-\frac{\theta_1}{2\pi N_1}\bb b_1-\frac{\theta_2}{2\pi N_2}\bb b_2\right)\cdot(l_1 N_1\bb a_1+l_2 N_2\bb a_2)}\right),
\nn
&=C\left(\sum_{\bR}e^{2\pi i\bk\cdot\bR}\ket{\bR,\alpha}\right)\left(\sum_{l_1,l_2}e^{2\pi i\left(\left(\frac{n_1}{N_1}-\frac{\theta_1}{2\pi N_1}\right)\bb b_1+\left(\frac{n_2}{N_2}-\frac{\theta_2}{2\pi N_2}\right)\bb b_2\right)\cdot(l_1 N_1\bb a_1+l_2 N_2\bb a_2)}\right),
\nn
&=C\left(\sum_{l_1,l_2}e^{i(l_1(2\pi n_1-\theta_1)+l_2(2\pi n_2-\theta_2))}\right)\sum_{\bR}e^{2\pi i\bk\cdot\bR}\ket{\bR,\alpha}.
\label{eq:momentum_quantization_derivation_supp}
\ea
In order for the equality to remain true, the quantity in the parentheses must be equal to $1$.
To do so, we impose $2\pi n_i - \theta_i\in 2\pi\mathbb{Z}$ for $i\in\{1,2\}$. 
Equivalently, we redefine $n_i\to n_i-\frac{\theta_i}{2\pi}$ in $\bk=\frac{n_1}{N_1}\bb b_1+\frac{n_2}{N_2}\bb b_2$.
This means,
\ba
\bk = \frac{1}{N_1}\left(n_1-\frac{\theta_1}{2\pi}\right)\bb b_1+\frac{1}{N_2}\left(n_2-\frac{\theta_2}{2\pi}\right)\bb b_2
\label{eq:momentum_quantization_tbc_supp_II}
\ea
which yields Eq.~\eqref{eq:momentum_quantization_tbc_supp}.
%%%%%%%

%%%%%%%
\section{Fully traced TCMs for arbitrary finite-size $C_n$-symmetric lattice utilizing twisted boundary conditions}
%%%%%%
Here, we review the implementation of fully traced TCMs for any $C_n$-symmetric lattice with finite dimensions $N_1\times N_2$ that \textit{do not} satisfy the constraint
\ba
& C_2: \, (N_1, N_2) = (0, 0) \pmod 2,
\nn
& C_4: \, N = 0 \pmod 2,
\nn
& C_{n=3,6}: \, N = 0 \pmod n,
\label{eq:perfect_constraint_supp_III}
\ea
using twisted boundary conditions.
Any lattice with dimensions $N_1\times N_2$ that satisfies Eq.~\eqref{eq:perfect_constraint_supp_III} will support a BZ with the complete set of high symmetry momenta.
This is a necessary condition to be able to express the Chern number, bulk polarization, and sector charge in terms of momentum-space rotation invariants, but limits the applicability to lattices of other dimensions.
When $(N_1,N_2)$ do not satisfy Eq.~\eqref{eq:perfect_constraint_supp_III}, the BZ supports a reduced set of high symmetry momenta when only periodic boundary conditions are considered.
Twisting the boundary conditions by introducing $\theta_1$ and $\theta_2$ adjusts the momentum quantization as shown in Eq.~\eqref{eq:momentum_quantization_tbc_supp}.
By choosing appropriate $\theta_1$ and $\theta_2$, different sets of high symmetry momenta can be accessed.
Thus, when a $C_n$-symmetric lattice with dimensions $N_1\times N_2$ does not satisfy Eq.~\eqref{eq:perfect_constraint_supp_III}, it is still possible to evaluate the Chern number, bulk polarization, and sector charge for such lattices by \textit{jointly} considering lattices with different sets of boundary conditions.
In this section, we will state the mapping of the fully traced TCMs corresponding to the lattice that satisfy Eq.~\eqref{eq:perfect_constraint_supp_III} to the fully traced TCMs corresponding to the $C_n$-symmetric lattices that do not satisfy Eq.~\eqref{eq:perfect_constraint_supp_III} by adding over the appropriate twisted boundary conditions.
Using this mapping, it is possible to extend the Chern number, bulk polarization, and sector charge for $C_n$-symmetric lattices with dimensions $N_1\times N_2$ that satisfy Eq.~\eqref{eq:perfect_constraint_supp_III}, to any $C_n$-symmetric lattice.
The fully traced TCM for a lattice with twisted boundary conditions is given by $\vev{c_n(\bb r_o)}_{F}^{(\theta_1,\theta_2)}$.
When $\theta_1=\theta_2=0$, which corresponds to periodic boundary conditions, this superscript will usually be suppressed, i.e., $\vev{c_n(\bb r_o)}_{F}^{(0,0)}\equiv\vev{c_n(\bb r_o)}_F$.
%%%%%%

%%%%%%
\tocless{\subsubsection{$C_2$ symmetry}}{\label{subsec:c2_tbc}}
%%%%%%
A $C_2$-symmetric lattice with dimensions $N_1\times N_2$ can support either $(N_1,N_2)=(0,0) \pmod{2}$, $(N_1,N_2)=(1,0) \pmod{2}$, $(N_1,N_2)=(0,1) \pmod{2}$, or $(N_1,N_2)=(1,1) \pmod{2}$.
For each of these dimensions, these are the corresponding set of invariant WPs in $\mc{X}[c_2(\bb r_o)]$,
\ba
&(N_1,N_2)=(0,0) \pmod{2}: \bb r_o\in\{\bx_{1a},\bx_{1b},\bx_{1c},\bx_{1d}\},
\nn
&(N_1,N_2)=(1,0) \pmod{2}: \bb r_o\in\{\bx_{1a},\bx_{1c}\},
\nn
&(N_1,N_2)=(0,1) \pmod{2}: \bb r_o\in\{\bx_{1a},\bx_{1b}\},
\nn
&(N_1,N_2)=(1,1) \pmod{2}: \bb r_o\in\{\bx_{1a}\}.
\label{eq:C2_Invariant_Positions_supp}
\ea
Hence, we can map the fully traced TCMs for the $(N_1,N_2)=(0,0) \pmod{2}$ lattice to the fully traced TCMs corresponding to the other lattice dimensions.
We will use the convention that the \textit{left-hand side} corresponds to $(N_1,N_2)=(0,0) \pmod{2}$ and the \textit{right-hand side} corresponds to $(N_1,N_2)\neq (0,0) \pmod{2}$.
For $(N_1,N_2)=(1,0) \pmod{2}$, this mapping is given as follows,
\ba
&\vev{c_2(\bx_{1a})}_F \leftrightarrow \sum_{\theta\in\{0,\pi\}}\vev{c_2(\bx_{1a})}_{F}^{(\theta,0)},
\nn
&\vev{c_2(\bx_{1b})}_F \leftrightarrow \sum_{\theta\in\{0,\pi\}}e^{i\theta}\vev{c_2(\bx_{1a})}_{F}^{(\theta,0)},
\nn
&\vev{c_2(\bx_{1c})}_F \leftrightarrow \sum_{\theta\in\{0,\pi\}}\vev{c_2(\bx_{1c})}_{F}^{(\theta,0)},
\nn
&\vev{c_2(\bx_{1d})}_F \leftrightarrow \sum_{\theta\in\{0,\pi\}}e^{i\theta}\vev{c_2(\bx_{1c})}_{F}^{(\theta,0)},
\label{eq:C2_Mapping_1_supp}
\ea
and for $(N_1,N_2)=(0,1) \pmod{2}$,
\ba
&\vev{c_2(\bx_{1a})}_F \leftrightarrow \sum_{\theta\in\{0,\pi\}}\vev{c_2(\bx_{1a})}_{F}^{(0,\theta)},
\nn
&\vev{c_2(\bx_{1b})}_F \leftrightarrow \sum_{\theta\in\{0,\pi\}}\vev{c_2(\bx_{1b})}_{F}^{(0,\theta)},
\nn
&\vev{c_2(\bx_{1c})}_F \leftrightarrow \sum_{\theta\in\{0,\pi\}}e^{i\theta}\vev{c_2(\bx_{1a})}_{F}^{(0,\theta)},
\nn
&\vev{c_2(\bx_{1d})}_F \leftrightarrow \sum_{\theta\in\{0,\pi\}}e^{i\theta}\vev{c_2(\bx_{1b})}_{F}^{(0,\theta)},
\label{eq:C2_Mapping_2_supp}
\ea
and finally, for $(N_1,N_2)=(1,1) \pmod{2}$,
\ba
&\vev{c_2(\bx_{1a})}_F \leftrightarrow \sum_{\theta_1\in\{0,\pi\}}\sum_{\theta_2\in\{0,\pi\}}\vev{c_2(\bx_{1a})}_{F}^{(\theta_1,\theta_2)},
\nn
&\vev{c_2(\bx_{1b})}_F \leftrightarrow \sum_{\theta_1\in\{0,\pi\}}\sum_{\theta_2\in\{0,\pi\}}e^{i\theta_1}\vev{c_2(\bx_{1a})}_{F}^{(\theta_1,\theta_2)},
\nn
&\vev{c_2(\bx_{1c})}_F \leftrightarrow \sum_{\theta_1\in\{0,\pi\}}\sum_{\theta_2\in\{0,\pi\}}e^{i\theta_2}\vev{c_2(\bx_{1a})}_{F}^{(\theta_1,\theta_2)},
\nn
&\vev{c_2(\bx_{1d})}_F \leftrightarrow \sum_{\theta_1\in\{0,\pi\}}\sum_{\theta_2\in\{0,\pi\}}e^{i(\theta_1+\theta_2)}\vev{c_2(\bx_{1a})}_{F}^{(\theta_1,\theta_2)}.
\label{eq:C2_Mapping_3_supp}
\ea
%%%%%%

%%%%%%
\tocless{\subsubsection{$C_3$ symmetry}}{\label{subsec:c3_tbc}}
%%%%%%
A $C_3$-symmetric lattice with dimensions $N\times N$ can support either $N=0 \pmod{3}$ or $N=\pm 1 \pmod{3}$.
For each of these dimensions, these are the corresponding set of invariant WPs in $\mc{X}[c_3(\bb r_o)]$,
\ba
&N=0 \pmod{3}: \bb r_o\in\{\bx_{1a},\bx_{1b},\bx_{1c}\},
\nn
&N=\pm 1 \pmod{3}: \bb r_o\in\{\bx_{1a}\}.
\label{eq:C3_Invariant_Positions_supp}
\ea
Hence, we can map the fully traced TCMs for the $N=0 \pmod{3}$ lattice to the fully traced TCMs corresponding to the other lattice dimensions.
We will use the convention that the \textit{left-hand side} corresponds to $N=0 \pmod{3}$ and the \textit{right-hand side} corresponds to $N=\pm 1 \pmod{3}$.
For $N=1 \pmod{3}$, this mapping is given as follows,
\ba
&\vev{c_3(\bx_{1a})}_F \leftrightarrow \sum_{\theta\in\left\{0,\pm\frac{2\pi}{3}\right\}}\vev{c_3(\bx_{1a})}_{F}^{(\theta,-\theta)},
\nn
&\vev{c_3(\bx_{1b})}_F \leftrightarrow \sum_{\theta\in\left\{0,\pm\frac{2\pi}{3}\right\}}e^{-i\theta}\vev{c_3(\bx_{1a})}_{F}^{(\theta,-\theta)},
\nn
&\vev{c_3(\bx_{1c})}_F \leftrightarrow \sum_{\theta\in\left\{0,\pm\frac{2\pi}{3}\right\}}e^{i\theta}\vev{c_3(\bx_{1a})}_{F}^{(\theta,-\theta)},
\label{eq:C3_Mapping_1_supp}
\ea
and for $N=-1 \pmod{3}$, 
\ba
&\vev{c_3(\bx_{1a})}_F \leftrightarrow \sum_{\theta\in\left\{0,\pm\frac{2\pi}{3}\right\}}\vev{c_3(\bx_{1a})}_{F}^{(\theta,-\theta)},
\nn
&\vev{c_3(\bx_{1b})}_F \leftrightarrow \sum_{\theta\in\left\{0,\pm\frac{2\pi}{3}\right\}}e^{i\theta}\vev{c_3(\bx_{1a})}_{F}^{(\theta,-\theta)},
\nn
&\vev{c_3(\bx_{1c})}_F \leftrightarrow \sum_{\theta\in\left\{0,\pm\frac{2\pi}{3}\right\}}e^{-i\theta}\vev{c_3(\bx_{1a})}_{F}^{(\theta,-\theta)},
\label{eq:C3_Mapping_2_supp}
\ea
%%%%%%

%%%%%%
\tocless{\subsubsection{$C_4$ symmetry}}{\label{subsec:c4_tbc}}
%%%%%%
A $C_4$-symmetric lattice with dimensions $N\times N$ can support either $N=0 \pmod{2}$ or $N=1 \pmod{2}$. For each of these dimensions, these are the corresponding set of WPs in $\mc{X}[c_4(\bb r_o)]$,
\ba
&N=0 \pmod{2}: \bb r_o\in\{\bx_{1a},\bx_{1b}\},
\nn
&N=1 \pmod{2}: \bb r_o\in\{\bx_{1a}\},
\label{eq:C4_Invariant_Positions_supp}
\ea
and the corresponding set of WPs in $\mc{X}[c_2(\bb r_o)]$,
\ba
&N=0 \pmod{2}: \bb r_o\in\{\bx_{1a},\bx_{1b},\bx_{2c}^{(1)},\bx_{2c}^{(2)}\},
\nn
&N=1 \pmod{2}: \bb r_o\in\{\bx_{1a}\}.
\label{eq:C2_C4_Invariant_Positions_supp}
\ea
where $\bx_{2c}^{(1)}=\frac{1}{2}\bb a_1$ and $\bx_{2c}^{(2)}=\frac{1}{2}\bb a_2$.
Hence, we can map the fully traced TCMs for the $N=0 \pmod{2}$ lattice to the fully traced TCMs corresponding to the other lattice dimensions.
We will use the convention that the \textit{left-hand side} corresponds to $N=0 \pmod{2}$ and the \textit{right-hand side} corresponds to $N=1 \pmod{2}$.
Therefore, the mapping is given as follows,
\ba
&\vev{c_4(\bx_{1a})}_F \leftrightarrow \sum_{\theta\in\{0,\pi\}}\vev{c_4(\bx_{1a})}_{F}^{(\theta,\theta)},
\nn
&\vev{c_4(\bx_{1b})}_F \leftrightarrow \sum_{\theta\in\{0,\pi\}}e^{i\theta}\vev{c_4(\bx_{1a})}_{F}^{(\theta,\theta)},
\nn
&\vev{c_2(\bx_{1a})}_F \leftrightarrow \sum_{\theta_1\in\{0,\pi\}}\sum_{\theta_2\in\{0,\pi\}}\vev{c_2(\bx_{1a})}_{F}^{(\theta_1,\theta_2)},
\nn
&\vev{c_2(\bx_{1b})}_F \leftrightarrow \sum_{\theta_1\in\{0,\pi\}}\sum_{\theta_2\in\{0,\pi\}}e^{i(\theta_1+\theta_2)}\vev{c_2(\bx_{1a})}_{F}^{(\theta_1,\theta_2)},
\nn
&\vev{c_2(\bx_{2b}^{(1)})}_F \leftrightarrow \sum_{\theta_1\in\{0,\pi\}}\sum_{\theta_2\in\{0,\pi\}}e^{i\theta_1}\vev{c_2(\bx_{1a})}_{F}^{(\theta_1,\theta_2)},
\nn
&\vev{c_2(\bx_{2b}^{(2)})}_F \leftrightarrow \sum_{\theta_1\in\{0,\pi\}}\sum_{\theta_2\in\{0,\pi\}}e^{i\theta_2}\vev{c_2(\bx_{1a})}_{F}^{(\theta_1,\theta_2)}.
\label{eq:C4_Mapping_supp}
\ea
%%%%%%

%%%%%%
\tocless{\subsubsection{$C_6$ symmetry}}{\label{subsec:c6_tbc}}
%%%%%%
A $C_6$-symmetric lattice with dimensions $N\times N$ can support either $N=0 \pmod{6}$, $N=\pm 1 \pmod{6}$, $N=\pm 2 \pmod{6}$, or $N=\pm 3 \pmod{6}$. For each of these dimensions, it is always true that the only WP in $\mc{X}[c_6(\bb r_o)]$ is $\bb r_o=\bx_{1a}$.
The corresponding set of WPs in $\mc{X}[c_3(\bb r_o)]$ for each set of lattice dimensions is given as follows,
\ba
&N=0,\pm 2 \pmod{6}: \bb r_o\in\{\bx_{1a},\bx_{2b}^{(1)},\bx_{2b}^{(2)}\},
\nn
&N=\pm 1,\pm 3 \pmod{6}: \bb r_o\in\{\bx_{1a}\},
\label{eq:C6_C3_Invariant_supp}
\ea
and the corresponding set of WPs in $\mc{X}[c_2(\bb r_o)]$ for each set of lattice dimensions is given as follows,
\ba
&N=0,\pm 3 \pmod{6}: \bb r_o\in\{\bx_{1a},\bx_{3c}^{(1)},\bx_{3c}^{(2)},\bx_{3c}^{(3)}\},
\nn
&N=\pm 1, \pm 2 \pmod{6}: \bb r_o\in\{\bx_{1a}\},
\label{eq:C6_C2_Invariant_supp}
\ea
where $\bx_{2b}^{(1)}=\frac{1}{3}(\bb a_1+\bb a_2)$, $\bx_{2b}^{(2)}=\frac{1}{3}(-\bb a_1+2\bb a_2)$, $\bb x_{3c}^{(1)}=\frac{1}{2}\bb a_1$, $\bb x_{3c}^{(2)}=\frac{1}{2}\bb a_2$, and $\bb x_{3c}^{(3)}=\frac{1}{2}(\bb a_2 - \bb a_1)$.
Hence, we can map the fully traced TCMs for the $N=0 \pmod{6}$ lattice to the fully traced TCMs corresponding to the other lattice dimensions.
We will use the convention that the \textit{left-hand side} corresponds to $N=0 \pmod{6}$ and the \textit{right-hand side} corresponds to $N\neq 0 \pmod{6}$.
For $N=1 \pmod{6}$, this mapping is given as follows,
\ba
&\vev{c_6(\bx_{1a})}_F \leftrightarrow \vev{c_6(\bx_{1a})}_F,
\nn
&\vev{c_3(\bx_{1a})}_F \leftrightarrow \sum_{\theta\in\left\{0,\pm\frac{2\pi}{3}\right\}}\vev{c_3(\bx_{1a})}_{F}^{(\theta,-\theta)},
\nn
&\vev{c_3(\bx_{2b}^{(1)})}_F \leftrightarrow \sum_{\theta\in\left\{0,\pm\frac{2\pi}{3}\right\}}e^{-i\theta}\vev{c_3(\bx_{1a})}_{F}^{(\theta,-\theta)},
\nn
&\vev{c_3(\bx_{2b}^{(2)})}_F \leftrightarrow \sum_{\theta\in\left\{0,\pm\frac{2\pi}{3}\right\}}e^{i\theta}\vev{c_3(\bx_{1a})}_{F}^{(\theta,-\theta)},
\nn
&\vev{c_2(\bx_{3c}^{(1)})}_F \leftrightarrow \sum_{\theta_1\in\{0,\pi\}}\sum_{\theta_2\in\{0,\pi\}}e^{i\theta_1}\vev{c_2(\bx_{1a})}_{F}^{(\theta_{1},\theta_{2})},
\nn
&\vev{c_2(\bx_{3c}^{(2)})}_F \leftrightarrow \sum_{\theta_1\in\{0,\pi\}}\sum_{\theta_2\in\{0,\pi\}}e^{i\theta_2}\vev{c_2(\bx_{1a})}_{F}^{(\theta_{1},\theta_{2})},
\nn
&\vev{c_2(\bx_{3c}^{(3)})}_F \leftrightarrow \sum_{\theta_1\in\{0,\pi\}}\sum_{\theta_2\in\{0,\pi\}}e^{i(\theta_1+\theta_2)}\vev{c_2(\bx_{1a})}_{F}^{(\theta_{1},\theta_{2})}.
\label{eq:C6_Mapping_1_supp}
\ea
and for $N=-1 \pmod{6}$, the mappings are identical to Eq.~\eqref{eq:C6_Mapping_1_supp} with the exception that the mappings for $\vev{c_3(\bx_{2b}^{(1)})}_F$ and $\vev{c_3(\bx_{2b}^{(2)})}_F$ are interchanged.
For $N=2 \pmod{6}$, the mapping is given by,
\ba
&\vev{c_6(\bx_{1a})}_F \leftrightarrow \vev{c_6(\bx_{1a})}_F,
\nn
&\vev{c_3(\bx_{1a})}_F \leftrightarrow \sum_{\theta\in\left\{0,\pm\frac{2\pi}{3}\right\}}\vev{c_3(\bx_{1a})}_F,
\nn
&\vev{c_3(\bx_{2b}^{(1)})}_F \leftrightarrow \sum_{\theta\in\left\{0,\pm\frac{2\pi}{3}\right\}}e^{-i\theta}\vev{c_3(\bx_{1a})}_{F}^{(\theta,-\theta)},
\nn
&\vev{c_3(\bx_{2b}^{(2)})}_F \leftrightarrow \sum_{\theta\in\left\{0,\pm\frac{2\pi}{3}\right\}}e^{i\theta}\vev{c_3(\bx_{1a})}_{F}^{(\theta,-\theta)},
\nn
&\vev{c_2(\bx_{3c})}_F \leftrightarrow \vev{c_2(\bx_{3c})}_F.
\label{eq:C6_Mapping_2_supp}
\ea
For $N=-2 \pmod{6}$, the mapping are identical to Eq.~\eqref{eq:C6_Mapping_2_supp} with the exception that the mappings for $\vev{c_3(\bx_{2b}^{(1)})}_F$ and $\vev{c_3(\bx_{2b}^{(2)})}_F$ are interchanged.
Finally, for $N=\pm 3 \pmod{6}$,
\ba
&\vev{c_6(\bx_{1a})}_F \leftrightarrow \vev{c_6(\bx_{1a})}_F,
\nn
&\vev{c_3(\bx_{1a})}_F \leftrightarrow \vev{c_3(\bx_{1a})}_F,
\nn
&\vev{c_3(\bx_{2b})}_F \leftrightarrow \vev{c_3(\bx_{2b})}_F,
\nn
&\vev{c_2(\bx_{3c}^{(1)})}_F \leftrightarrow \sum_{\theta_1\in\{0,\pi\}}\sum_{\theta_2\in\{0,\pi\}}e^{i\theta_1}\vev{c_2(\bx_{1a})}_{F}^{(\theta_{1},\theta_{2})},
\nn
&\vev{c_2(\bx_{3c}^{(2)})}_F \leftrightarrow \sum_{\theta_1\in\{0,\pi\}}\sum_{\theta_2\in\{0,\pi\}}e^{i\theta_2}\vev{c_2(\bx_{1a})}_{F}^{(\theta_{1},\theta_{2})},
\nn
&\vev{c_2(\bx_{3c}^{(3)})}_F \leftrightarrow \sum_{\theta_1\in\{0,\pi\}}\sum_{\theta_2\in\{0,\pi\}}e^{i(\theta_1+\theta_2)}\vev{c_2(\bx_{1a})}_{F}^{(\theta_{1},\theta_{2})}.
\label{eq:C6_Mapping_3_supp}
\ea
%

%%%%%%Refs
%\bibliographystyle{apsrev}
\bibliography{ref.bib}
%%%%%%